\documentclass[pre,groupedaddress,preprintnumbers]{revtex4-2} %add twocolumn
\usepackage{graphicx}% Include figure files
\usepackage{dcolumn}% Align table columns on decimal point
\usepackage{bm}% bold math
\usepackage{hyperref}% add hypertext capabilities
\usepackage{xcolor}
\usepackage{amsmath}
\usepackage{amsfonts}
\usepackage[font=small,labelfont=bf]{caption} % Required for specifying captions to tables and figures
\DeclareUnicodeCharacter{2212}{-}
\usepackage{ulem}

\begin{document}

%\preprint{}
\preprint{Phys. Rev. E}

\title{Gramian Angular Fields for leveraging pre-trained computer vision models\\ with anomalous diffusion trajectories}

\author{Òscar Garibo-i-Orts}
\email{oscar.garibo@campusviu.es}
\affiliation{GRID - Grupo de Investigación en Ciencia de Datos\\ Valencian International University - VIU, Carrer Pintor Sorolla 21, 46002 València, Spain.}% VRAIN -  Valencian Research Institute for Artificial Intelligence\\  Universitat Polit\`ecnica de Val\`encia, 46022 València, Spain.\\}
\altaffiliation[Also at ]{\\VRAIN -  Valencian Research Institute for Artificial Intelligence\\  Universitat Polit\`ecnica de Val\`encia, Camí de Vera s/n, 46022 València, Spain.}%Lines break automatically or can be forced with \\

\author{Nicolás Firbas}%
\email{Nicolas.Firbas@gmail.com}
\affiliation{%
DBS - Department of Biological Sciences,
National University of Singapore\\ 16 Science Drive 4, Singapore 117558, Singapore.
}%

\author{Laura Sebastiá}
\email{lsebastia@upv.es}
\affiliation{VRAIN -  Valencian Research Institute for Artificial Intelligence\\  Universitat Polit\`ecnica de Val\`encia, Camí de Vera s/n, 46022 València, Spain}

\author{J. Alberto Conejero}
\email{aconejero@upv.es}
\affiliation{Instituto Universitario de Matem\'atica Pura y Aplicada\\ 
Universitat Polit\`ecnica de Val\`encia, Camí de Vera s/n, 46022 Val\`encia, Spain.}

\date{\today}

\begin{abstract}
Anomalous diffusion is present at all scales, from atomic to large scales. Some exemplary systems are; ultra-cold atoms, telomeres in the nucleus of cells, moisture transport in cement-based materials, the free movement of arthropods, and the migration patterns of birds. The characterization of the diffusion gives critical information about the dynamics of these systems and provides an interdisciplinary framework with which to study diffusive transport.
Thus, the problem of identifying underlying diffusive regimes and inferring the anomalous diffusion exponent $\alpha$ with high confidence is critical to physics, chemistry, biology, and ecology.\medskip

Classification and analysis of raw trajectories combining machine learning techniques with statistics extracted from them have widely been studied in the Anomalous Diffusion Challenge (Muñoz-Gil et al., 2021). Here we present a new data-driven method for working with diffusive trajectories. This method utilizes Gramian Angular Fields (GAF) to encode one-dimensional trajectories as images (Gramian Matrices), while preserving their spatiotemporal structure for input to computer-vision models. This allows us to leverage two well-established pre-trained computer-vision models, ResNet and MobileNet, to characterize the underlying diffusive regime, and infer the anomalous diffusion exponent $\alpha$. Short raw trajectories, of lengths between 10 and 50, are commonly encountered in single-particle tracking experiments and are the most difficult to characterize. We show that by using GAF images, we can outperform the current state-of-the-art while increasing accessibility to machine learning methods in an applied setting.
\end{abstract}

%\keywords{}

\maketitle


\section{Introduction}

By themselves, computer vision models are unsuitable for sequence processing since the temporal structure of sequential data has few similarities with static images. An image of a dog represents a dog irrespective of its orientation. On the contrary, when dealing with sequences, like natural language, the order of words is highly relevant to the meaning of a sentence. Similarly, in a diffusive trajectory, the order in which features appear is highly relevant to the type of diffusion. For instance, imagine that we have a sequence of confined diffusion, where the existence of barriers slows down the particle's motion, which results in subdiffusion. If we play this sequence backward, it would appear as super diffusion because the particle would increase its mobility with time as it escapes the barriers. \medskip

Recurrent neural networks (RNNs) are widely used to combat the loss of temporal information when working with sequential data \cite{cheng2019novel_process,poznyak2019background_dynamical,hadian2021application_artificial,wu2020smart_manufacturing}. However, RNNs have a key disadvantage since they must be trained sequentially. This greatly increases model training time, particularly when working with very large training data sets. Following Vaswani \textit{et al.} 2017 \cite{vaswani2017attention}, Transformers have largely taken the place of RNNs in natural language processing (NLP), with increased performance over RNNs and the ability to be trained in parallel \cite{wolf2019huggingface}. However, there exists limited work in tokenizing and positionally encoding diffusive trajectories for use with Transformers, which prevents easy deployment of these models with one-dimensional data.\medskip

The difficulty and impracticality of designing, training, and tuning a custom machine learning (ML) model for a limited use scenario, where no pre-trained models are available, often discard the use of ML models in experimental settings. This has traditionally been true in diffusive studies, where the entry barrier to deploying ML models to characterize anomalous diffusion largely kept them out of the field. In fact, until the Anomalous Diffusion Challenge (AnDi Challenge\footnote{http://www.andi-challenge.org}) 2020 \cite{munoz-gil2021objective}, it was widely speculated that traditional statistical analysis could outperform ML-based methods. For instance, it took several weeks for our group to develop our ML model (ConvLSTM) based on convolutional neural networks (CNN) \cite{lecun1990handwritten}, and Long Short-Term Memory networks (LSTM) \cite{hochreiter1997LSTM}. While our model produced excellent results, it is easy to see how such a time-consuming process could prevent the deployment of a similar solution.\medskip

The AnDi challenge showed that short trajectories were the most difficult to work with both in classifying the underlying diffusive model and in inferring the anomalous diffusion exponent $\alpha$. In this work, we will address this difficulty directly by working with short trajectories of lengths between 10 and 50. By working with these short trajectories, we will show that using Gramian Angular Fields (GAFs) to leverage pre-trained computer vision models natively available in Keras. We can outperform state-of-the-art custom models in a way that is accessible to those without an extensive machine learning background.\medskip

The paper is organized as follows: In Section \ref{sec:anomalous_diffusion} we review some preliminaries about anomalous diffusion, give examples in which it can be appreciated in experiments, and summarize some existing machine learning models used for analysing the diffusive nature of the trajectories. In Section \ref{sec:Gramian_Angular_Fields}, we show how to convert a trajectory into different image representations through Gramian Angular Fields. Then, in Section \ref{sec:methodology} we describe the generation of the trajectories for training and validation of the models. Later, in Section \ref{sec:results}, we compare the results of the new GAF fed computer vision models to our previous ConvLSTM model presented in the Andi Challenge \cite{garibo-i-orts2021efficient}, and we will benchmark the best of these new models using the AnDi Interactive tool \footnote{\url{http://andi-challenge.org/interactive-tool/}}. Finally, we draw some conclusions in Section \ref{sec:conclusions}.\medskip

\section{Anomalous Diffusion}
\label{sec:anomalous_diffusion}
\subsection{Introduction to Anomalous Diffusion}
\label{sec:intro-anomalous-diffusion}
In 1827, Brown discovered that pollen grains placed in a fluid would move randomly and diffuse throughout the medium \cite{brown1828brief}. Due to the stochasticity of the movement, the probability $P(x,t)$ of finding a particle at time $t$ and position $x \in \mathbb{R}^d,\; d = 1, 2, 3$ is used to determine the dynamics of the particle. A traditional metric, relying on the underlying stochastic nature of diffusion, is the \textit{Mean Square Displacement} (MSD). When multiples trajectories are available, the MSD is defined as the average width or variance of the trajectories $\text{MSD}(0, t) \equiv \langle x(t)-x(0)\rangle^2$ with respect to two points in time. The MSD is assumed to be taken with respect to a time $t$ and the initial time $t=0$, as such, it is commonly abbreviated as $\langle x^2 \rangle $. However, if no ensemble of trajectories is available, and the available trajectory(s) is/are long, then the MSD can be approximated by the time average MSD (taMSD) along each trajectory. Where taMSD for a time interval $\tau$ is given by 

\begin{equation}
    \text{taMSD}(\tau) = \lim_{T \to \infty} \frac{1}{T - \tau} \int_{0}^{T-\tau} (x(t + \tau) - x(t))^2 dt.
    \label{eq:taMSD}
\end{equation}

If MSD grows linearly with time, $\langle x^2 \rangle \sim t$, then we say that the system diffuses normally. In contrast, anomalous diffusion happens when the MSD does not grow linearly with time, $\langle x^2 \rangle \sim t^\alpha$  with $\alpha \neq 1$, where $\alpha$ is known as the anomalous diffusion exponent.\medskip

The diffusive behavior of a particle is known to vary greatly with $\alpha$. As such, it has commonly been used to characterize anomalous diffusion. There are two kinds of anomalous diffusion; sub-diffusion, when $0< \alpha <1$, and super-diffusion, when $\alpha > 1$. At the lower end of sub-diffusion, $\alpha$ close to zero, we have immobile trajectories. When $\alpha$ is small, the width of the probability density function (PDF) $P(x,t)$ governing a particle's displacements becomes small. With the subsequent loss of variance between displacements, we have particle arrestation. At the upper end of the super-diffusive exponent range, we have what is referred to as ballistic motion, for $\alpha = 2$, and hyper-ballistic motion for $\alpha > 2$ \cite{flekkoy2021_hyperballistic}. Ballistic motion is characterized by unimpeded movement in a single line. In this work, we will only consider the case $0 < \alpha < 2$.\medskip

The exponent $\alpha$ alone cannot be used to characterize a trajectory. While a change in $\alpha$ does indicate a different diffusive pattern, it is possible to have different underlying diffusive behaviors with the same anomalous diffusion exponent. For example, messenger RNA (mRNA) trajectories in a living \textit{E. coli} cell can have very similar anomalous diffusion exponents while having distinct trajectories \cite{metzler2014anomalous}. As such, there exists a need to be able to further describe a trajectory beyond its $\alpha$.\medskip

The need to further describe the underlying movement of a subject undergoing diffusion inspired the classification task of the AnDi challenge \cite{munoz-gil2021objective}. Classification of a trajectory based on similarity to a well-understood underlying diffusive regime can give more detailed information, such as its ergodicity, than knowing the MSD or anomalous diffusion exponent $\alpha$ alone. In the context of diffusion, a process is said to be ergodic if the behavior of a single trajectory summarizes the whole system. 
This informs us about a possible difference between the local behavior and global behavior of a particle. If a trajectory is non-ergodic, researchers may find it important to investigate the local behavior of the particle further to discover the source of the ergodicity breaking. In movement ecology, ergodicity breaking can result from variation among individuals, changes in behavior in the same individual, or the inherent heterogeneity of the landscape where resources are not evenly distributed; see for instance \cite{Vilk2022_ergodicity_breaking_avian}.\medskip

Following the precedent set by the AnDi Challenge, and in order to facilitate benchmarking, we will consider five underlying diffusive regimes, which we have summarized below. Further details regarding the computational implementation of these models can be found in \cite{munoz-gil2021etai, munoz-gil2021objective}.

\begin{itemize}
    \item \textit{Continuous-Time Random Walk} (CTRW) happens when a particle motion can be described as a sequence of displacements sampled from a Gaussian distribution with zero mean. The waiting times between displacements are sampled from a power-law distribution $\psi(t)=t^{-\sigma}$ \cite{scher1975anomalous}.
    \item \textit{Lévy Walk} (LW) can be considered a special case of CTRW where dispersal lengths are correlated with waiting times. The probability density function (PDF) that describes the random time intervals between successive jumps is a power-law distribution $\psi(t)\sim t^{-\sigma-1}$ (as in CTRW), and the probability of a dispersal of length $\Delta x $ at time $t$ is $\Psi(\Delta x, t) = \frac{1}{2} \delta(|\Delta x| - vt)\psi(t)$ \cite{munoz-gil2021objective, klafter1994levy}.
    \item \textit{Annealed Transient Time Motion} (ATTM) occurs when a particle undergoes Brownian motion, with Diffusivity coefficient $D_i$ for an interval $t_i$. The length of the interval $t_{i}$ varies with $D_i$ such that $P_{t_i}(t_i|D_i)$ has mean $E[t_i|D_i] = D^{-\gamma}$. Thus, for $\gamma>0$, we would expect to see longer periods of low dispersal Brownian motion punctuated by shorter periods of high dispersal \cite{massignan2014nonergodic}.    
    \item \textit{Fractional Brownian Motion} (FBM) is defined by the Langevin equation, which is the stochastic differential equation governing the movement of a single particle with stochastic noise driving its movement. In this case noise is not white (or fractional Gaussian noise) and follows a normal distribution with zero mean and power-law correlations so that it has two regimes depending on the sign of the correlation. It is subdiffusive for negative correlations ($0 < \alpha < 1$), superdiffusive for positive correlations ($1 < \alpha < 2$), and converges to Brownian motion if no correlation occurs ($\alpha = 1$) \cite{mandelbrot1968fractional,jeon2010fractional}.
    \item \textit{Scaled Brownian Motion} (SBM) also derives from the Langevin equation, but in this case, diffusivity depends on time, even with white Gaussian noise \cite{lim2002self-similar}.
\end{itemize}

\subsection{Anomalous diffusion in experiments}
\label{sec:anomalous diffusion_experiments}
Anomalous diffusion happens in a broad range of experimental situations at all scales \cite{oliveira2019anomalous,klafter2005anomalous}. At the lowest scale, anomalous diffusion has been stated in experiments with ultra-cold atoms \cite{sagi2012observation,dechant2019continous, kindermann2017nonergodic}. Experiments also show anomalous diffusion in biological systems such as the telomers' motion in the cell's nucleus, where transient anomalous diffusion happens \cite{krapf2019spectral,bronstein2009transient,stadler2017non-equilibrium}. A revolution in cell biology has been boosted by the developments achieved in single particle tracking techniques \cite{manzo2015review} with experiments finding anomalous diffusion in the cytoplasm \cite{caspi2000enhanced,weber2010bacterial,regner2013anomalous} and in the plasma membrane \cite{weigel2011ergodic,manzo2015weak}.\medskip

One can also find anomalous diffusion in bigger systems such as living yeast cells \cite{tolic-norrelykke2004anomalous}, worm-like micellar solutions \cite{jeon2013anomalous}, water in porous biological tissues \cite{ozarslan2006anomalous,magin2013anomalous}, and cement based materials \cite{zhang2020dual-permeability}. Anomalous diffusion can even be observed in the migration patterns of storks between Africa and Europe. Tail winds on the storks' return to Africa speed up their journey giving a higher $\alpha$ for the trajectory from Europe to Africa than from Africa to Europe \cite{vilk2021unravelling}. Thus, given an experiment, it is of paramount importance to characterize the model behind the data that best explain it and to infer the associated exponent $\alpha$. For example, there is an ongoing discussion about the ergodicity and diffusion models at experiments in \cite{tolic-norrelykke2004anomalous,magdziarz2009fractional, golding2006physical,magdziarz2011anomalous,he2008random,molina-garcia2016fractional}.\medskip 

%change the order of subsequent paragraphs put ergodicty explanation here

In the context of diffusion, %\sout{a process is said to be ergodic if the behavior of a single trajectory can be said to summarize the whole system. More concretely} 
a system is ergodic if $\text{taMSD}(\tau) = \text{MSD}(t, t+\tau)$. In experiments, the possibility to perform time averages when taking the limit as $T$ tends to $\infty$ in \eqref{eq:taMSD} does not exist, therefore we consider a discretized approximation to Eq. \ref{eq:taMSD} as 
\begin{equation}
taMSD(\tau) \approx \frac{1}{N-\tau} \sum_{t=1}^{N-\tau} (x(t + \tau) - x(t))^2, 
\end{equation}
which will always sum to a large enough fixed value $N$.\medskip

It should be noted that for non-stationary processes, the position of a particle at time $t$, $x(t)$, depends on $t$. By considering dispersal over an interval, as in taMSD, we remove the dependence on $t$. As a result, taMSD$(\tau)$ cannot converge to all MSD$(t, t+\tau)$ and is said to be trivially non-ergodic \cite{thiel2013-weak-ergodicity-breaking}.Another special case of ergodicity breaking is where taMSD and MSD differ by a constant factor. This is referred to as ultra-weak ergodicity breaking\cite{godec2013finite-time,godec2013linear}. As we can see, though the concept of ergodicity is simple, determining whether ergodicity is present in an experimental setting is complicated by the lack of data availability, which can be limited both in terms of the number of trajectories and their lengths. As such, using our ML models to classify trajectories based on their underlying diffusive regime, with studied characteristics such as ergodicity, can be convenient to obtain more information about a process under real-world constraints.\medskip

The five diffusive processes explained in Section \ref{sec:intro-anomalous-diffusion} have different ergodic properties. We know that CTRW, ATTM, and SBM show weak ergodicity breaking \cite{massignan2014nonergodic,manzo2015weak}, whereas Brownian motion and FBM are ergodic (the ergodicity of FBM requires a closer analysis \cite{deng2009ergodic,schwarzl2017quantifying,mardoukhi2020spurious}). Thus, understanding the underlying diffusive model of a process is important as it can help direct further study. For instance, if a trajectory follows ATTM, it will be locally Brownian, with sharp changes in the diffusivity coefficient $D$. In the context of research, it may be of interest to study the cause of the shifts in $D$, as this may shed light on an important behavior.
\medskip 

With regard to signals resulting from Single Particle Tracking (SPT), it is important to remember that these experimental signals are inherently noisy and have localization error \cite{chenouard2014objective}. This noise is problematic, as it has been known to hide non-ergodic behavior \cite{jeon2013noisy} and interfere with statistical analyses. More broadly, the difficulties resulting from the realities of experimentation --short trajectories, noisy trajectories, and few replications-- necessitate versatile tools, like those based on ML, to characterize diffusion in experiments.\medskip

\subsection{Machine Learning and Anomalous Diffusion}
The difficulty of characterizing anomalous diffusion has given rise to diverse statistical methodologies to infer the anomalous exponent for a given trajectory. To name a few: Bayesian estimation for FBM processes \cite{benmehdi2011bayesian,hinsen2016communication}; a statistical inference approach for finding interactions between moving particles \cite{agliari2020statistical}; a method based on fractionally integrated moving averages \cite{burnecki2015estimating}; and a method based in the information contained in the power spectral density of a trajectory \cite{krapf2018power,krapf2019spectral}. Similarly, statistical methods have been used to discriminate between different diffusion models. For example, Bayesian methods to distinguish between Brownian motion, SBM, and FBM in \cite{thapa2018bayesian}, and other methodologies to distinguish between FBM and CTRW can be found in  \cite{magdziarz2009fractional, jeon2010analysis}.\medskip

%\sout{The disadvantage of such statistical methods, both for the inference of $\alpha$ and determination of the underlying regime, is that they are specific to a particular type of diffusion. Machine learning-based inference and characterization methodology can be more flexible in that, since they can be applied to trajectories without prior knowledge of the underlying diffusion regime. However, such flexibility comes with the cost of lower interpretability and explainability (black box effect). Thus, when possible, we should strive to use well-established statistical methodology as we are able to make additional inferences based on these metrics}\medskip

The new wave of ML methods can be seen as a direct response to the realities of experimental design, current Single Particle Tracking (SPT) technology, and the inherent noise that exists in nature. Some of the first works to use ML for the characterization of anomalous diffusion are based on random forests (RFs), which allowed for some model interpretability. In \cite{wagner2016classification} RFs were used to discriminate between direct motion, normal, and anomalous diffusion. Also, RFs were able to classify trajectories as CTRW, ATTM, FBM, or LW and to infer the exponent $\alpha$ \cite{munoz-gil2020single}, and RFs together with gradient boosting trees were used in \cite{janczura2020classification,loch-olszewska2020impact} to classify among normal, super and sub-diffusive trajectories. In \cite{kowalek2019classification}, convolutional neural networks were used to classify trajectories as normal diffusion, anomalous diffusion, directed motion, or confined motion and these results were compared with RFs and gradient boosting trees.\medskip

Beyond explanatory machine learning approximations, we also find works which use a combination of classical statistics analysis and supervised deep learning (a deep feed-forward neural network to cluster parameters extracted from the statistical features of individual trajectories) are used to classify among the aforementioned five diffusive models, and to infer the anomalous exponent $\alpha$ \cite{gentili2021characterization}. Some other approaches to both problems lay in using deep learning methods, mainly based on LSTMs \cite{bo2019measurement, argun2021classification, garibo-i-orts2021efficient}. The role of the LSTM layers can be exchanged with Transformers, as it has been recently shown in \cite{firbas2022transformers} with very competitive results for short trajectories.\medskip

%%%%%%%%%%%%%%%%%%%%%%%%%%%%%%%%%%%%%%%%%%%%%%%%%%%%%%%%%%%%%%%%%%%%%%%%%%%%%%%%%%%%%%%%%%%
\section{Gramian Angular Fields}
\label{sec:Gramian_Angular_Fields}
First proposed by J.P. Gram \cite{gram1883ueber}, Gramian Angular Fields (GAFs) provide a methodology for converting a time-series, sequence, or vectors, to a matrix representation while retaining the existing spatial and temporal relations between the terms. GAFs were first used, together with Markov Transition Fields, to feed computer vision models in Wang et al. 2015, where they were used for time series classification, and imputation \cite{wang2015imaging}. Since then, GAFs have been extensively used to perform diverse tasks such as: Forecasting day-ahead solar irradiation \cite{hong2020day-ahead}, prediction of the myocardial infarction risk from electroencephalogram signals \cite{Zhang2019Automated}, epilepsy detection
\cite{Palani2020Implementation},
classification of human activity from sensor data \cite{Xu2020HumanActivity}, analysis of near infra-red spectroscopy signals \cite{Wickramaratne2021ADeep,LIU2022Determination}, and removal of motion artifacts in photoplethysmograph sensors \cite{Liu2020MotionArtifact}.
However, to the best of our knowledge, this is the first use of GAF using computer vision models with anomalous diffusion trajectories.\medskip

\begin{figure*}
\includegraphics[width =.8\textwidth]{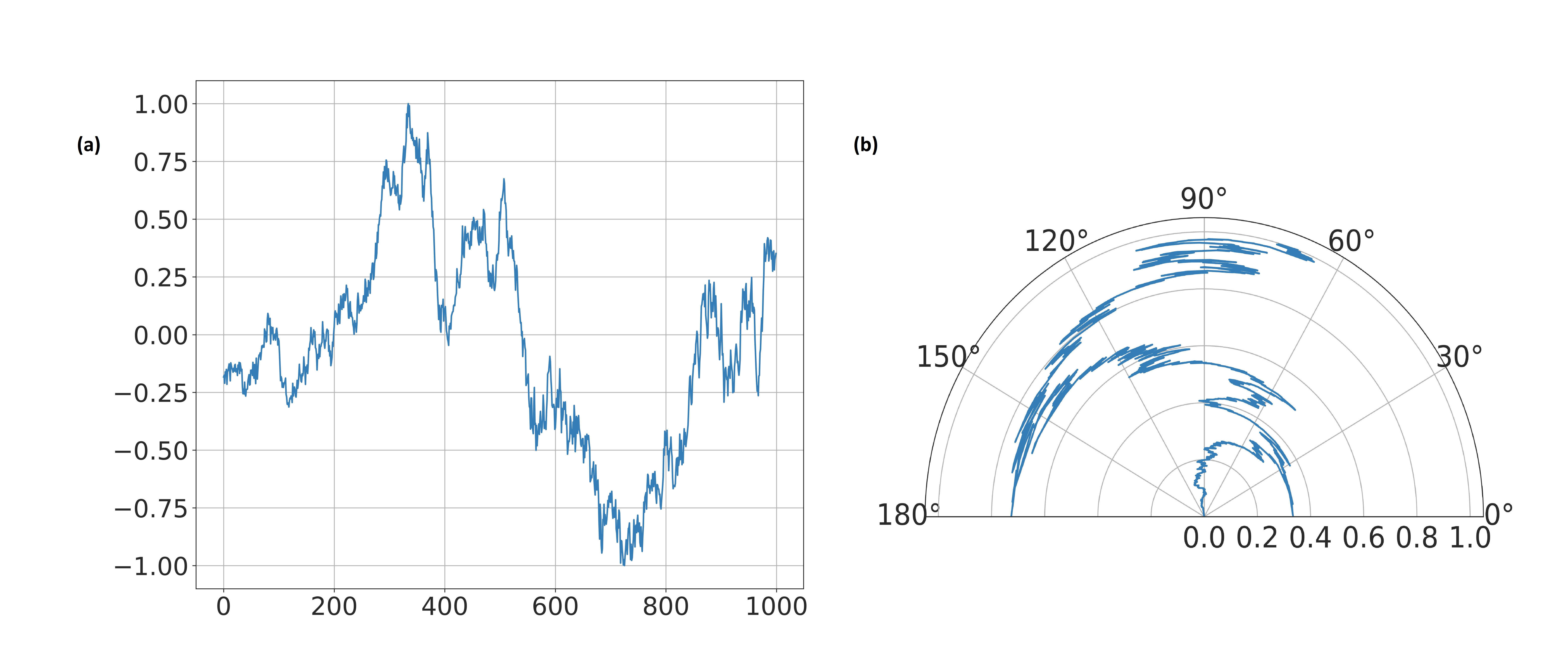}
\caption{A normalized super diffusive L\'evy flight trajectory (a). (b) The polar encode of the trajectory in (a).}
\label{fig:polar-encode}
\end{figure*}

In order to encode a time series ${\bf x} = \{x_{1}, x_{2}, x_{3}, \ldots, x_N\}$ of length $N$,  we first normalize ${\bf x}$, as  ${\bf \tilde x}$ such that all its values belong to $[-1, 1]$, see Eq. \ref{eq:normalization}.\medskip

\begin{equation}
    \label{eq:normalization}
    {\tilde x_i} = \frac{(x_{i} - \max({\bf x})) + (x_i - \min({\bf x}))}{\max ({\bf x}) − \min ({\bf x})},\; 1\le i\le N.
\end{equation}

Once normalized, we convert each element in the sequence to polar coordinates. These are encoded by the angular cosine and are stored in $\boldsymbol{\phi}$. Likewise, the temporal positions of each value are stored in the radius $\boldsymbol{r}$ of the polar coordinates\medskip

\begin{equation}
    \label{eq:polar_coordenates}
    \begin{cases}
    \phi_i = \arccos(\tilde{x_{i}}),\;\tilde{x_i} \in {\bf \tilde x} \\ 
    r_i = \frac{i}{N},\; 1\le i\le N,
    \end{cases}
\end{equation}
where $N$ ensures that all $r_i \in [0,1]$. As time increases, the values twist around the origin, as seen in Figure \ref{fig:polar-encode}(b). This way of encoding a time series has two important properties: (1) it is bijective since $\cos(\phi_i))$ is monotonic when $\phi_i \in [0, \pi]$ and (2) the polar coordinates system preserves absolute temporal relations in contrast to Cartesian coordinates.\medskip

We can benefit from the angular perspective by considering the trigonometric sum/difference between each pair of points in the sequence to identify the temporal correlation within different time intervals. This yields two representations known as \textit{Gramian Angular Summation Field} (GASF) \ref{eq:gasf} and \textit{Gramian Angular Difference Field} (GADF) \ref{eq:gadf}

\begin{equation}\textnormal{GASF} =
\begin{bmatrix} 
    \cos(\phi_1 + \phi_1)  & \dots & \cos(\phi_1 + \phi_N)\\
    \cos(\phi_2 + \phi_1)  & \dots & \cos(\phi_2 + \phi_N)\\
    \vdots & \ddots & \vdots\\
    \cos(\phi_N + \phi_1) & \dots  & \cos(\phi_N + \phi_N)
\end{bmatrix}
\label{eq:gasf}
\end{equation}

\begin{equation}\textnormal{GADF} =
\begin{bmatrix} 
    \sin(\phi_1 - \phi_1)  & \dots & \sin(\phi_1 - \phi_N)\\
    \sin(\phi_2 - \phi_1)  & \dots & \sin(\phi_2 - \phi_N)\\
    \vdots & \ddots & \vdots\\
    \sin(\phi_N - \phi_1) & \dots  & \sin(\phi_N - \phi_N)
\end{bmatrix}.
\label{eq:gadf}
\end{equation}

\begin{figure}[htbp]
 \centering
 \includegraphics[width=0.34\linewidth]{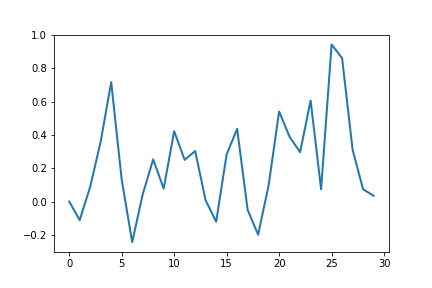}
% \text{(a). Raw trajectory of length 40.}
 \includegraphics[width=0.2\linewidth]{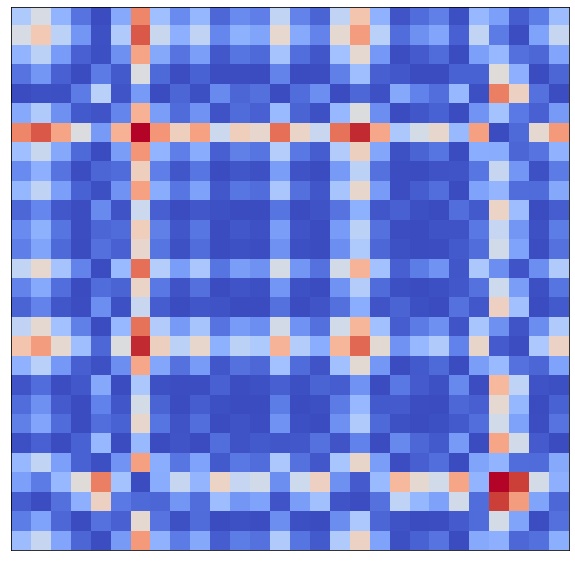}
% \text{(b). GASF representation of raw trajectory} % with $\alpha = 0.2$}
 \includegraphics[width=0.2\linewidth]{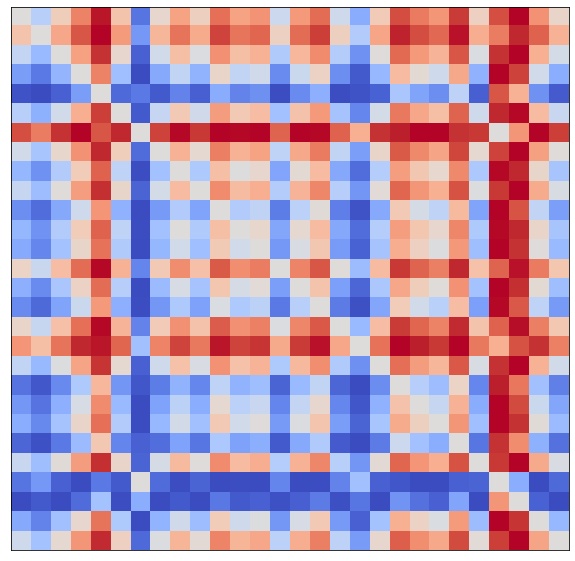}
% \text{(c). GADF representation of raw trajectory}
 %\centering
 \caption{A one-dimensional trajectory of length time 30 units (in a certain dimensionalized system scale) of a particle that is initially located in origin (left). The time steps from 0 to 29  are indicated on the x-axis and the displacement of the points in the trajectory is on the y-axis. In the GASF representation (middle), the peak on the first time step in the trajectory is converted into one strong vertical and one horizontal lines (times $t=3$ and $t=4$), as a result of the sum. In the GADF (right), we also have these stronger lines as a result of the difference, but in the rest of the picture, these differences are almost 0.}
 \label{fig:sequence_GAF_represenations}
 \end{figure}

\begin{figure}[htbp]
 \centering
 \includegraphics[width=.4\linewidth]{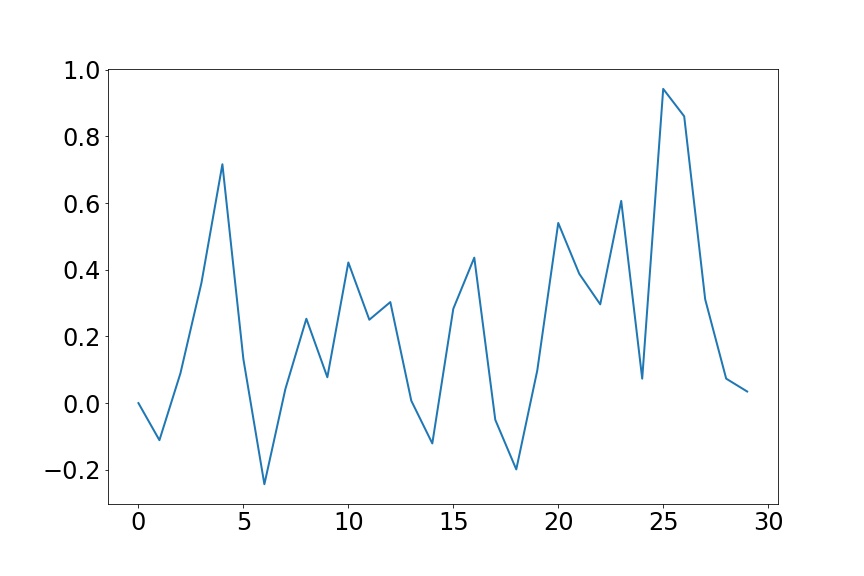}
% \caption{(a). CTRW trajectory with $\alpha = 0.2$}
% \text{(a). CTRW trajectory with $\alpha = 0.2$}
\hspace{0.2cm}
 \includegraphics[width=.4\linewidth]{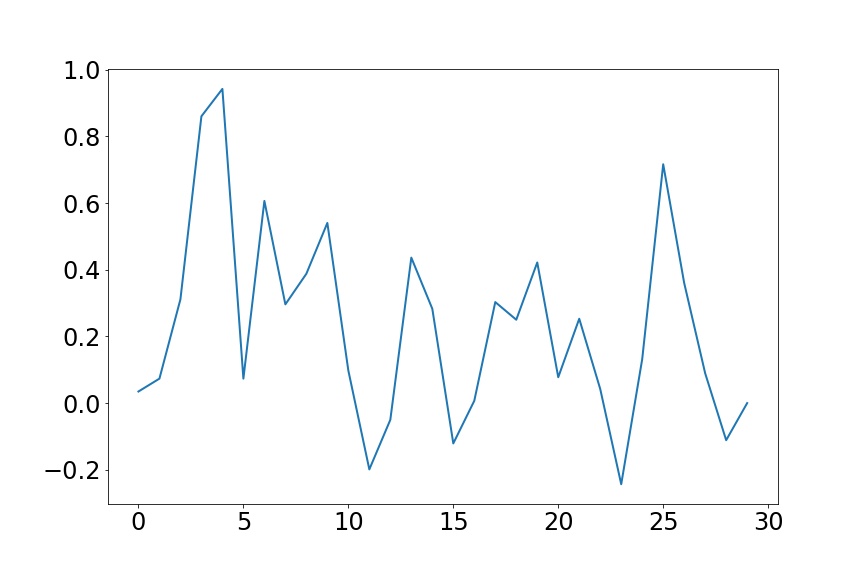} \\
% \text{(b). Reversed CTRW trajectory} % with $\alpha = 0.2$}
 \includegraphics[width=.32\linewidth]{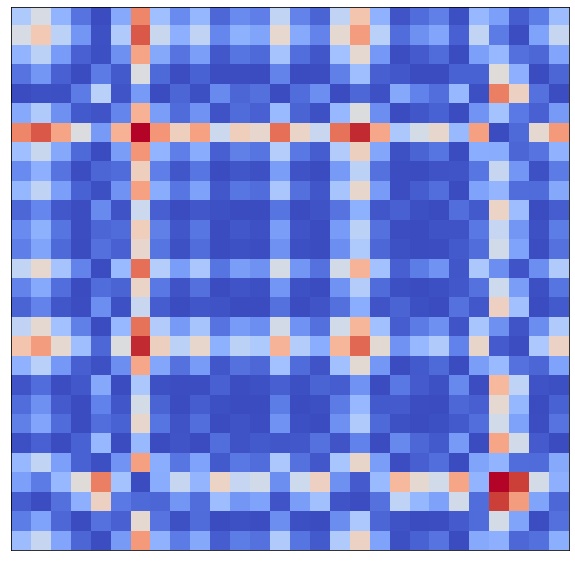}\hspace{1.6cm}
% \text{(c). GASF representation of (a).}
 \includegraphics[width=.32\linewidth]{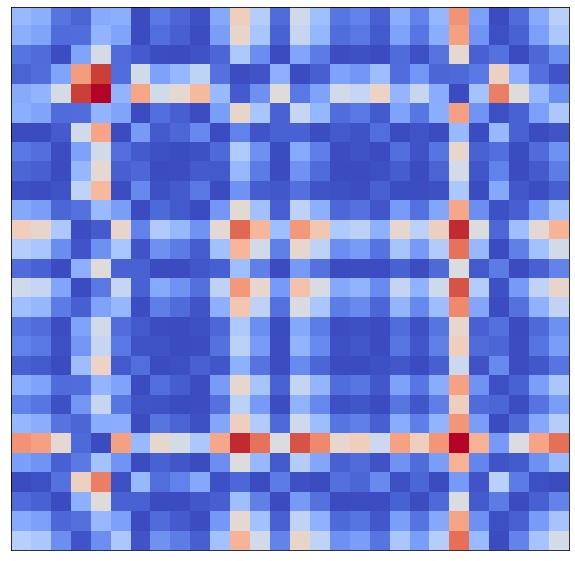}
 %\text{(d). GASF representation of (b).}
 %\centering
 \\
 \caption{In (upper left) we have a CTRW trajectory of length 30-time units and $\alpha = 0.2$. In (upper right), we have the same trajectory but reversed. In (lower left) and (lower right) we have the GASF image associated with the trajectory and its reversed copy. We see that these images are symmetric with respect to the main diagonal. Furthermore, we see that image (lower right) is a $\pi$ rotation of the image (lower right), indicating that time is encoded along the same main diagonal.}
 \label{fig:ctrw_and_reversed}
 \end{figure}

In Figure \ref{fig:sequence_GAF_represenations}, we show a CTRW in its raw format (left), and the GASF (middle) and GADF (right) representations of that sequence. These GASF and GADF representations are then used to train two computer vision models: ResNet and MobileNet. To show how GAF representations maintain spatiotemporal relations, in Figure \ref{fig:ctrw_and_reversed} we show a trajectory forward and backward in time with corresponding GASF representations. We can see that both images are symmetric with respect to the main diagonal and that the GASF matrices are 180-degree rotations of one another. This means that, for GAF representations, time is encoded along the main diagonal of the GAF matrix and that the rows and columns contain information about the spatial relations between the sequence terms.\medskip

To show how these representations maintain spatiotemporal relations, in Figure \ref{fig:ctrw_and_reversed}, we can see an example of a CTRW sequence (A), the same sequence backward (B), and their GASF representations (C) and (D) respectively. Figure \ref{fig:ctrw_and_reversed}(D) is a 180 degree rotation of Figure \ref{fig:ctrw_and_reversed}(C). Additionally, both images are symmetric with respect to the main diagonal. This tells us that, for both GASF and GADF, time is encoded along the main diagonal of the GAF matrix and the x and y axes contain the spatial relations between terms of the sequence.\medskip

%%%%%%%%%%%%%%%%%%%%%%%%%%%%%%%%%%%%%%%%%%%%%%%%%%%%%%%%%%%%%%%%%%%%%%%%%%%
\section{Methodology}
\label{sec:methodology}

In this work, we use Gramian Angular Summation/Difference Fields (GASF/GADF) to represent one-dimensional trajectories as images to facilitate the use of computer vision models for two tasks: (i) The classification task, which consists of classifying trajectories by one out of five, aforementioned, generating diffusion models ATTM, CTRW, FBM, LW, and SBM), and (ii) the regression task where we infer the anomalous diffusion exponent $\alpha$ of each trajectory. Each task is conducted independently. In particular, we will focus on short trajectories of lengths 10 to 50, since these trajectories are the hardest ones to classify \cite{munoz-gil2021objective} and should prove most useful for those who wish to apply our methodology.\medskip

The Python package \textit{pyts} \cite{faouzi2020pyts} provides a variety of different tools for time series classification. Of interest to us, is the \texttt{pyts.image} module which  includes the \textit{GramianAngularFields} that transforms times series into Gramian Angular Fields images. One can select to produce Gramian Angular Summation or Difference Fields just by assigning the correct value to the parameter \textit{method} when calling the function. The complete Python code to generate the images, train the models and obtain the predictions can be found on GitHub (\url{https://github.com/OscarGariboiOrts/Gramian-Angular-Fields}).\medskip

With the GASF/GADF images, we will train and validate two well-known models for dealing with the images ResNet \cite{he2015resnet} and MobileNet \cite{howard2017mobilenet}. ResNet was created to address a loss in accuracy as convectional networks become deeper \cite{he2015resnet}. On the other hand, as the name implies, MobileNet is a small and efficient convolutional architecture that was designed to work well in mobile computer vision deployments. MobileNet is composed of depth wise separable convolutions and has two hyper-parameters, a width multiplier and resolution multiplier \cite{howard2017mobilenet}.\medskip

\begin{figure}
	  	\centering	
	  	\includegraphics[width = .95\linewidth]{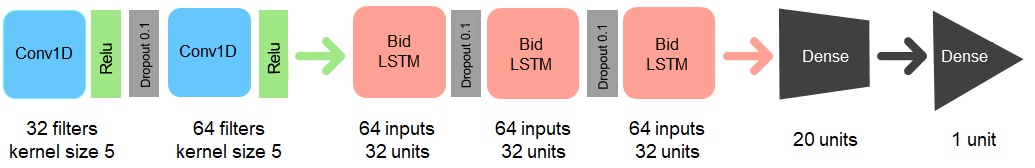}
	  	\caption{ConvLSTM model architecture used for anomalous diffusion analysis in \cite{garibo-i-orts2021efficient,munoz-gil2021objective}.}
	  	\label{fig:model}
	 \end{figure}

In order to benchmark our GASF/GADF fed Resnet and Mobilenet networks, we will primarily use the ConvLSTM method presented in \cite{garibo-i-orts2021efficient}, which placed in the top two models at the 2020 AnDi Challenge both in underlying diffusion model classification and $\alpha$ regression tasks in one dimension \cite{munoz-gil2021objective}. In short, this method combines two convolutional layers, three bidirectional LSTM layers, and a last dense layer (Figure \ref{fig:model}). Other models that present very good results can be found in \cite{gentili2021characterization}, \cite{argun2021classification} , \cite{kowalek2022boosting_performance}, \cite{gajowczyk2021detecting_anomalous} and \cite{firbas2022transformers}.\medskip

\subsection{Generation of training and validation datasets}

The quantity and quality of a data training quality greatly determine the performance of supervised machine-learning techniques. Thus, we require that our training set has enough samples of each of the five models and, simultaneously, covers the whole range of the anomalous exponent $\alpha$. In order to generate trajectories, we used the code provided by the organizers of the AnDi Challenge \cite{munoz-gil2021etai}, which is publicly available on GitHub at \url{https://github.com/AnDiChallenge}. Further details on how these trajectories are generated can be found in \cite{munoz-gil2021etai,munoz-gil2021objective}.\medskip

In order to train our GASF/GADF models, we generated two different training data sets consisting of $4\cdot 10^6$ trajectories each, of lengths ranging between 10 and 50. To ensure that image size is consistent across our GASF/GADF representations all trajectories were padded with zeros at the beginning of the trajectory to make them all length 50. The first data set was used for classification purposes and considered the five aforementioned classes (ATTM, CTRW, FBM, LW, and SBM) as labels. The second one was built for the $\alpha$ exponent regression with $\alpha \in [0.05, 1.95]$ with increments of 0.05. Not all underlying models are defined for the same range of $\alpha$. Trajectories of type FBM and SBM span the entire range of tested alpha values ($\alpha \in (0, 2)$) from sub to super-diffusion, ATTM and CTRW trajectories are sub to normally diffusive ($\alpha \in (0,1]$, finally, LW trajectories are normal and super-diffusive ($\alpha \in [1, 2)$). 
 In both classification and regression, we have considered trajectories with Gaussian noise. Let Gaussian noise have a standard deviation $\sigma_{noise}$, which is some portion of the standard deviation of the trajectory displacements $\sigma_D$. We then define the signal-to-noise ratio (SNR) of a trajectory as $\rm{SNR} = \frac{\sigma_D}{\sigma_{{\rm noise}}}$.\medskip
 %\sout{We will consider noisy trajectories with SNR = 1, that is $\sigma_{\text{noise}} = \sigma_D$}.\medskip

In each one of the regression and classification tasks, the training data sets were independently split into training (95\%) and validation (5\%) at each epoch. The models were trained until we got no improvement after ten consecutive epochs. Although these rates break the well-known 80/20 splitting rule of machine learning, the abundance of noisy trajectories prevents us from over-fitting and permits us to deploy a more robust model. Finally, we have also generated two other data sets of $10^4$ trajectories for testing the models and presenting the results.

%%%%%%%%%%%%%%%%%%%%%%%%%%%%%%%%%%%%%%%%%%%%%%%%%%%%%%%%%%%%%%%%%%%%%%%%%%%%%

\section{Results}
\label{sec:results}

\subsection{Diffusion model classification}
\label{sec:classification_results}
We recall that the classification task consists in predicting which model better explains each trajectory among five different classes (ATTM, CTRW, FBM, LW and SBM) detailed in Section \ref{sec:methodology}. 
%\sout{It is worth mentioning that the trajectories following models ATTM and CTRW are always sub-diffusive, with $\alpha$ in the interval $(0,1]$, while all the trajectories from model LW are super-diffusive, with $\alpha$ in \textcolor{red}{$[1,2)$}. Finally, trajectories from FBM and SBM models take $\alpha$ in the full range $(0,2)$}. 
Gaussian noise was added to each trajectory to investigate the effect of noise in the classification.
%\sout{Throughout the discussion of the classification problem, we will just show the results for $\rm{SNR}=1$. As expected, results will always be better for ${\rm SNR} = 2, (\sigma_{\text{noise}} = 0.5)$ than for ${\rm SNR} = 1, (\sigma_{\text{noise}} = 1)$. That is, trajectories with lower noise are more accurately identified than noisier trajectories.}
%\sout{The results and conclusions on the models' performance for $\rm{SNR}=2$, are similar and can be found in the Appendix.} \medskip

In order to study the classification performance of each of the tested models, we consider the \textit{F1-score} that is defined in Eq. \eqref{eq:F1-score}, as the harmonic mean of precision and recall.\medskip

\begin{equation}
    \textnormal{F1-score} = 2* \frac{{\rm precision} * {\rm recall}}{{\rm{precision} + \rm{recall}}
    \label{eq:F1-score}}
\end{equation}

We remember that the \textit{precision} measures the classifier's ability to correctly label positive samples; see Eq. \eqref{eq:precision},\medskip

\begin{equation}
{\rm precision} = \frac{\text{{\rm $\sharp$ correctly predicted instances}}}{\text{{\rm $\sharp$ predicted labels}}}
\label{eq:precision}
\end{equation}
and the \textit{recall} measures the classifier's ability to find all the positive samples, see Eq. \eqref{eq:recall}

\begin{equation}
{\rm recall} = \frac{\text{$\sharp$ correctly predicted instances}}{\text{$\sharp$ labels in the gold standard}}
\label{eq:recall}
\end{equation}

\begin{figure}[htbp]
 \centering
\includegraphics[width=7.5cm]{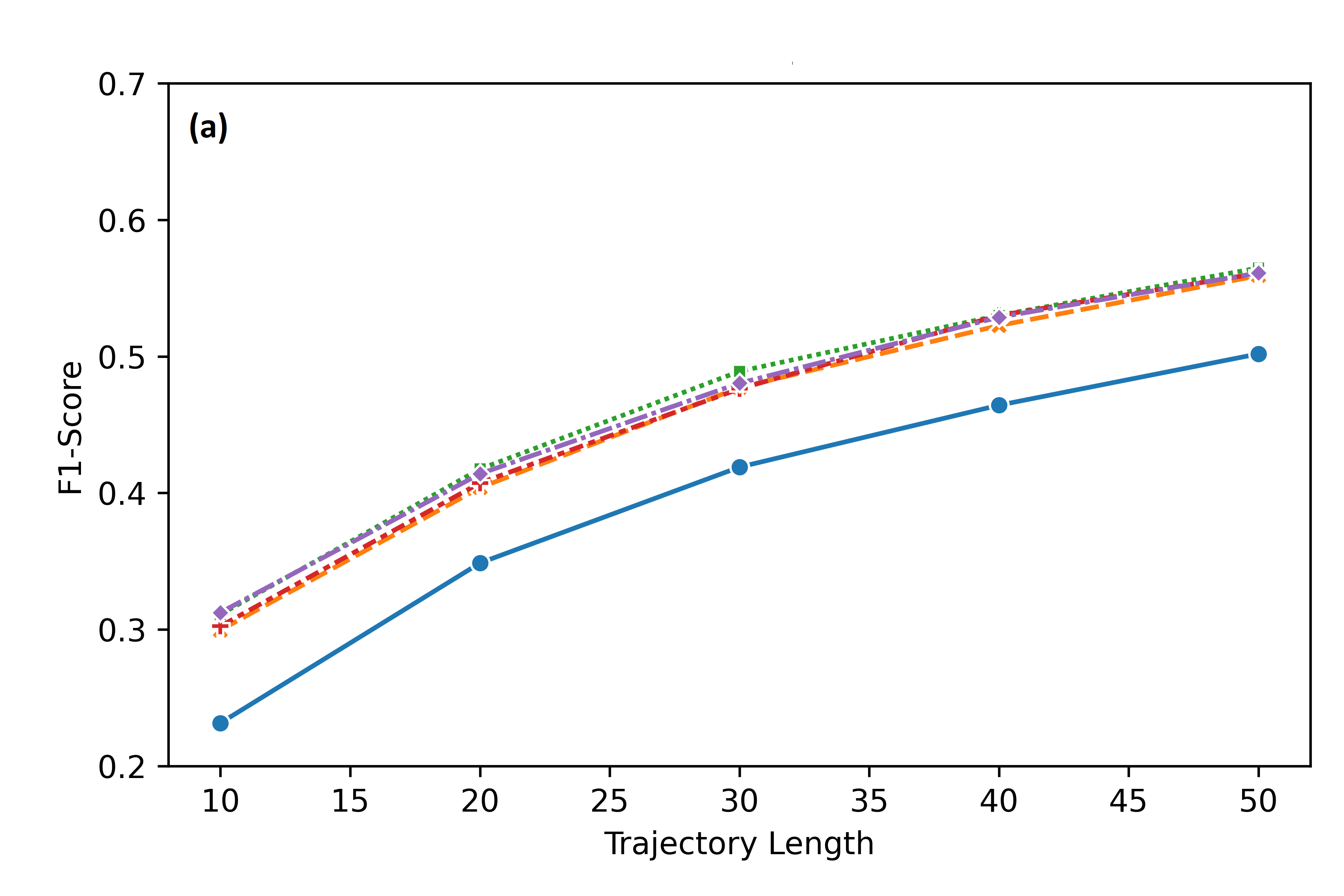}
\includegraphics[width=7.5cm]{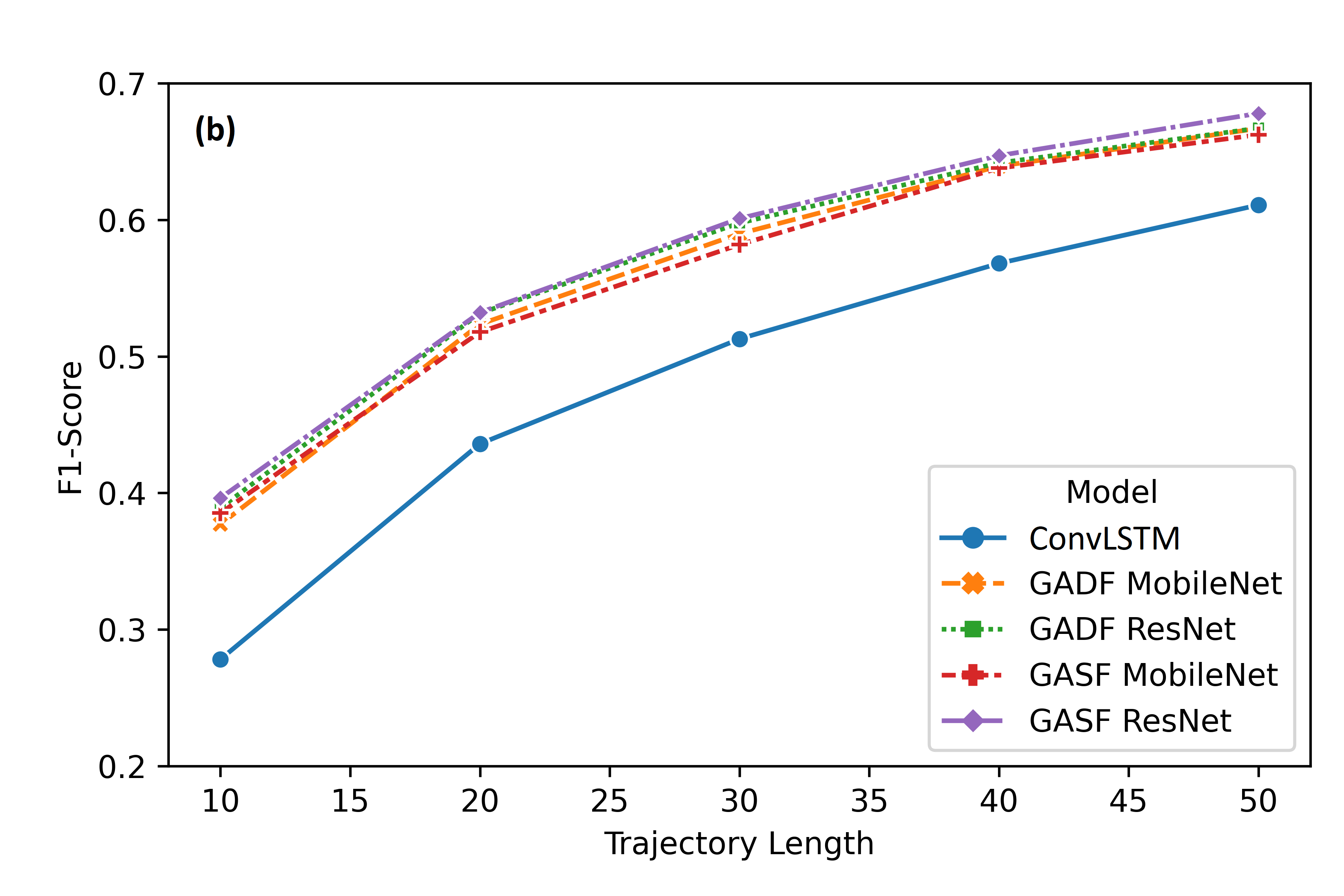}
\caption{F1-score is plotted against the trajectory length, with line color indicating performance for each of the ML models tested. In (left) we see the results for $\rm{SNR} = 1 $ and in (right) and the results for $\rm{SNR} = 2$.}
 \label{fig:classification_convlstm_vs_rest_snr1}
\end{figure}

It is worth mentioning that any of the ResNet/MobileNet combinations with GASF/GADF outperforms the benchmark ConvLSTM model in classification, for every trajectory length and noise level, as it can be seen in Figure \ref{fig:classification_convlstm_vs_rest_snr1}.
%\sout{and in the Appendix}. 
Performance increases with decreased noise, with no significant interactions. Thus, for the classification and regression tasks, we have chosen to focus primarily on our high noise (SNR $= 1$), that is $\sigma_{\text{noise}} = \sigma_D$, as this is more representative of a real-life deployment.\medskip

%formerly fig5 and fig5b

%\sout{In Figure \ref{fig:classification_convlstm_vs_rest_snr1} we plot the F1-score as a function of trajectory length and the ML model used. }

\begin{figure}[htbp]
 \centering
 % \begin{subfigure}{.48\textwidth}
 \includegraphics[width=7.5cm]{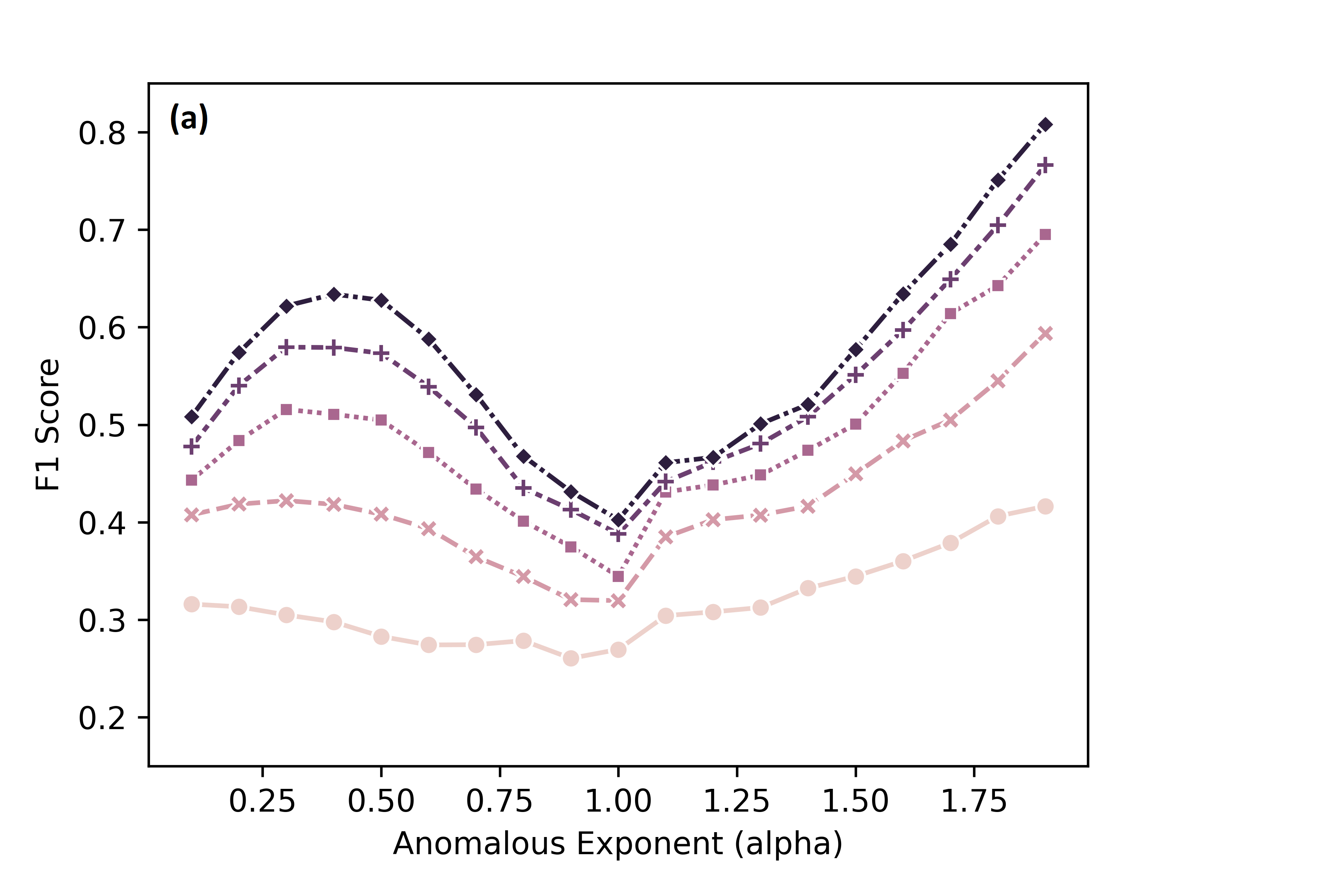}
%classification_figs/F1vsAlpha_hue_trajLength_SNR1_GASF ResNet.png
 %\caption{F1-scores for the GASF/Resnet model.}
% \end{subfigure}
% \begin{subfigure}{.48\textwidth}
 \includegraphics[width=7.5cm ]{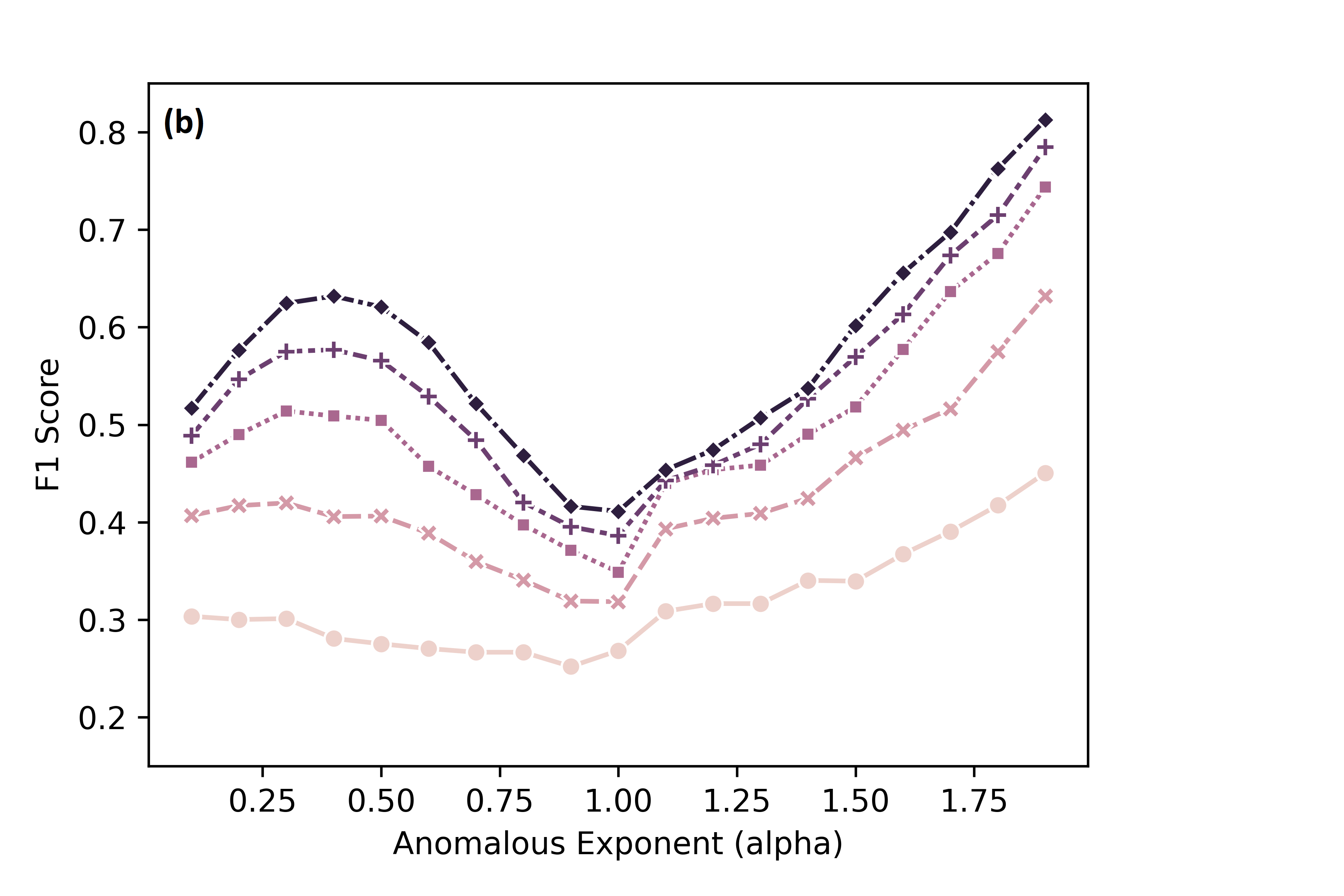}
%classification_figs/F1vsAlpha_hue_trajLength_SNR1_GADF ResNet.png
 %\caption{F1-scores for the GADF/Resnet model.}
% \end{subfigure}
%  \hspace{-1.33cm}
%  \begin{subfigure}{.48\textwidth}
 \includegraphics[width= 7.5cm]{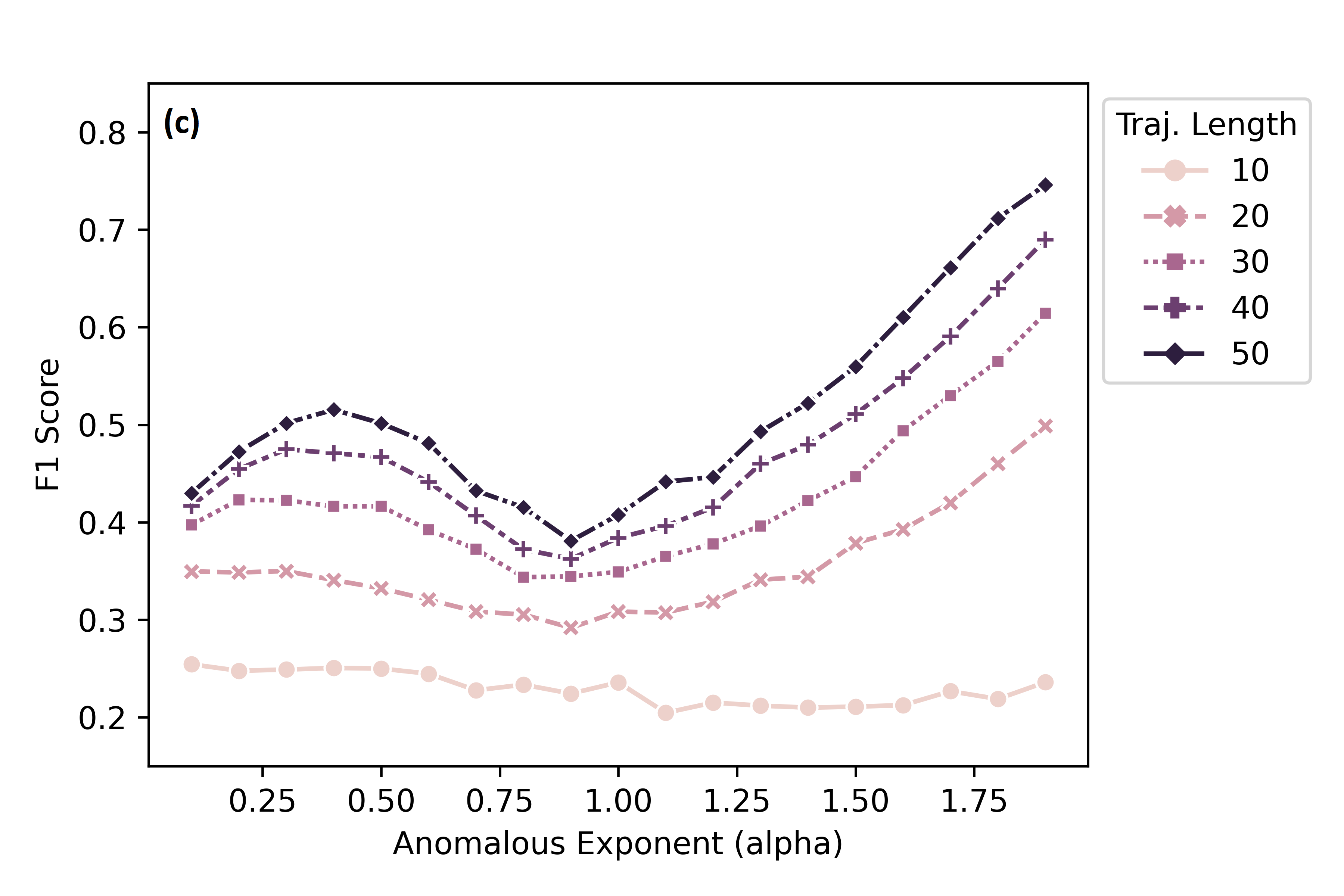}
%classification_figs/F1vsAlpha_hue_trajLength_SNR1_ConvLSTM.png 
% \end{subfigure}
 \caption{F1-scores for the GASF/ResNet (upper left), GADF/ResNet (upper right), and ConvLSTM (bottom) models for different trajectory lengths and values of the anomalous exponent $\alpha$ for noisy trajectories with ${\rm SNR} = 1$.}
 \label{fig:gadsf_lengths_snr1}
 \end{figure}

\begin{figure}[htbp]
 \centering
% \begin{subfigure}{width = .48\textwidth}
 \includegraphics[width= 7.5cm]{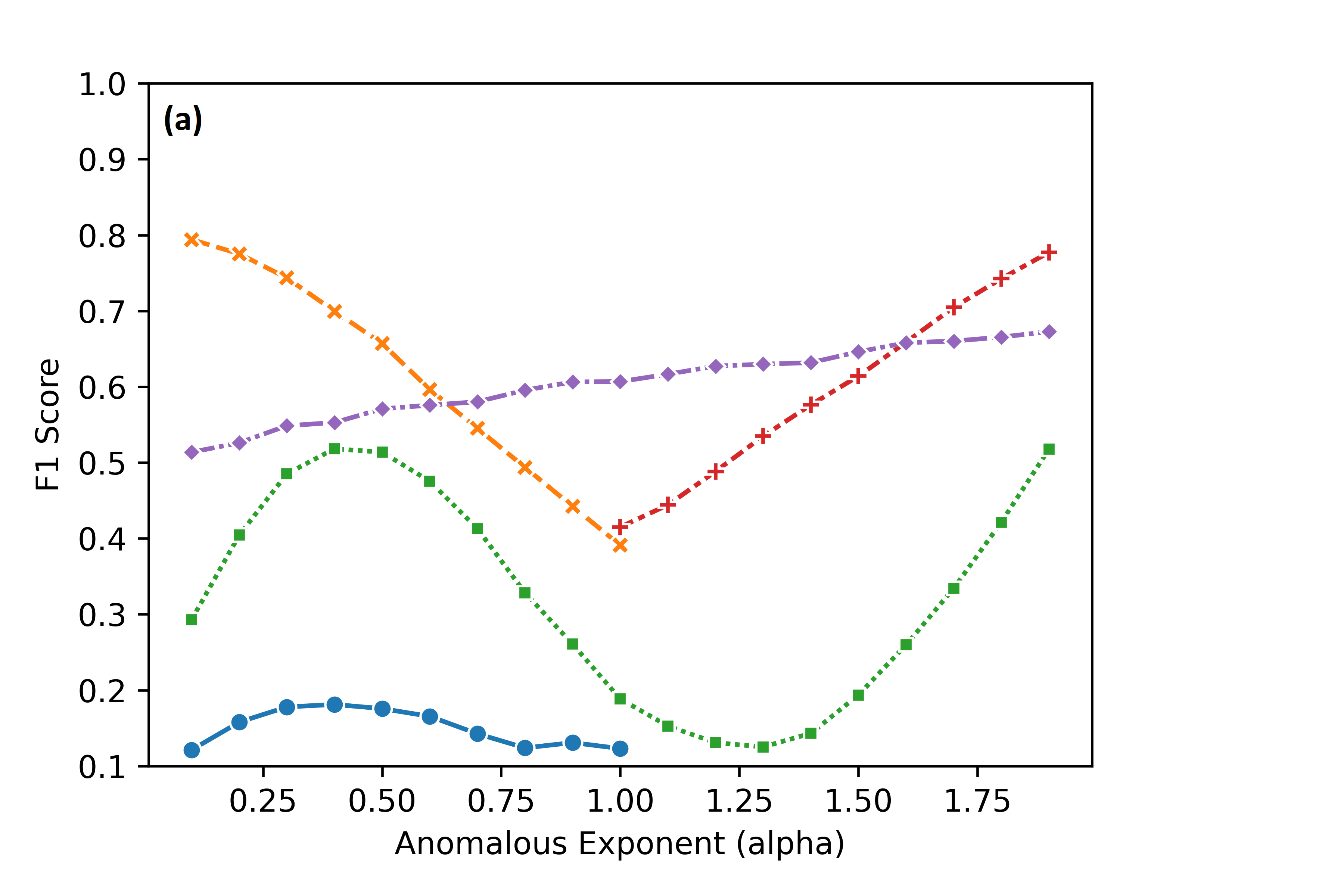}
% classification_figs/F1vsAlpha_hue_trajType_SNR1_GASF ResNet.png
 %\caption{F1-scores for the GASF/Resnet model.}
% \end{subfigure}
% \begin{subfigure}{width = .48\textwidth}
 \includegraphics[width=7.5cm]{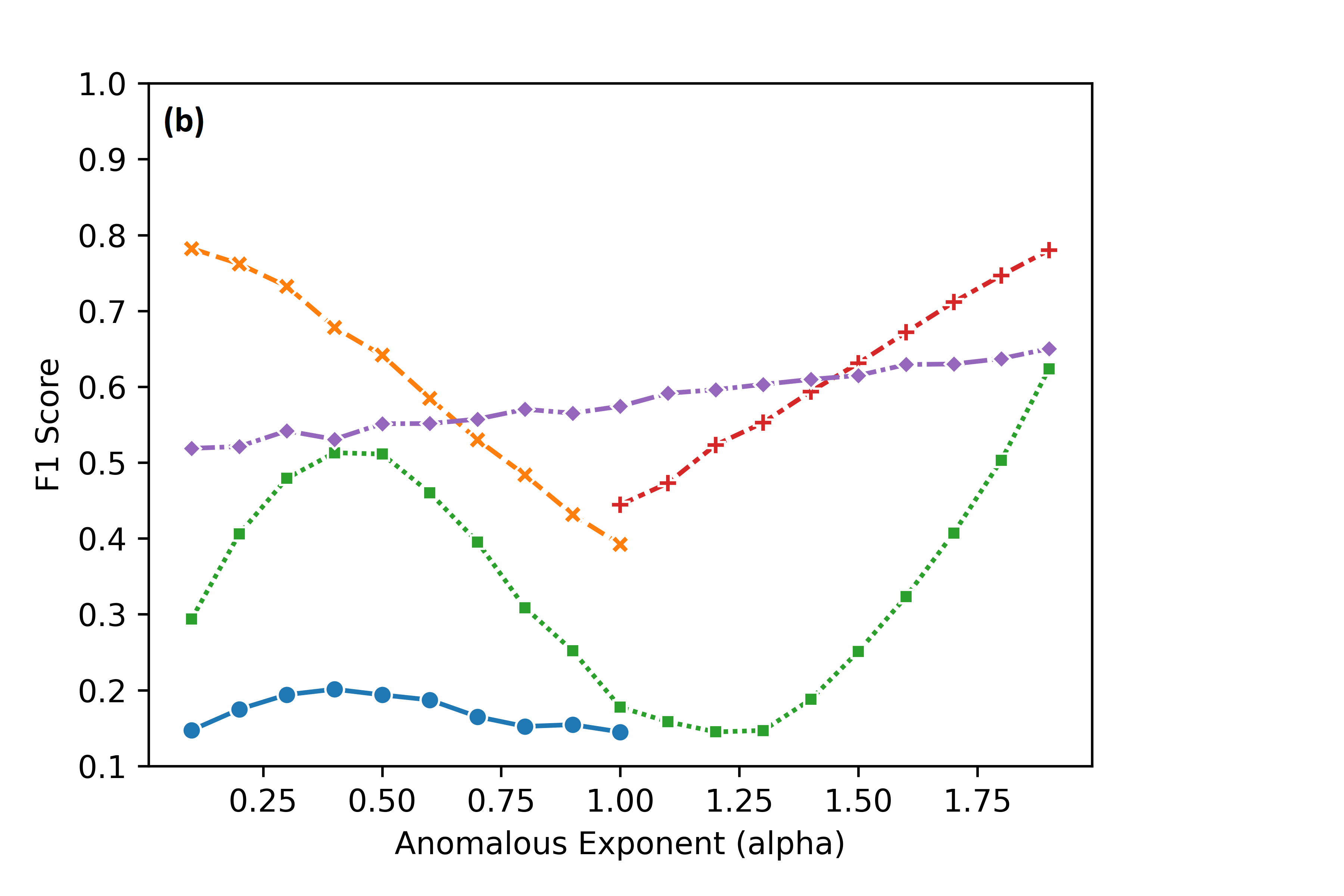}
 %\caption{F1-scores for the GADF/Resnet model.}
% \end{subfigure}
% \begin{subfigure}{width = .48\textwidth}
 \includegraphics[width=7.5cm]{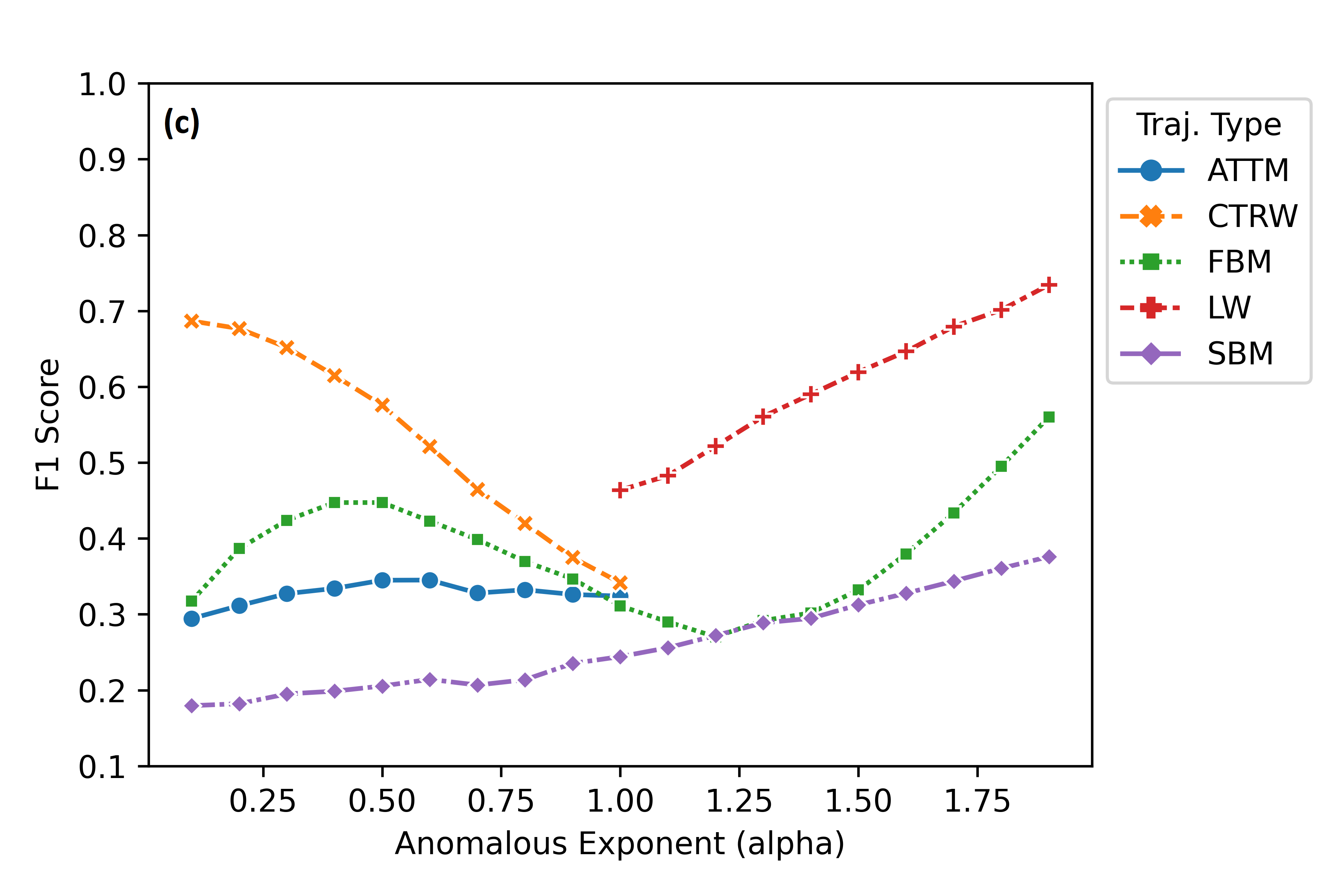}
% classification_figs/F1vsAlpha_hue_trajType_SNR1_ConvLSTM.png
% \end{subfigure}
 \caption{F1-scores for the different models on noisy trajectories, ${\rm SNR} = 1$, of all the five classes (ATTM, CTRW, FBM, LW, and SBM) with GASF/ResNet (upper left), GADF/ResNet (upper right), and ConvLSTM (bottom).}
 \label{fig:gadsf_traj_class_snr1}
 \end{figure}

Results from Figure \ref{fig:classification_convlstm_vs_rest_snr1} show that models fed with GAF images outperform ConvLSTM in classification at any trajectory length. These results are very impressive as the ConvLSTM was the best overall model in one-dimensional classification at all trajectory lengths and at restricted lengths [10, 50] during the AnDi challenge, as we will see later. 
On average, the GASF-GADF/ResNet models achieve the best results among all the GAF fed models.
So that, we will compare them against the ConvLSTM in detail. When we focus on F1-score as a function of the anomalous diffusion exponent $\alpha$ at different trajectory length, as in Figure \ref{fig:gadsf_lengths_snr1}, we can see that F1-score improves with increased length, across all values of $\alpha$. Though, if we focus on the interaction of length and $\alpha$ we can see that increasing length has a disproportionate effect on the F1-score for $\alpha$ close to 1.9 and for $\alpha$ around $0.3$. The source of this interaction is unclear. Though, as we can see in Figure \ref{fig:gadsf_traj_class_snr1} there are several reasons behind: in CTRW F1-score falls as $\alpha$ value tends to 1, and in LW also the worst values for F1-score are obtained for $\alpha$ close to 1. Also, FBM performs dramatically better at $\alpha$ around $0.4$ and $\alpha$ greater than $1.75$.\medskip

 %formerly figs 6a, 6b, 6c

Upon closer inspection of model performance by trajectory type (Figure \ref{fig:gadsf_traj_class_snr1}), we see that the GASF and GADF ResNet models perform roughly the same. However, the difference in architecture between our GAF fed computer vision models and our ConvLSTM is immediately evident when we compare performance across the different trajectory types. Most notably, the classification performance of SBM trajectories nearly doubles from the ConvLSTM to any GAF fed ResNet Model.\medskip

%formerly had figs 7a 7b and 7c
% \begin{figure}[htbp]
% \centering
% \includegraphics[width=0.4\linewidth]{Fig8a.png}
% \includegraphics[width=0.4\linewidth]{Fig8b.png}
% \caption{F1-scores of GASF/ResNet (a) and GADF/ResNet (b) models for $\rm{SNR}=1,2$.}
% \label{fig:gadsf_snrs}
% \end{figure}

While performance across different trajectory types was found to deviate between the GAF fed ResNet models and the ConvLSTM, the rough shape of the performance curves by model stayed the same. For instance, the F1-score curve of FBM  vs. $\alpha$ is roughly sinusoidal, the performance of CTRW and LW are approximately mirrored images of each other , and SBM scales positively with $\alpha$ across all models. In particular, the shape of the CTRW and LW F1-score $\sim \alpha$ curves was also observed when using the ConvTransformer model \cite{firbas2022transformers}. Similar performance curve shapes across different ML architectures seem to indicate that certain values of $\alpha$ make some diffusive regimes appear as others, as we can also note in \cite[Fig. 3]{munoz-gil2021objective}.\medskip

While overall performance was significantly improved with our GASF and GADF ResNet models over the ConvLSTM, the greatest improvements from the GAF fed ResNet over the ConvLSTM came from $\alpha \in [0.25, 0.6]$ and $\alpha \ge 1.5$. Moreover, there are also small improvements in the classification of trajectories with $\alpha \approx 1$ (Figure \ref{fig:gadsf_lengths_snr1}). It should be noted that there are two cases where GADF/GASF ResNet classification performance was worse than in the ConvLSTM. Most notably, the ResNet models have great difficulty identifying ATTM trajectories, with F1-scores dropping below 0.2, which is what one would expect from guessing. Then the sinusoidal shape of F1-score vs. $\alpha$ of FBM trajectories is exacerbated in the ResNet models leading to a loss of performance around $1.25$, with performance gains elsewhere (Figure \ref{fig:gadsf_traj_class_snr1}).\medskip

\begin{figure}[htbp]
 \centering
 \includegraphics[width=7.5cm]{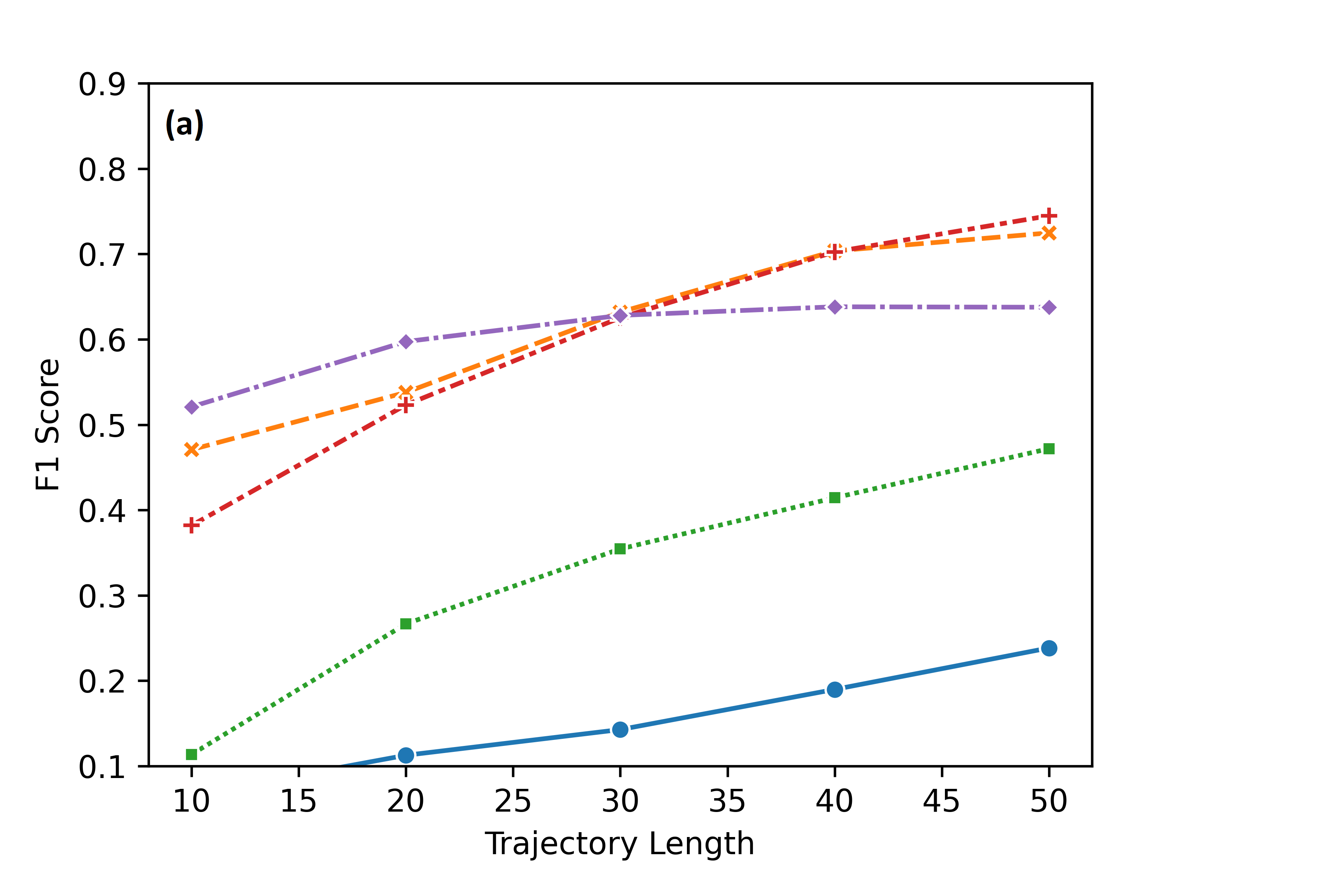}
 % classification_figs/F1vsTrajLen_hue_trajType_SNR1_GASF ResNet.png
 %\caption{F1-scores for the GASF/ResNet model.}
 \includegraphics[width=7.5cm]{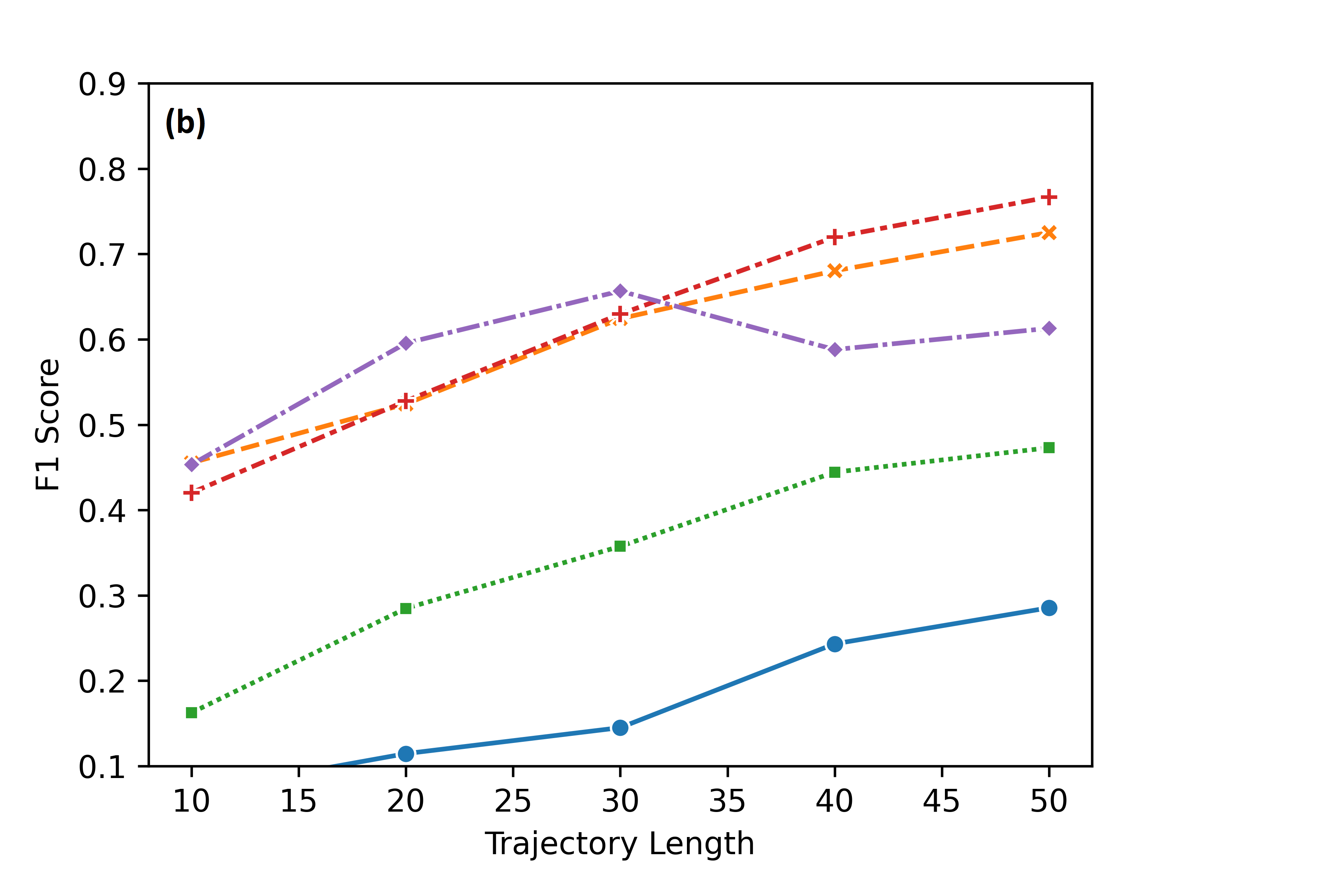}
 \includegraphics[width=7.5cm ]{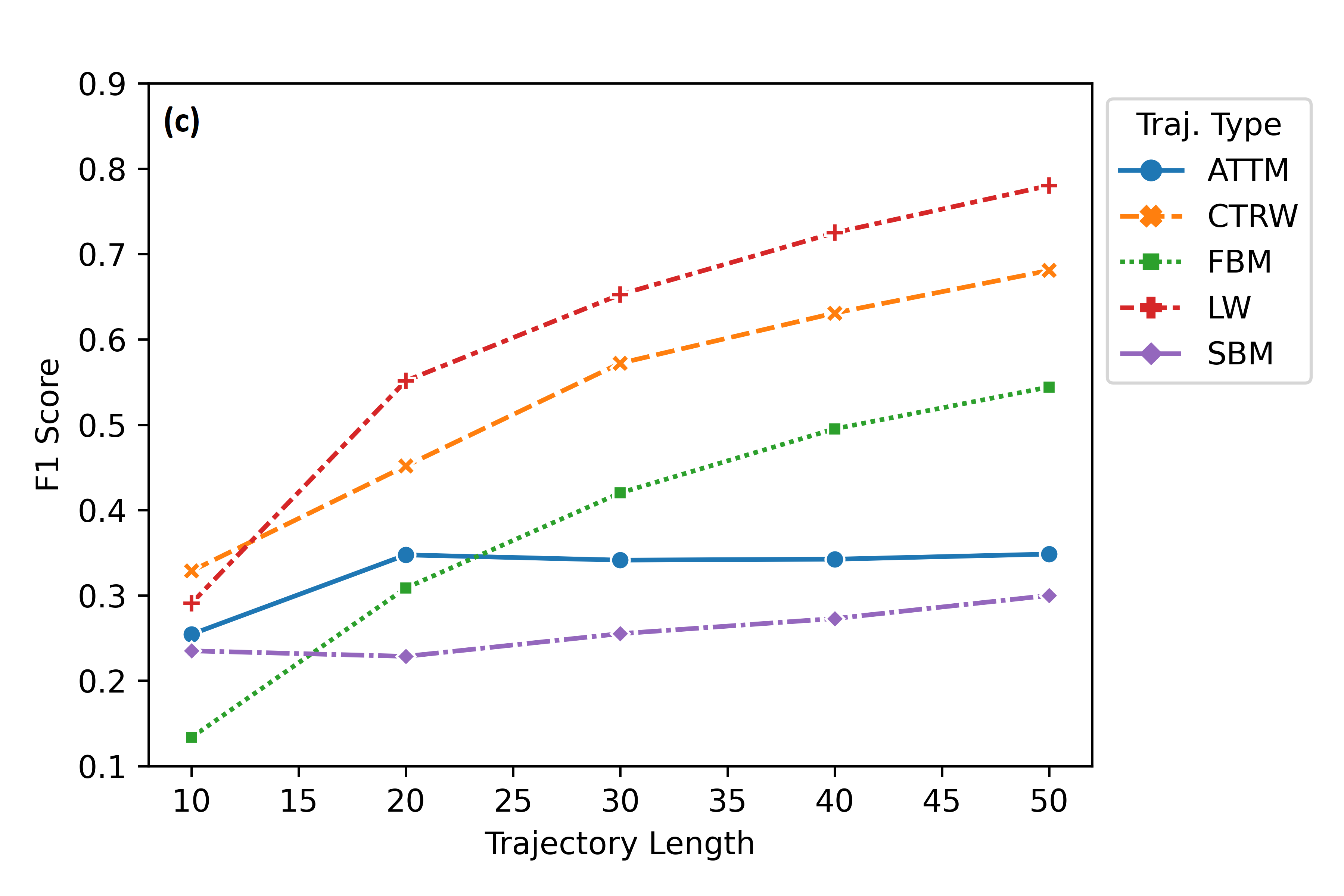}
 \caption{F1-scores of noisy trajectories, with $\rm{SNR}=1$, by trajectory length for GASF/ResNet (upper left), GADF/ResNet (upper right), and ConvLSTM (bottom) for each model (ATTM, CTRW, FBM, LW, and SBM)}
 \label{fig:gadsf_length_classes_snr1}
\end{figure}

To gain further insight into classification performance by underlying diffusive regime we plotted F1-Score as a function of trajectory length by underlying diffusive model in Figure \ref{fig:gadsf_length_classes_snr1}. We can see that across the three ML models, with some variation due to stochasticity, performance scales positively with the F1-score. The exception is the classification performance of SBM trajectories by the GADF ResNet model  
(\ref{fig:gadsf_length_classes_snr1}), which decreases from length 30 to 40, probably due to a stochastic artifact.
%\sout{We are not sure what the cause of the behavior is, however the same behavior repeats itself for $SNR = 2$ testing data as well *reference snr2 graph from appendix*}. 
It should be noted that this behavior disappears when we pool performance across the different underlying diffusive models as in Figure \ref{fig:gadsf_lengths_snr1}, where we can clearly appreciate that the GAF Resnet models clearly outperform our ConvLSTM benchmark for trajectory length greater than 30.\medskip

%\sout{ We have separately analyzed each diffusion type in terms of the trajectory length; see Figure \ref{fig:gadsf_length_classes_snr1}. The GADF/ResNet model significantly outperforms the benchmark for any length and anomalous diffusion exponent $\alpha$ for all models. This can be clearly appreciated for LW and ATTM trajectories of lengths greater or equal to 30. The unique exceptions are ATTM and FBM trajectories of lengths higher or equal to 20. It seems that ResNet models try to concentrate on increasing the accuracy of CTRW, LW, and SBM diffusion types in comparison with ConvLSTM, which is more distributed within the five diffusion models.\medskip}

%formerly fig 8a 8b and 8c

\begin{figure}[htbp]
 \centering
 \includegraphics[width=0.4\linewidth]{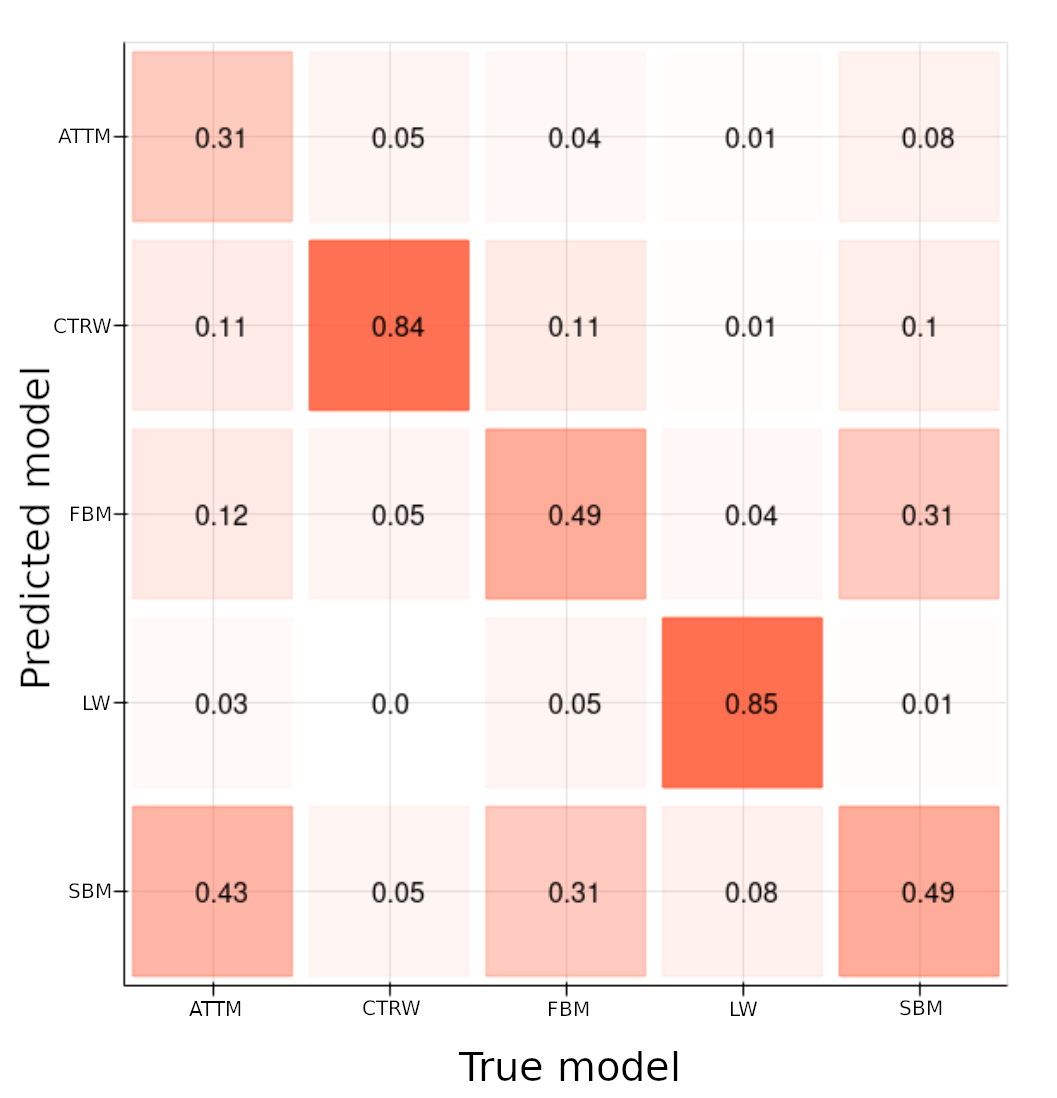}
 %interactive-tool-figs/Task2-GAF-edited.png
 \caption{Resulting confusion matrix Results of the GASF/ResNet model under the andi interactive tool.}
  \label{fig:classification_gasf_resnet_andi_tool}
 \end{figure}

Finally, we have bench-marked our GASF ResNet model using the AnDi interactive tool for trajectories of lengths from 10 to 50. The resulting confusion matrix can be seen in Fig. \ref{fig:classification_gasf_resnet_andi_tool} and it largely summarizes the results we discussed in this section. Namely that the GAF models are best at identifying CTRW and LW trajectories, with difficulty identifying ATTM trajectories. From the outputs of the AnDi interactive tool we can verify that our GADF and GASF ResNet models significantly outperform the previous top models from the AnDi Challenge 2020 in overall F1-score and AUC (Table \ref{tab:comparison_classification}).\medskip

\begin{table}[htbp]
\centering
\begin{tabular}{|c|c|c|c|}
\hline
\textbf{Team} & \textbf{F1-score} & \textbf{AUC} & \textbf{Model Type}\\
\hline
eduN & 0.499 & 0.82 & RNN + Dense NN \cite{argun2021classification}\\
FCI & 0,525 & 0,86 & CNN \cite{bai2018empirical, granik2019single-particle}\\
UPV-MAT & 0.560 & 0.87 & CNN + biLSTM\cite{garibo-i-orts2021efficient}\\
\hline
GASF/ResNet & \textbf{0.581} & \textbf{0.89} & GASF/ResNet\\
\hline
\end{tabular}
\caption{Performance comparison of GASF/ResNet model with best AnDi Challenge models at the classification of the underlying diffusive regime in trajectory lengths 10 to 50.}
\label{tab:comparison_classification}
\end{table}

We were able to achieve these performance increases, while gaining ease of implementation in order to maintain our goal of increasing accessibility to ML methods for the characterization of anomalous diffusion. In fact, the GASF and GADF models were significantly easier to deploy than, the previous state-of-the-art ConvLSTM model and as such should be very useful in an applied setting. 

%formerly fig9

\subsection{Inference of the anomalous diffusion exponent}
\label{sec:inference_results}
The anomalous diffusion exponent reflects the relationship of dispersal and time, as such, it highly conditions diffusion characteristics. In this section, we will analyze and compare the performance of our GASF and GADF ResNet Models to the ConvLSTM. In order to assess model accuracy, we will use the \textit{mean absolute error} (MAE) between the exponent used to generate each trajectory $\alpha_{{\rm truth}}$ and the predicted exponent value $\alpha_{{\rm pred}}$. Given a data set containing \textit{N} samples, the general MAE is defined as:

\begin{equation}
    \rm{MAE} = \frac{1}{N} \sum_{j=1}^{N} | \alpha_{j,{\rm pred}} - \alpha_{j, {\rm truth}} |.
\end{equation}

\begin{figure}[htbp]
    \centering
 %   \includegraphics{}
 %   \caption{Caption}
 %   \label{fig:my_label}
 %\centering
 \includegraphics[width=7.5cm]{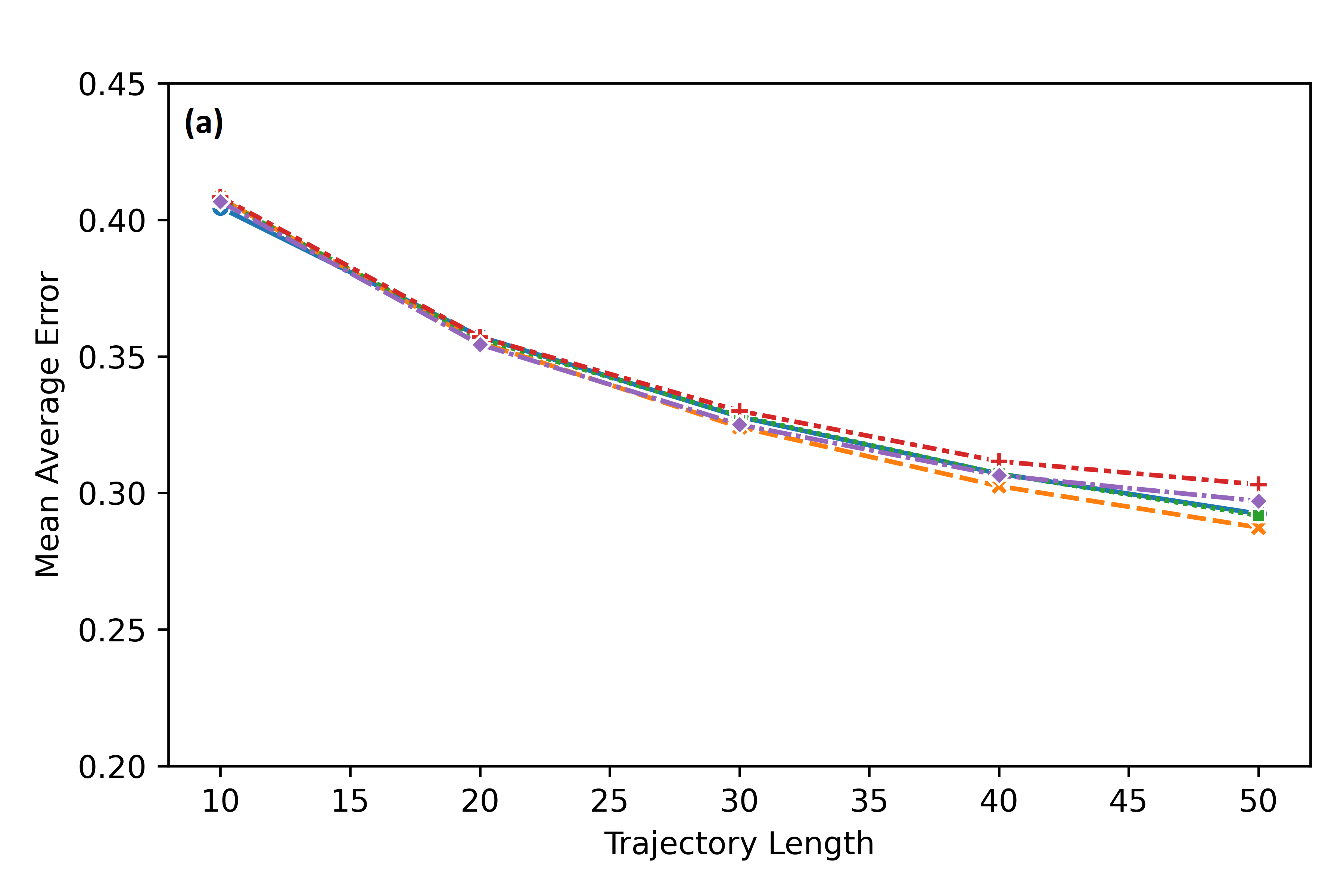}
 %regression_figs/MAEvsTrajLen_hue_model_SNR1.png
  \includegraphics[width=7.5cm]{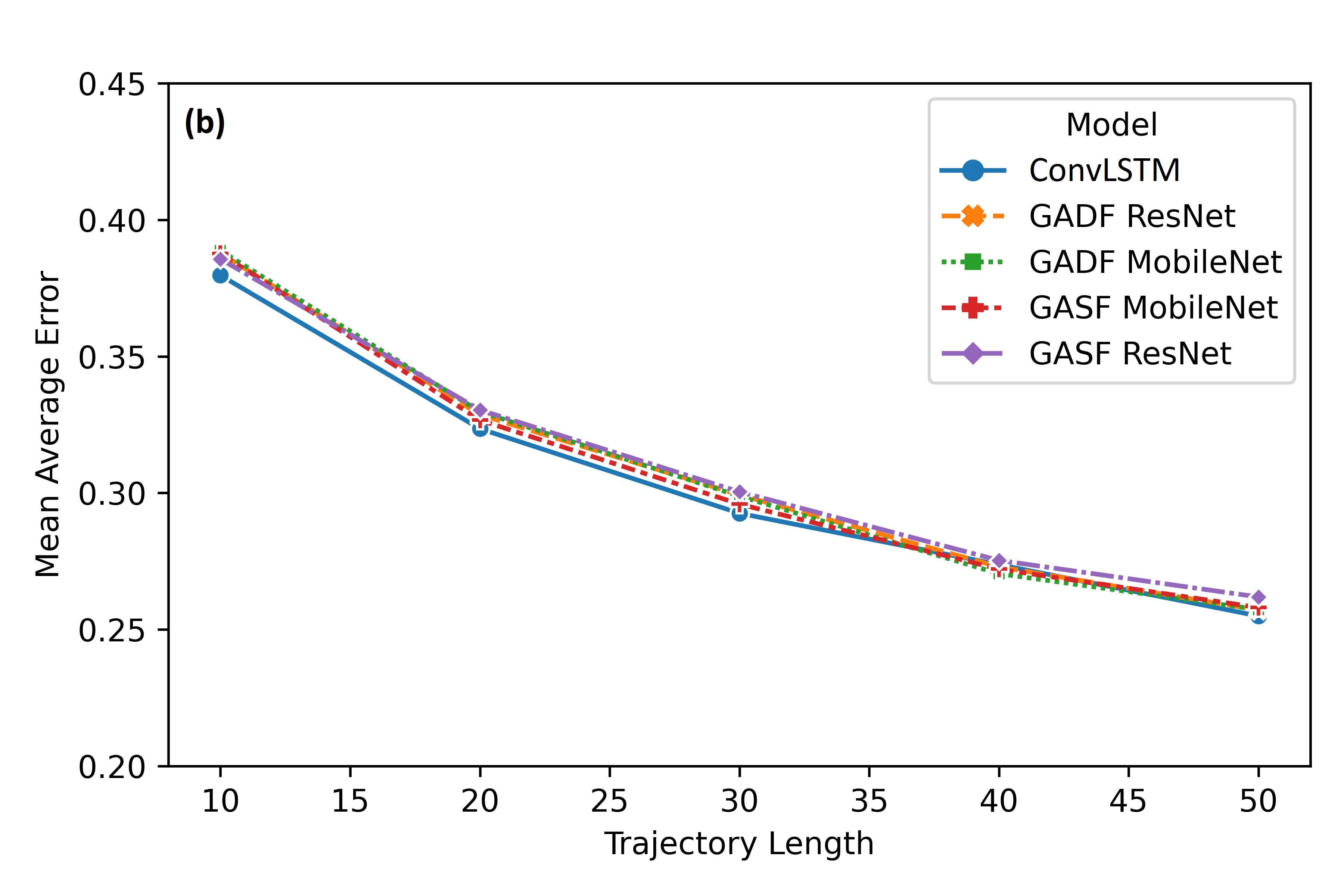}
%regression_figs/MAEvsTrajLen_hue_model_SNR2.png
\caption{Models' performance comparison for noisy trajectories with  $\rm{SNR} = 1$ (left) and $\rm{SNR} = 2$ (right).}
 \label{fig:regression_convlstm_vs_rest_snr}
\end{figure}

As with the classification task, we will compare the performance for the regression of the anomalous exponent $\alpha$ of our GAF models against the ConvLSTM.  Once again, GASF and GADF fed ResNet models performed the best for noisy trajectories with $\rm{SNR}=1$. For trajectories with $\rm{SNR}=2$, the performance of all five models tested is remarkably close, but the ConvLSTM provides better performance at shorter trajectory lengths (Figure \ref{fig:regression_convlstm_vs_rest_snr}). However, the difference in performance between GAF models and the ConvLSTM is quite marginal, and we would still favor the GASF and GADF ResNets for their ease of deployment and marginally better performance in noisier trajectories. When we consider trajectory length, we observe the same global behavior as in the Classification task, where across models performance improves (decreased MAE) with increased length (Figure \ref{fig:regression_convlstm_vs_rest_snr}). \medskip.

%\sout{MAE is inversely proportional to length with all diffusive regimes following the same trend of increased performance (lower MAE) with longer trajectory length across GASF/GADF ResNet models and the ConvLSTM (Figure \ref{fig:mae_gadsf_resnet_vs_convlstm_snr1_by_type})}\medskip

%formerly fig 10 a and b

\begin{figure}[htpb]
 \centering
 \includegraphics[width=7.5cm]{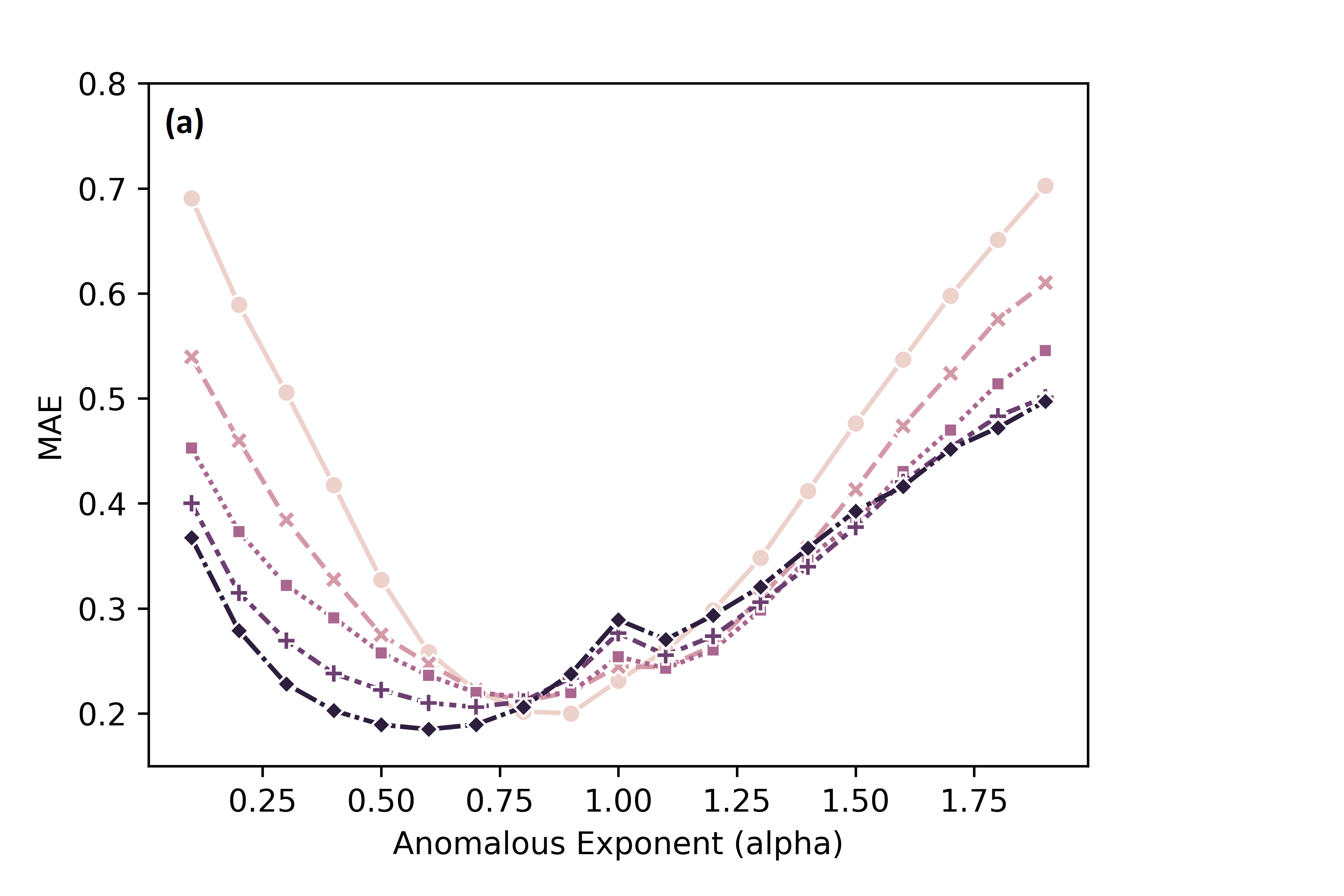}
 %regression_figs/MAEvsAlpha_hue_TrajLen_SNR1_GASF ResNet.png
 \includegraphics[width=7.5cm]{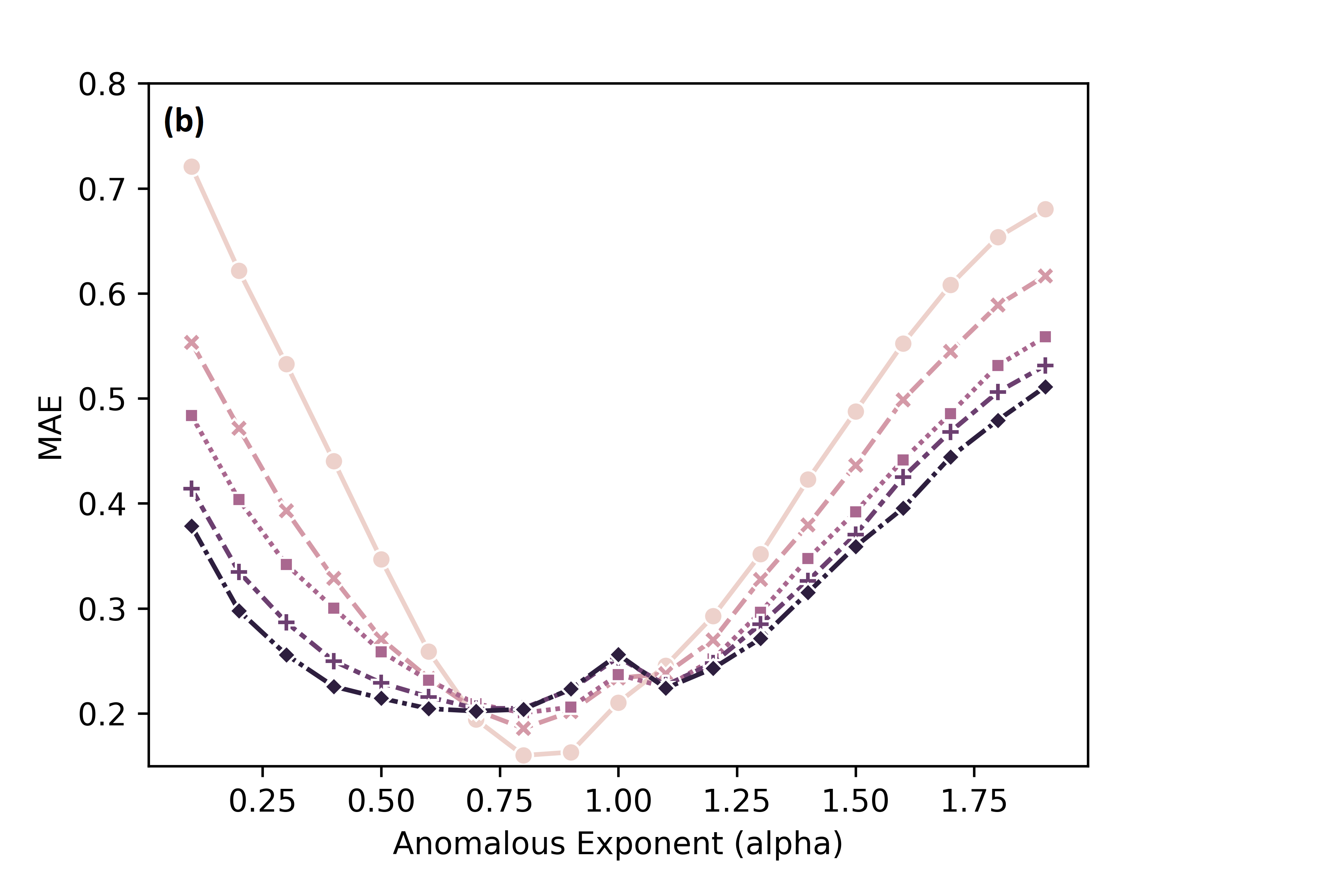}
%regression_figs/MAEvsAlpha_hue_TrajLen_SNR1_GADF ResNet.png
 \includegraphics[width=7.5cm]{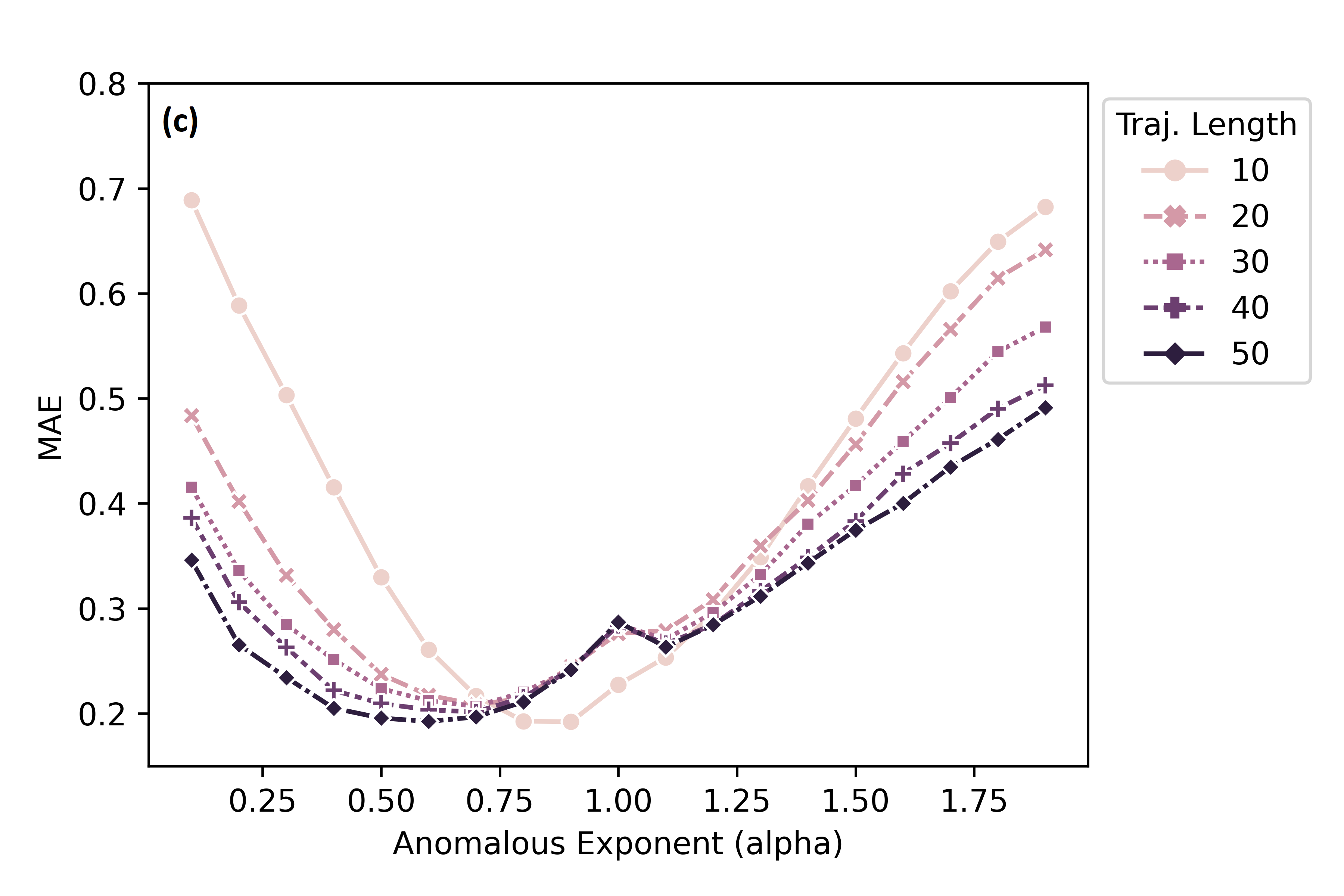}
% regression_figs/MAEvsAlpha_hue_TrajLen_SNR1_ConvLSTM.png
 \caption{MAE of GASF/ResNet (upper left), GADF/ResNet (upper right) and ConvLSTM (bottom) for different trajectory lengths and values of the anomalous exponent $\alpha$ on noisy trajectories with $\rm{SNR} = 1$.}
 \label{fig:mae_gadf_resnet_vs_convlstm_snr1_by_length}
 \end{figure}

With regards to model performance as a function of the anomalous diffusion exponent $\alpha$, we can see that all three models are best able to infer $\alpha$ when it lays in the interval $\alpha \in [0.5, 0.9]$ (Fig. \ref{fig:mae_gadf_resnet_vs_convlstm_snr1_by_length}). This is in opposition to the classification task, where all models struggled to classify roughly normal trajectories. We believe this indicates that the different diffusive regimes behave similarly at $\alpha \approx 1$, which is equivalent to increasing the amount of available training data. Here, our models no longer have to account for which of the five models it is looking at in order to infer the $\alpha$ exponent,  which pools together the training data for each of the models.\medskip

%formerly fig 11a 11b and 11c

\begin{figure}[htbp]
 \centering
 \includegraphics[width=7.5cm]{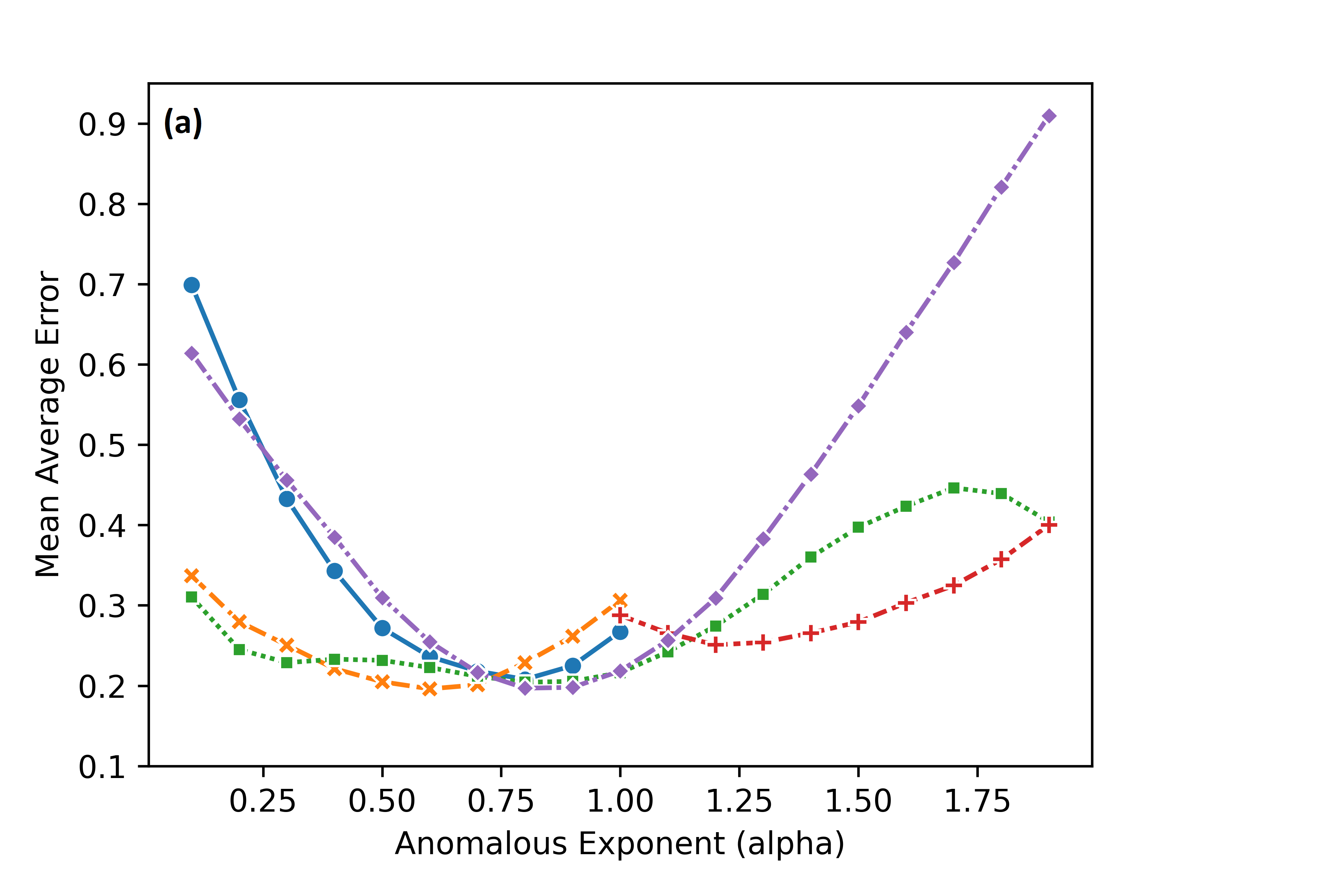}
 %regression_figs/MAEvsAlpha_hue_TrajType_SNR1_GASF ResNet.png
 \includegraphics[width=7.5cm]{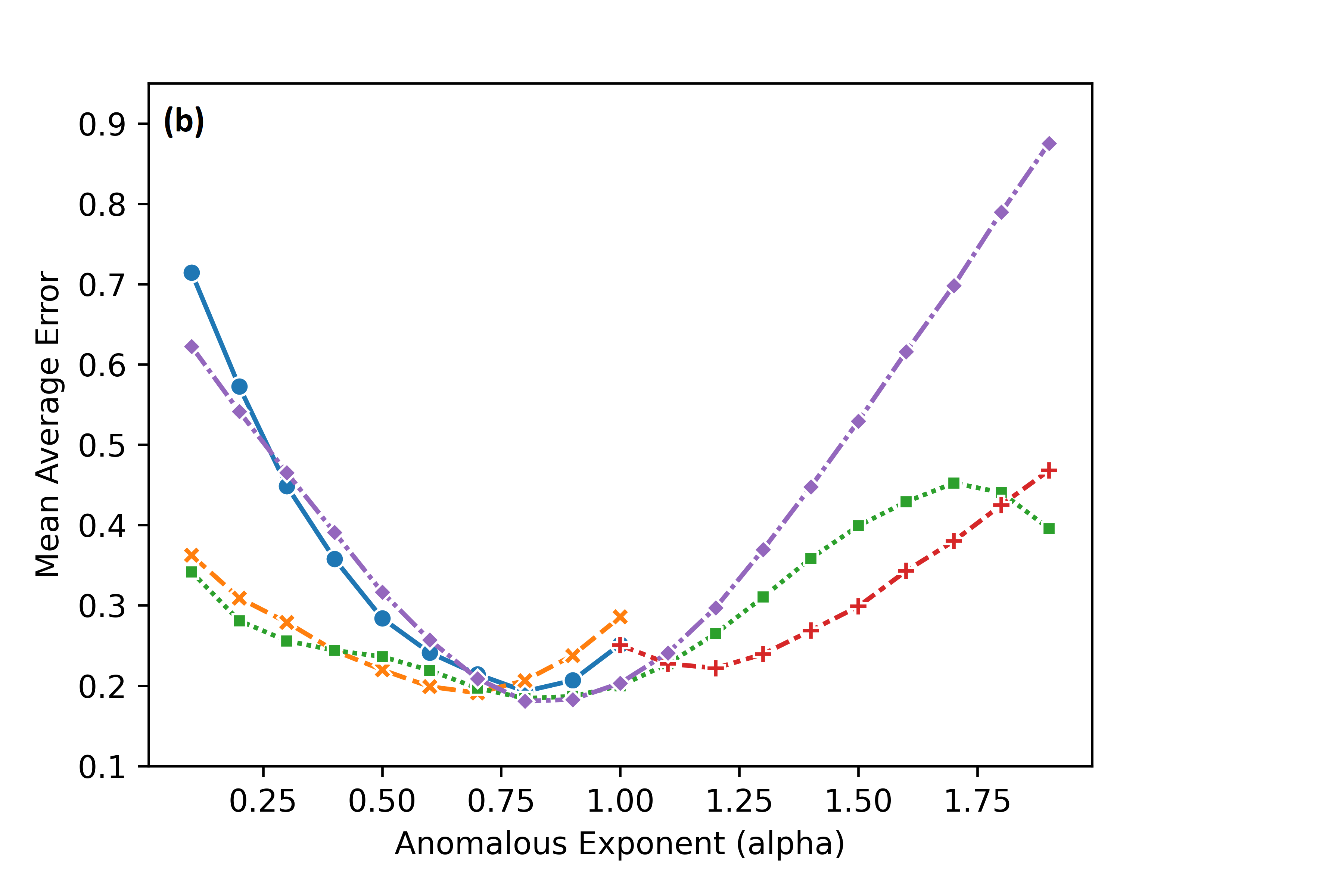}
 %regression_figs/MAEvsAlpha_hue_TrajType_SNR1_GADF ResNet.png
 \includegraphics[width=7.5cm]{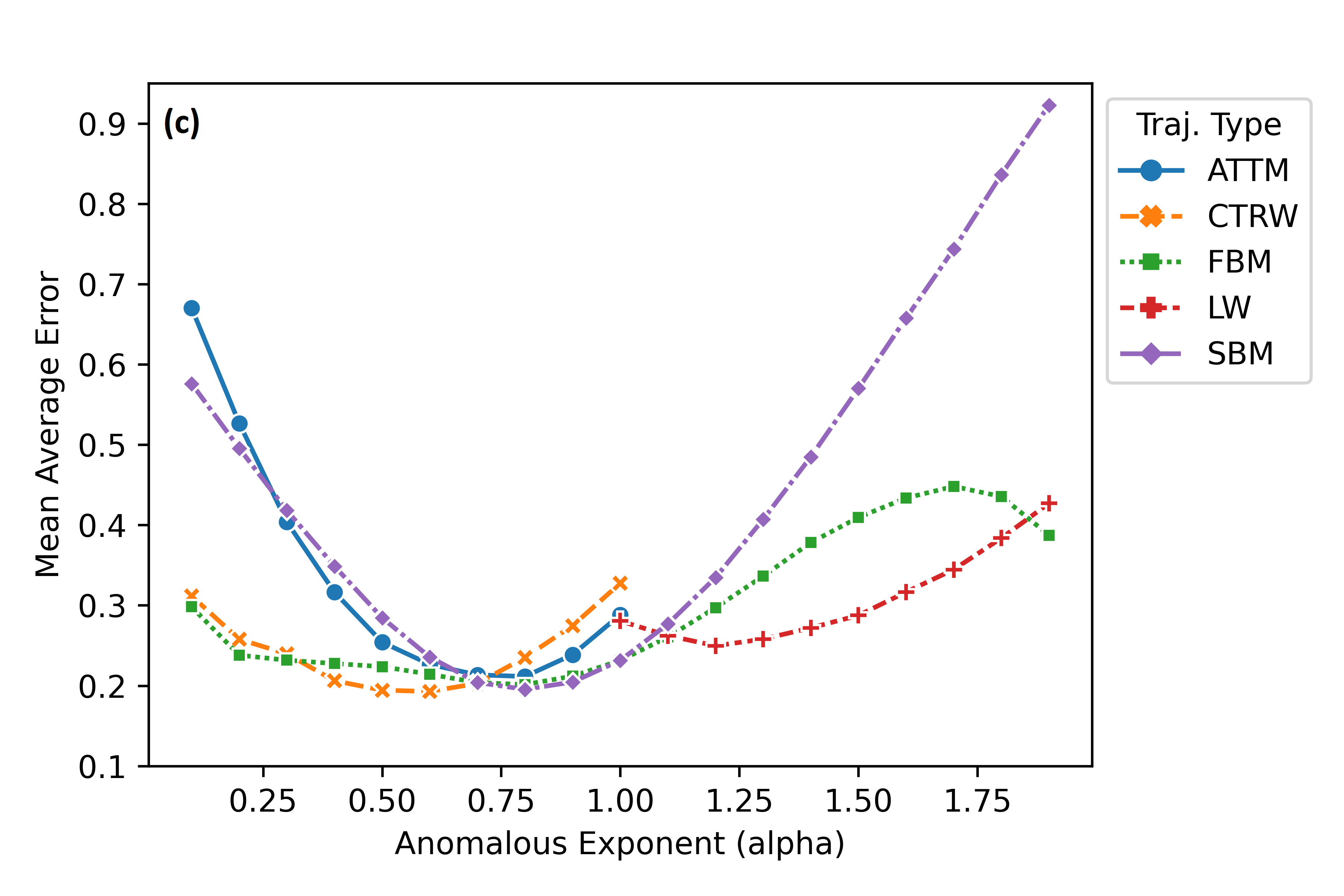}
 %regression_figs/MAEvsAlpha_hue_trajType_SNR1_ConvLSTM.png
 \caption{MAE of GASF/ResNet (upper left), GADF/ResNet (upper right) and ConvLSTM (bottom) for different anomalous exponents $\alpha$ and (ATTM, CTRW, FBM, LW, and SBM) models of noisy trajectories with $\rm{SNR} = 1$.}
\label{fig:mae_gadsf_resnet_vs_convlstm_snr1_by_type}
 \end{figure}

Again, in contrast to what we observed in the classification task, the worst performance was achieved at both extremes of each trajectory $\alpha$ domain. For instance, MAE for CTRW was highest in all three models when $\alpha \approx 0.1$ and $\alpha \approx 1.0$. Likewise, SBM performed worst at the most super and sub-diffusive extremes (Figure \ref{fig:mae_gadsf_resnet_vs_convlstm_snr1_by_type}). Again we believe that this is because diffusive regimes are at their most distinct (easiest to classify) at the extremes of their $\alpha$ domain. This, in turn, confuses our models as they must implicitly learn specifically to infer $\alpha$ for a specific underlying regime. This effect would be equivalent to decreasing the amount of available training data.\medskip

%formerly fig 12a 12b and 12c
%\begin{figure}[htbp]
%\centering
%\includegraphics[width=7.5cm]{regression_figs/MAEvsAlpha_hue_TrajType_SNR1_GASF ResNet.png}
%\includegraphics[width=7.5cm]{regression_figs/MAEvsAlpha_hue_TrajType_SNR1_GADF ResNet.png}
%\includegraphics[width=7.5cm]{regression_figs/MAEvsAlpha_hue_TrajType_SNR1_ConvLSTM.png}
%\caption{MAE of GASF/ResNet (upper left), GADF/ResNet (upper right) and ConvLSTM (bottom) for different trajectory lengths of noisy trajectories with $\rm{SNR} = 1$.}
%\label{fig:mae_gadsf_resnet_vs_convlstm_snr1_by_type}
%\end{figure}
%formerly 13a 13b and 13c

\begin{figure}[htbp]
 \centering
 \includegraphics[width=7.5cm]{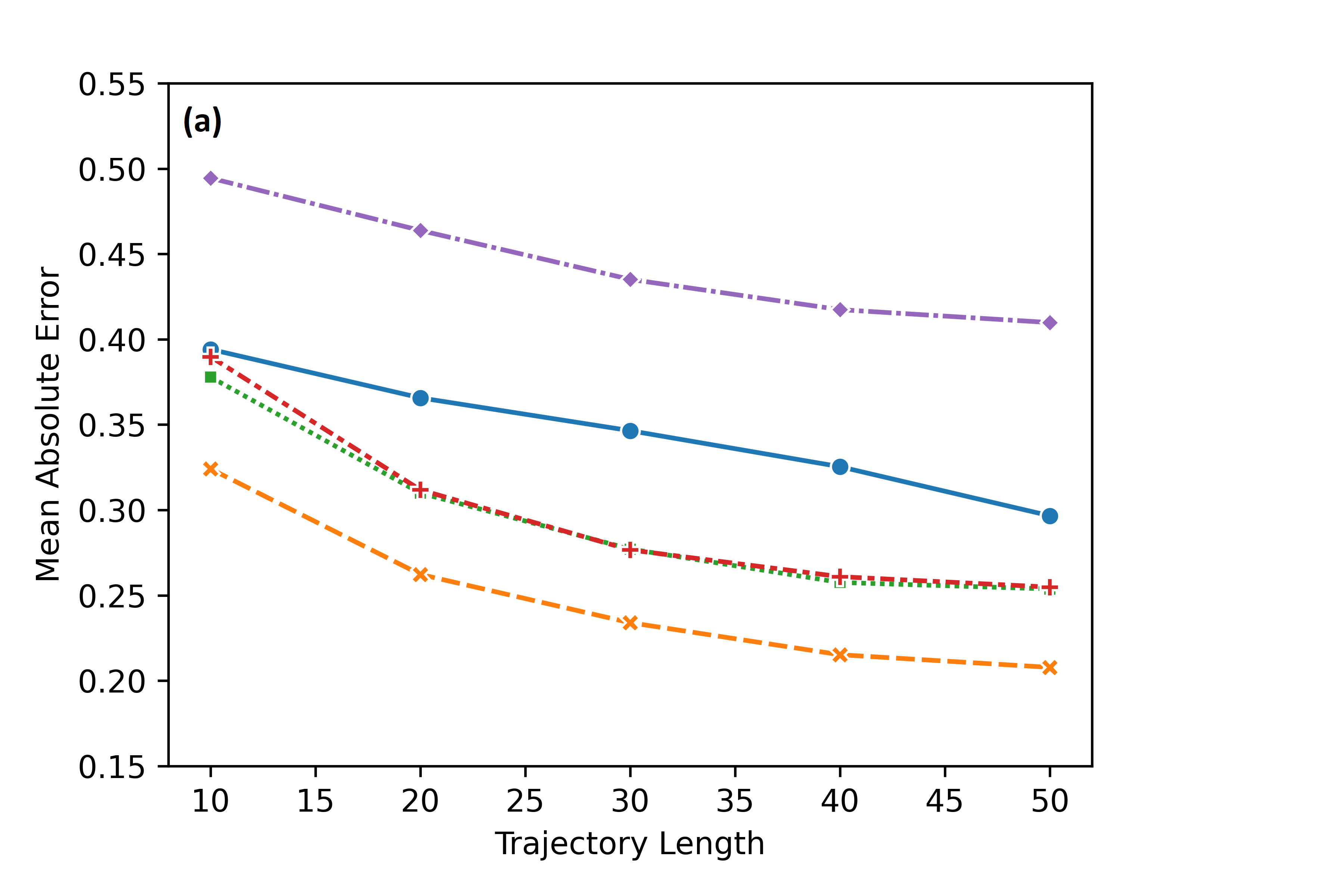}
 \includegraphics[width=7.5cm]{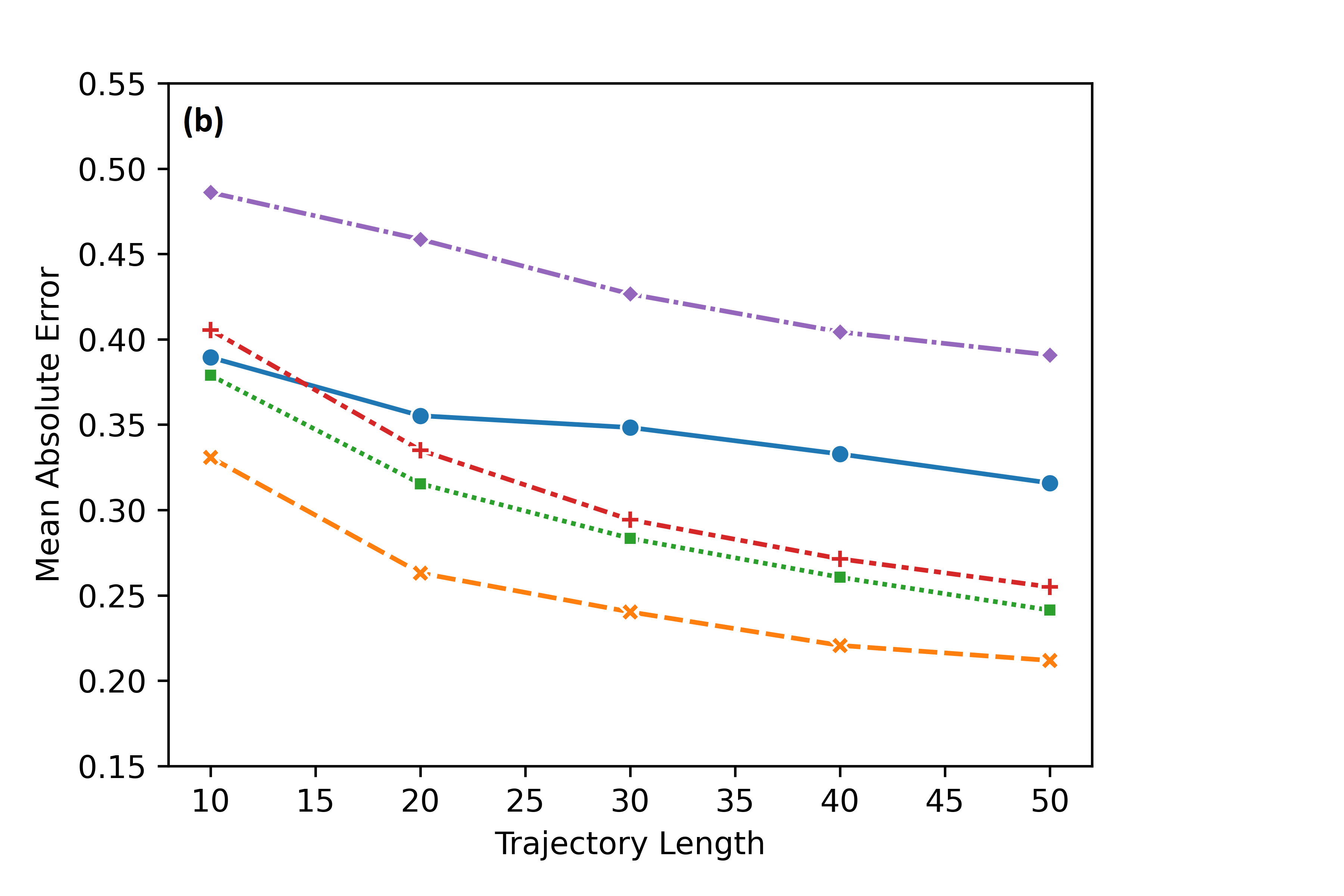}
 \includegraphics[width=7.5cm]{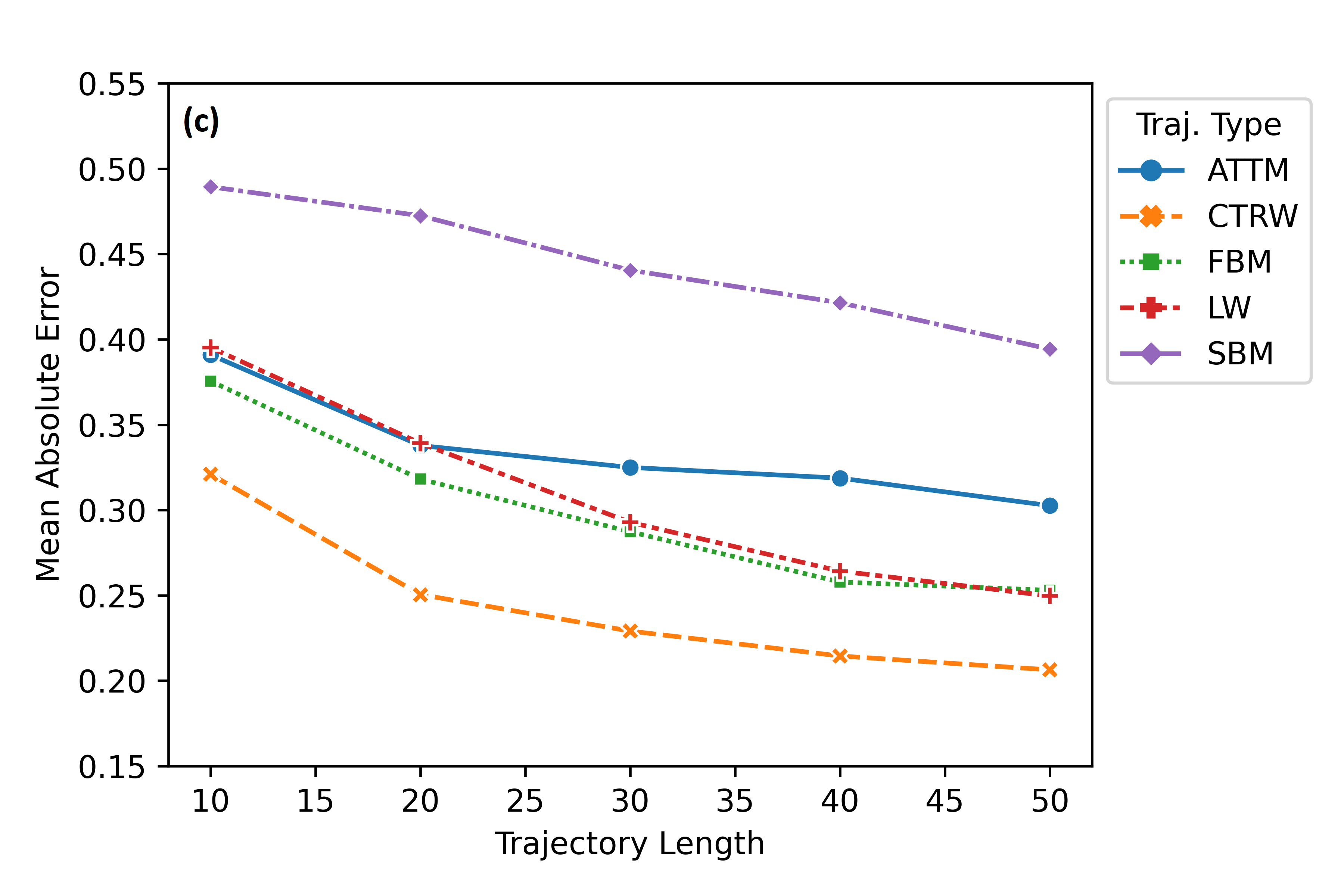}
 \caption{MAE of GASF/ResNet (a), GADF/ResNet (b) and
ConvLSTM (c) for different trajectory lengths of noisy trajectories
with SNR = 1.}
\label{fig:mae_gadsf_resnet_vs_convlstm_snr1_by_length}
 \end{figure}

Looking again at the MAE in terms of the trajectory length, we show the results separated for each diffusion model. The comparison is more illustrative than in Figure \ref{fig:mae_gadf_resnet_vs_convlstm_snr1_by_length}, since the models have different ranges for the anomalous diffusion exponent, which makes it more difficult to compare them all together. Again, we show how results improve as the trajectory length increases (Figure \ref{fig:mae_gadsf_resnet_vs_convlstm_snr1_by_length}). The most significant change is given when increasing the trajectory length from 10 to 20. From 20 onwards, the MAE improvement for the ATTM is relatively small compared to the rest.

\begin{figure}[htpb]
 \centering
 \includegraphics[width=.45\linewidth]{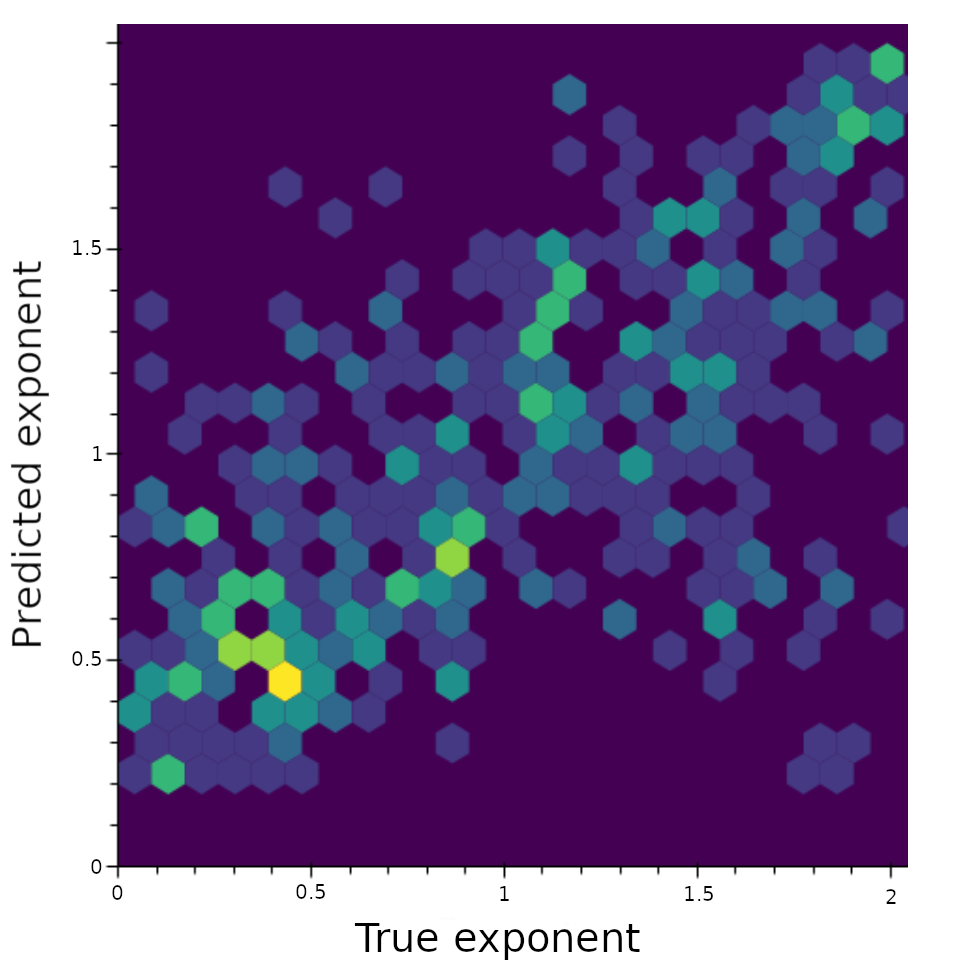}
 \caption{Results of the GASF/ResNet model under the AnDi interactive tool for trajectory lengths 10 to 50.}
  \label{fig:inference_gasf_resnet_andi_tool}
 \end{figure}

\begin{figure}[htbp]
 \centering
 \includegraphics[width=0.7\linewidth]{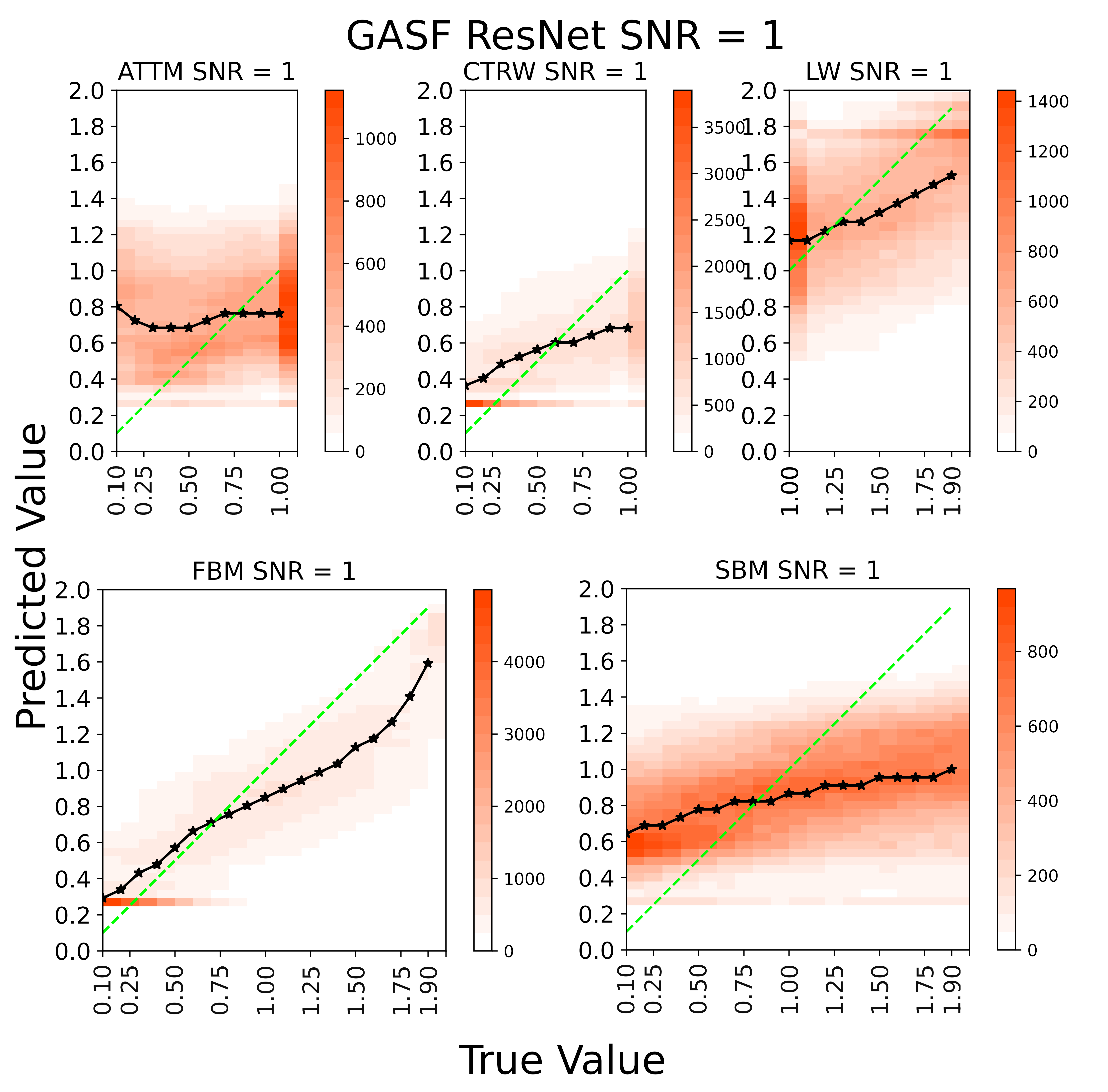}
 % \caption{GASF ResNet $\alpha$ inference results.}
 \caption{Heatmaps with the results for the GASF ResNet model in the inference of the anomalous diffusion exponent $\alpha$. The black star line represents the median prediction for each of the true values and the green dashed line represents the true value line.}
 \label{fig:gadf_gasf_compiled_heatmaps}
 \end{figure}

Lastly, we show a heat map of the inference results generated using the AnDi interactive tool (Fig. \ref{fig:inference_gasf_resnet_andi_tool}). The heat map shows the distribution of the predicted $\alpha$ as a function of the ground truth for the GASF ResNet model. Additional heat maps, that predicted show $\alpha$ as a function of true $\alpha$ as well as the underlying diffusive regime can be seen in Figure \ref{fig:gadf_gasf_compiled_heatmaps}. 

 %formerly fig 14

Furthermore, Table \ref{tab:comparison_regression} summarizes the output of the AnDi interactive tool for the three best-performing models in the AnDi challenge 2020 compared to our GADF ResNet. We can see that our GADF/ResNet outperforms the majority of them, except of UPV-MAT, where the difference in performance is negligible. We can say that both models behave roughly the same, with the GASF ResNet model being far easier to implement, especially considering that the Conv5LSTM for inference of $\alpha$ is a compilation of 12 individually trained ConvLSTMs \cite{garibo-i-orts2021efficient}.\medskip

 \begin{table}[htpb]
\centering
\begin{tabular}{|c|c|c|}
\hline
\textbf{Team} & \textbf{MAE} & \textbf{Method} \\
\hline
eduN & 0.385 & RNN + Dense NN \cite{argun2021classification}\\
FCI & 0.369 & CNN \cite{bai2018empirical, granik2019single-particle}\\
UCL & 0.367 & feature engineering + NN \cite{gentili2021characterization}\\
UPV-MAT & \textbf{0.326} & CNN + biLSTM\cite{garibo-i-orts2021efficient}\\
\hline
GADF/ResNet & 0.33 & GADF fed ResNet\\
\hline
\end{tabular}
\caption{Regression: performance comparison of GASF/ResNet model with best AnDi Challenge models.}
\label{tab:comparison_regression}
\end{table}

%formerly figures Fig15a and Fig15b
\section{Conclusions}
\label{sec:conclusions}

We set out with the objective of matching or improving the current methodology for the characterization of short and noisy diffusive trajectories while increasing accessibility to ML methods by decreasing the difficulty of deployment. 

We rely on the use of GAFs for converting trajectories into images to feed machine learning pre-trained models for image processing. Our GASF and GADF ResNet Models outperformed the current state of the art by a wide margin in trajectory classification, and were negligibly worse than the best method for regression of the anomalous diffusion exponent $\alpha$. These results, especially for classification, exceeded our hopes, particularly considering the ease of implementation. ResNet can be natively implemented by using Keras in Python, and the \textit{pyts.image} package allows one to convert time series to Gramian matrices with a single function. Since GAFs represent the existence of temporal correlations between pairs of points in the trajectory, it will be interesting to explore alternative approaches to compute the existence of such correlations. We have also indicated the symmetric properties of GAF images. It will also be nice to see how they succeed in studying trajectories obtained from confined diffusion, where there is a potential absence of time-reversal symmetry.

%closing paragraph sobre la importancia de anomalous diffusion en experimental setting.
Diffusion is all around us, and perturbations to a normal diffusive regime, such as confining movement to a room, diffusion on a fractal, or even a breeze have the capability to make diffusion anomalous. Traditionally, particularly in a more applied setting like ecology, the lack of tools and data availability led researchers to assume that movement was normal. The lens of diffusion provides an interdisciplinary framework for the characterization of movement, which should increase collaborations between traditionally insular fields. We hope that the rise of flexible tractable ML models, like our GASF/GADF fed ResNet, facilitate a new wave of applied interdisciplinary diffusive studies, which advance our knowledge of diffusion regardless of scale. 

\subsection*{Acknowledgements}
{We thank M.A. García-March for helpful comments and discussions on the topic. NF is supported by the National University of Singapore through the Singapore International Graduate Student Award (SINGA) program. OGO and LS acknowledge funding from
MINECO project, grant TIN2017-88476-C2-1-R.
JAC acknowledges funding from grant PID2021-124618NB-C21 funded by MCIN/AEI/ 10.13039/501100011033 and by ``ERDF A way of making Europe'', by the ``European Union''.}

\bibliography{references}

\begin{thebibliography}{88}%
\makeatletter
\providecommand \@ifxundefined [1]{%
 \@ifx{#1\undefined}
}%
\providecommand \@ifnum [1]{%
 \ifnum #1\expandafter \@firstoftwo
 \else \expandafter \@secondoftwo
 \fi
}%
\providecommand \@ifx [1]{%
 \ifx #1\expandafter \@firstoftwo
 \else \expandafter \@secondoftwo
 \fi
}%
\providecommand \natexlab [1]{#1}%
\providecommand \enquote  [1]{``#1''}%
\providecommand \bibnamefont  [1]{#1}%
\providecommand \bibfnamefont [1]{#1}%
\providecommand \citenamefont [1]{#1}%
\providecommand \href@noop [0]{\@secondoftwo}%
\providecommand \href [0]{\begingroup \@sanitize@url \@href}%
\providecommand \@href[1]{\@@startlink{#1}\@@href}%
\providecommand \@@href[1]{\endgroup#1\@@endlink}%
\providecommand \@sanitize@url [0]{\catcode `\\12\catcode `\$12\catcode
  `\&12\catcode `\#12\catcode `\^12\catcode `\_12\catcode `\%12\relax}%
\providecommand \@@startlink[1]{}%
\providecommand \@@endlink[0]{}%
\providecommand \url  [0]{\begingroup\@sanitize@url \@url }%
\providecommand \@url [1]{\endgroup\@href {#1}{\urlprefix }}%
\providecommand \urlprefix  [0]{URL }%
\providecommand \Eprint [0]{\href }%
\providecommand \doibase [0]{https://doi.org/}%
\providecommand \selectlanguage [0]{\@gobble}%
\providecommand \bibinfo  [0]{\@secondoftwo}%
\providecommand \bibfield  [0]{\@secondoftwo}%
\providecommand \translation [1]{[#1]}%
\providecommand \BibitemOpen [0]{}%
\providecommand \bibitemStop [0]{}%
\providecommand \bibitemNoStop [0]{.\EOS\space}%
\providecommand \EOS [0]{\spacefactor3000\relax}%
\providecommand \BibitemShut  [1]{\csname bibitem#1\endcsname}%
\let\auto@bib@innerbib\@empty
%</preamble>
\bibitem [{\citenamefont {Cheng}\ and\ \citenamefont
  {Zhao}(2019)}]{cheng2019novel_process}%
  \BibitemOpen
  \bibfield  {author} {\bibinfo {author} {\bibfnamefont {F.}~\bibnamefont
  {Cheng}}\ and\ \bibinfo {author} {\bibfnamefont {J.}~\bibnamefont {Zhao}},\
  }\bibfield  {title} {\bibinfo {title} {A novel process monitoring approach
  based on feature points distance dynamic autoencoder},\ }in\ \href
  {https://doi.org/https://doi.org/10.1016/B978-0-12-818634-3.50127-2} {\emph
  {\bibinfo {booktitle} {29th European Symposium on Computer Aided Process
  Engineering}}},\ \bibinfo {series} {Computer Aided Chemical Engineering},
  Vol.~\bibinfo {volume} {46},\ \bibinfo {editor} {edited by\ \bibinfo {editor}
  {\bibfnamefont {A.~A.}\ \bibnamefont {Kiss}}, \bibinfo {editor}
  {\bibfnamefont {E.}~\bibnamefont {Zondervan}}, \bibinfo {editor}
  {\bibfnamefont {R.}~\bibnamefont {Lakerveld}},\ and\ \bibinfo {editor}
  {\bibfnamefont {L.}~\bibnamefont {Özkan}}}\ (\bibinfo  {publisher}
  {Elsevier},\ \bibinfo {year} {2019})\ pp.\ \bibinfo {pages}
  {757--762}\BibitemShut {NoStop}%
\bibitem [{\citenamefont {Poznyak}\ \emph {et~al.}(2019)\citenamefont
  {Poznyak}, \citenamefont {{Chairez Oria}},\ and\ \citenamefont
  {Poznyak}}]{poznyak2019background_dynamical}%
  \BibitemOpen
  \bibfield  {author} {\bibinfo {author} {\bibfnamefont {T.~I.}\ \bibnamefont
  {Poznyak}}, \bibinfo {author} {\bibfnamefont {I.}~\bibnamefont {{Chairez
  Oria}}},\ and\ \bibinfo {author} {\bibfnamefont {A.~S.}\ \bibnamefont
  {Poznyak}},\ }\bibfield  {title} {\bibinfo {title} {Chapter3 - background on
  dynamic neural networks},\ }in\ \href
  {https://doi.org/https://doi.org/10.1016/B978-0-12-812847-3.00012-3} {\emph
  {\bibinfo {booktitle} {Ozonation and Biodegradation in Environmental
  Engineering}}},\ \bibinfo {editor} {edited by\ \bibinfo {editor}
  {\bibfnamefont {T.~I.}\ \bibnamefont {Poznyak}}, \bibinfo {editor}
  {\bibfnamefont {I.}~\bibnamefont {{Chairez Oria}}},\ and\ \bibinfo {editor}
  {\bibfnamefont {A.~S.}\ \bibnamefont {Poznyak}}}\ (\bibinfo  {publisher}
  {Elsevier},\ \bibinfo {year} {2019})\ pp.\ \bibinfo {pages}
  {57--74}\BibitemShut {NoStop}%
\bibitem [{\citenamefont {Hadian}\ \emph {et~al.}(2021)\citenamefont {Hadian},
  \citenamefont {Saryazdi}, \citenamefont {Mohammadzadeh},\ and\ \citenamefont
  {Babaei}}]{hadian2021application_artificial}%
  \BibitemOpen
  \bibfield  {author} {\bibinfo {author} {\bibfnamefont {M.}~\bibnamefont
  {Hadian}}, \bibinfo {author} {\bibfnamefont {S.~M.~E.}\ \bibnamefont
  {Saryazdi}}, \bibinfo {author} {\bibfnamefont {A.}~\bibnamefont
  {Mohammadzadeh}},\ and\ \bibinfo {author} {\bibfnamefont {M.}~\bibnamefont
  {Babaei}},\ }\bibfield  {title} {\bibinfo {title} {Chapter 11 - application
  of artificial intelligence in modeling, control, and fault diagnosis},\ }in\
  \href {https://doi.org/https://doi.org/10.1016/B978-0-12-821092-5.00006-1}
  {\emph {\bibinfo {booktitle} {Applications of Artificial Intelligence in
  Process Systems Engineering}}},\ \bibinfo {editor} {edited by\ \bibinfo
  {editor} {\bibfnamefont {J.}~\bibnamefont {Ren}}, \bibinfo {editor}
  {\bibfnamefont {W.}~\bibnamefont {Shen}}, \bibinfo {editor} {\bibfnamefont
  {Y.}~\bibnamefont {Man}},\ and\ \bibinfo {editor} {\bibfnamefont
  {L.}~\bibnamefont {Dong}}}\ (\bibinfo  {publisher} {Elsevier},\ \bibinfo
  {year} {2021})\ pp.\ \bibinfo {pages} {255--323}\BibitemShut {NoStop}%
\bibitem [{\citenamefont {Wu}\ and\ \citenamefont
  {Christofides}(2020)}]{wu2020smart_manufacturing}%
  \BibitemOpen
  \bibfield  {author} {\bibinfo {author} {\bibfnamefont {Z.}~\bibnamefont
  {Wu}}\ and\ \bibinfo {author} {\bibfnamefont {P.~D.}\ \bibnamefont
  {Christofides}},\ }\bibfield  {title} {\bibinfo {title} {Chapter 14 - smart
  manufacturing: Machine learning-based economic mpc and preventive
  maintenance},\ }in\ \href
  {https://doi.org/https://doi.org/10.1016/B978-0-12-820028-5.00014-X} {\emph
  {\bibinfo {booktitle} {Smart Manufacturing}}},\ \bibinfo {editor} {edited by\
  \bibinfo {editor} {\bibfnamefont {M.}~\bibnamefont {Soroush}}, \bibinfo
  {editor} {\bibfnamefont {M.}~\bibnamefont {Baldea}},\ and\ \bibinfo {editor}
  {\bibfnamefont {T.~F.}\ \bibnamefont {Edgar}}}\ (\bibinfo  {publisher}
  {Elsevier},\ \bibinfo {year} {2020})\ pp.\ \bibinfo {pages}
  {477--497}\BibitemShut {NoStop}%
\bibitem [{\citenamefont {Vaswani}\ \emph {et~al.}(2017)\citenamefont
  {Vaswani}, \citenamefont {Shazeer}, \citenamefont {Parmar}, \citenamefont
  {Uszkoreit}, \citenamefont {Jones}, \citenamefont {Gomez}, \citenamefont
  {Kaiser},\ and\ \citenamefont {Polosukhin}}]{vaswani2017attention}%
  \BibitemOpen
  \bibfield  {author} {\bibinfo {author} {\bibfnamefont {A.}~\bibnamefont
  {Vaswani}}, \bibinfo {author} {\bibfnamefont {N.}~\bibnamefont {Shazeer}},
  \bibinfo {author} {\bibfnamefont {N.}~\bibnamefont {Parmar}}, \bibinfo
  {author} {\bibfnamefont {J.}~\bibnamefont {Uszkoreit}}, \bibinfo {author}
  {\bibfnamefont {L.}~\bibnamefont {Jones}}, \bibinfo {author} {\bibfnamefont
  {A.~N.}\ \bibnamefont {Gomez}}, \bibinfo {author} {\bibfnamefont
  {L.}~\bibnamefont {Kaiser}},\ and\ \bibinfo {author} {\bibfnamefont
  {I.}~\bibnamefont {Polosukhin}},\ }\bibfield  {title} {\bibinfo {title}
  {Attention is all you need},\ }in\ \href@noop {} {\emph {\bibinfo {booktitle}
  {Proceedings of the 31st International Conference on Neural Information
  Processing Systems}}},\ \bibinfo {series and number} {NIPS'17}\ (\bibinfo
  {publisher} {Curran Associates Inc.},\ \bibinfo {address} {Red Hook, NY,
  USA},\ \bibinfo {year} {2017})\ p.\ \bibinfo {pages}
  {6000–6010}\BibitemShut {NoStop}%
\bibitem [{\citenamefont {Wolf}\ \emph {et~al.}(2020)\citenamefont {Wolf},
  \citenamefont {Debut}, \citenamefont {Sanh}, \citenamefont {Chaumond},
  \citenamefont {Delangue}, \citenamefont {Moi}, \citenamefont {Cistac},
  \citenamefont {Rault}, \citenamefont {Louf}, \citenamefont {Funtowicz},\ and\
  \citenamefont {Brew}}]{wolf2019huggingface}%
  \BibitemOpen
  \bibfield  {author} {\bibinfo {author} {\bibfnamefont {T.}~\bibnamefont
  {Wolf}}, \bibinfo {author} {\bibfnamefont {L.}~\bibnamefont {Debut}},
  \bibinfo {author} {\bibfnamefont {V.}~\bibnamefont {Sanh}}, \bibinfo {author}
  {\bibfnamefont {J.}~\bibnamefont {Chaumond}}, \bibinfo {author}
  {\bibfnamefont {C.}~\bibnamefont {Delangue}}, \bibinfo {author}
  {\bibfnamefont {A.}~\bibnamefont {Moi}}, \bibinfo {author} {\bibfnamefont
  {P.}~\bibnamefont {Cistac}}, \bibinfo {author} {\bibfnamefont
  {T.}~\bibnamefont {Rault}}, \bibinfo {author} {\bibfnamefont
  {R.}~\bibnamefont {Louf}}, \bibinfo {author} {\bibfnamefont {M.}~\bibnamefont
  {Funtowicz}},\ and\ \bibinfo {author} {\bibfnamefont {J.}~\bibnamefont
  {Brew}},\ }in\ \href {https://doi.org/10.18653/v1/2020.emnlp-demos.6} {\emph
  {\bibinfo {booktitle} {Proceedings of the 2020 Conference on Empirical
  Methods in Natural Language Processing: System Demonstrations}}}\ (\bibinfo
  {publisher} {Association for Computational Linguistics},\ \bibinfo {address}
  {Online},\ \bibinfo {year} {2020})\ pp.\ \bibinfo {pages}
  {38--45}\BibitemShut {NoStop}%
\bibitem [{Note1()}]{Note1}%
  \BibitemOpen
  \bibinfo {note} {Http://www.andi-challenge.org}\BibitemShut {NoStop}%
\bibitem [{\citenamefont {Mu{\~n}oz-Gil}\ \emph
  {et~al.}(2021{\natexlab{a}})\citenamefont {Mu{\~n}oz-Gil}, \citenamefont
  {Volpe}, \citenamefont {Garcia-March}, \citenamefont {Aghion}, \citenamefont
  {Argun}, \citenamefont {Hong}, \citenamefont {Bland}, \citenamefont {Bo},
  \citenamefont {Conejero}, \citenamefont {Garibo-i Orts},\ and\ \citenamefont
  {et~al.}}]{munoz-gil2021objective}%
  \BibitemOpen
  \bibfield  {author} {\bibinfo {author} {\bibfnamefont {G.}~\bibnamefont
  {Mu{\~n}oz-Gil}}, \bibinfo {author} {\bibfnamefont {G.}~\bibnamefont
  {Volpe}}, \bibinfo {author} {\bibfnamefont {M.}~\bibnamefont {Garcia-March}},
  \bibinfo {author} {\bibfnamefont {E.}~\bibnamefont {Aghion}}, \bibinfo
  {author} {\bibfnamefont {A.}~\bibnamefont {Argun}}, \bibinfo {author}
  {\bibfnamefont {C.}~\bibnamefont {Hong}}, \bibinfo {author} {\bibfnamefont
  {T.}~\bibnamefont {Bland}}, \bibinfo {author} {\bibfnamefont
  {S.}~\bibnamefont {Bo}}, \bibinfo {author} {\bibfnamefont {N.}~\bibnamefont
  {Conejero}, \bibfnamefont {J.A.and~Firbas}}, \bibinfo {author} {\bibfnamefont
  {O.}~\bibnamefont {Garibo-i Orts}},\ and\ \bibinfo {author} {\bibnamefont
  {et~al.}},\ }\bibfield  {title} {\bibinfo {title} {Objective comparison of
  methods to decode anomalous diffusion},\ }\bibfield  {journal} {\bibinfo
  {journal} {Nature Commun.}\ }\textbf {\bibinfo {volume} {12}},\ \href
  {https://doi.org/10.1038/s41467-021-26320-w} {10.1038/s41467-021-26320-w}
  (\bibinfo {year} {2021}{\natexlab{a}})\BibitemShut {NoStop}%
\bibitem [{\citenamefont {Cun}\ \emph {et~al.}(1990)\citenamefont {Cun},
  \citenamefont {Boser}, \citenamefont {Denker}, \citenamefont {Howard},
  \citenamefont {Habbard}, \citenamefont {Jackel},\ and\ \citenamefont
  {Henderson}}]{lecun1990handwritten}%
  \BibitemOpen
  \bibfield  {author} {\bibinfo {author} {\bibfnamefont {Y.}~\bibnamefont
  {Cun}}, \bibinfo {author} {\bibfnamefont {B.}~\bibnamefont {Boser}}, \bibinfo
  {author} {\bibfnamefont {J.}~\bibnamefont {Denker}}, \bibinfo {author}
  {\bibfnamefont {R.}~\bibnamefont {Howard}}, \bibinfo {author} {\bibfnamefont
  {W.}~\bibnamefont {Habbard}}, \bibinfo {author} {\bibfnamefont
  {L.}~\bibnamefont {Jackel}},\ and\ \bibinfo {author} {\bibfnamefont
  {D.}~\bibnamefont {Henderson}},\ }\bibinfo {title} {Handwritten digit
  recognition with a back-propagation network},\ in\ \href@noop {} {\emph
  {\bibinfo {booktitle} {Advances in Neural Information Processing Systems
  2}}}\ (\bibinfo  {publisher} {Morgan Kaufmann Publishers Inc.},\ \bibinfo
  {address} {San Francisco, CA, USA},\ \bibinfo {year} {1990})\ p.\ \bibinfo
  {pages} {396–404}\BibitemShut {NoStop}%
\bibitem [{\citenamefont {Hochreiter}\ and\ \citenamefont
  {Schmidhuber}(1997)}]{hochreiter1997LSTM}%
  \BibitemOpen
  \bibfield  {author} {\bibinfo {author} {\bibfnamefont {S.}~\bibnamefont
  {Hochreiter}}\ and\ \bibinfo {author} {\bibfnamefont {J.}~\bibnamefont
  {Schmidhuber}},\ }\bibfield  {title} {\bibinfo {title} {Long short-term
  memory},\ }\href {https://doi.org/10.1162/neco.1997.9.8.1735} {\bibfield
  {journal} {\bibinfo  {journal} {Neural Comput.}\ }\textbf {\bibinfo {volume}
  {9}},\ \bibinfo {pages} {1735} (\bibinfo {year} {1997})}\BibitemShut
  {NoStop}%
\bibitem [{\citenamefont {Garibo-i Orts}\ \emph {et~al.}(2021)\citenamefont
  {Garibo-i Orts}, \citenamefont {Baeza-Bosca}, \citenamefont {Garcia-March},\
  and\ \citenamefont {Conejero}}]{garibo-i-orts2021efficient}%
  \BibitemOpen
  \bibfield  {author} {\bibinfo {author} {\bibfnamefont {O.}~\bibnamefont
  {Garibo-i Orts}}, \bibinfo {author} {\bibfnamefont {A.}~\bibnamefont
  {Baeza-Bosca}}, \bibinfo {author} {\bibfnamefont {M.}~\bibnamefont
  {Garcia-March}},\ and\ \bibinfo {author} {\bibfnamefont {J.}~\bibnamefont
  {Conejero}},\ }\bibfield  {title} {\bibinfo {title} {Efficient recurrent
  neural network methods for anomalously diffusing single particle short and
  noisy trajectories},\ }\href@noop {} {\bibfield  {journal} {\bibinfo
  {journal} {J. Phys. A: Math. Theor.}\ }\textbf {\bibinfo {volume} {54}},\
  \bibinfo {pages} {504002} (\bibinfo {year} {2021})}\BibitemShut {NoStop}%
\bibitem [{Note2()}]{Note2}%
  \BibitemOpen
  \bibinfo {note} {\protect \url
  {http://andi-challenge.org/interactive-tool/}}\BibitemShut {NoStop}%
\bibitem [{\citenamefont {Brown}(1828)}]{brown1828brief}%
  \BibitemOpen
  \bibfield  {author} {\bibinfo {author} {\bibfnamefont {R.}~\bibnamefont
  {Brown}},\ }\bibfield  {title} {\bibinfo {title} {A brief account of
  microscopical observations made in the months of june, july, and august 1827,
  on the particles contained in the pollen of plants; and on the general
  existence of active molecules in organic and inorganic bodies},\ }\href
  {https://doi.org/10.1080/14786442808674769} {\bibfield  {journal} {\bibinfo
  {journal} {Philosoph. Mag.}\ }\textbf {\bibinfo {volume} {4}},\ \bibinfo
  {pages} {161} (\bibinfo {year} {1828})}\BibitemShut {NoStop}%
\bibitem [{\citenamefont {Flekk{\o}}\ \emph {et~al.}(2021)\citenamefont
  {Flekk{\o}}, \citenamefont {Hansen},\ and\ \citenamefont
  {Baldelli}}]{flekkoy2021_hyperballistic}%
  \BibitemOpen
  \bibfield  {author} {\bibinfo {author} {\bibfnamefont {E.}~\bibnamefont
  {Flekk{\o}}}, \bibinfo {author} {\bibfnamefont {A.}~\bibnamefont {Hansen}},\
  and\ \bibinfo {author} {\bibfnamefont {B.}~\bibnamefont {Baldelli}},\
  }\bibfield  {title} {\bibinfo {title} {Hyperballistic superdiffusion and
  explosive solutions to the non-linear diffusion equation},\ }\bibfield
  {journal} {\bibinfo  {journal} {Front. Phys.}\ }\textbf {\bibinfo {volume}
  {9}},\ \href {https://doi.org/10.3389/fphy.2021.640560}
  {10.3389/fphy.2021.640560} (\bibinfo {year} {2021})\BibitemShut {NoStop}%
\bibitem [{\citenamefont {Metzler}\ \emph {et~al.}(2014)\citenamefont
  {Metzler}, \citenamefont {Jeon}, \citenamefont {Cherstvy},\ and\
  \citenamefont {Barkai}}]{metzler2014anomalous}%
  \BibitemOpen
  \bibfield  {author} {\bibinfo {author} {\bibfnamefont {R.}~\bibnamefont
  {Metzler}}, \bibinfo {author} {\bibfnamefont {J.-H.}\ \bibnamefont {Jeon}},
  \bibinfo {author} {\bibfnamefont {A.}~\bibnamefont {Cherstvy}},\ and\
  \bibinfo {author} {\bibfnamefont {E.}~\bibnamefont {Barkai}},\ }\bibfield
  {title} {\bibinfo {title} {Anomalous diffusion models and their properties:
  non-stationarity, non-ergodicity, and ageing at the centenary of single
  particle tracking},\ }\href@noop {} {\bibfield  {journal} {\bibinfo
  {journal} {Phys. Chem. Chem. Phys.}\ }\textbf {\bibinfo {volume} {16}},\
  \bibinfo {pages} {24128} (\bibinfo {year} {2014})}\BibitemShut {NoStop}%
\bibitem [{\citenamefont {Vilk}\ \emph
  {et~al.}(2022{\natexlab{a}})\citenamefont {Vilk}, \citenamefont {Orchan},
  \citenamefont {Charter}, \citenamefont {Ganot}, \citenamefont {Toledo},
  \citenamefont {Nathan},\ and\ \citenamefont
  {Assaf}}]{Vilk2022_ergodicity_breaking_avian}%
  \BibitemOpen
  \bibfield  {author} {\bibinfo {author} {\bibfnamefont {O.}~\bibnamefont
  {Vilk}}, \bibinfo {author} {\bibfnamefont {Y.}~\bibnamefont {Orchan}},
  \bibinfo {author} {\bibfnamefont {M.}~\bibnamefont {Charter}}, \bibinfo
  {author} {\bibfnamefont {N.}~\bibnamefont {Ganot}}, \bibinfo {author}
  {\bibfnamefont {S.}~\bibnamefont {Toledo}}, \bibinfo {author} {\bibfnamefont
  {R.}~\bibnamefont {Nathan}},\ and\ \bibinfo {author} {\bibfnamefont
  {M.}~\bibnamefont {Assaf}},\ }\bibfield  {title} {\bibinfo {title}
  {Ergodicity breaking in area-restricted search of avian predators},\ }\href
  {https://doi.org/10.1103/PhysRevX.12.031005} {\bibfield  {journal} {\bibinfo
  {journal} {Phys. Rev. X}\ }\textbf {\bibinfo {volume} {12}},\ \bibinfo
  {pages} {031005} (\bibinfo {year} {2022}{\natexlab{a}})}\BibitemShut
  {NoStop}%
\bibitem [{\citenamefont {Mu{\~n}oz-Gil}\ \emph
  {et~al.}(2021{\natexlab{b}})\citenamefont {Mu{\~n}oz-Gil}, \citenamefont
  {Volpe}, \citenamefont {Garc{\'\i}a-March}, \citenamefont {Metzler},
  \citenamefont {Lewenstein},\ and\ \citenamefont {Manzo}}]{munoz-gil2021etai}%
  \BibitemOpen
  \bibfield  {author} {\bibinfo {author} {\bibfnamefont {G.}~\bibnamefont
  {Mu{\~n}oz-Gil}}, \bibinfo {author} {\bibfnamefont {G.}~\bibnamefont
  {Volpe}}, \bibinfo {author} {\bibfnamefont {M.}~\bibnamefont
  {Garc{\'\i}a-March}}, \bibinfo {author} {\bibfnamefont {R.}~\bibnamefont
  {Metzler}}, \bibinfo {author} {\bibfnamefont {M.}~\bibnamefont
  {Lewenstein}},\ and\ \bibinfo {author} {\bibfnamefont {c.}~\bibnamefont
  {Manzo}},\ }\bibfield  {title} {\bibinfo {title} {{The Anomalous Diffusion
  challenge: objective comparison of methods to decode anomalous diffusion}},\
  }in\ \href {https://doi.org/10.1117/12.2595716} {\emph {\bibinfo {booktitle}
  {Emerging Topics in Artificial Intelligence (ETAI) 2021}}},\ Vol.\ \bibinfo
  {volume} {11804},\ \bibinfo {editor} {edited by\ \bibinfo {editor}
  {\bibnamefont {G.Volpe}}, \bibinfo {editor} {\bibfnamefont {J.}~\bibnamefont
  {Pereira}}, \bibinfo {editor} {\bibnamefont {D.Brunner}},\ and\ \bibinfo
  {editor} {\bibfnamefont {A.}~\bibnamefont {Ozcan}}},\ \bibinfo {organization}
  {Int. Soc. Opt. Photonics}\ (\bibinfo  {publisher} {SPIE},\ \bibinfo {year}
  {2021})\ p.\ \bibinfo {pages} {1180416}\BibitemShut {NoStop}%
\bibitem [{\citenamefont {Scher}\ and\ \citenamefont
  {Montroll}(1975)}]{scher1975anomalous}%
  \BibitemOpen
  \bibfield  {author} {\bibinfo {author} {\bibfnamefont {H.}~\bibnamefont
  {Scher}}\ and\ \bibinfo {author} {\bibfnamefont {E.}~\bibnamefont
  {Montroll}},\ }\bibfield  {title} {\bibinfo {title} {Anomalous transit-time
  dispersion in amorphous solids},\ }\href
  {https://doi.org/10.1103/PhysRevB.12.2455} {\bibfield  {journal} {\bibinfo
  {journal} {Phys. Rev. B}\ }\textbf {\bibinfo {volume} {12}},\ \bibinfo
  {pages} {2455} (\bibinfo {year} {1975})}\BibitemShut {NoStop}%
\bibitem [{\citenamefont {Klafter}\ and\ \citenamefont
  {Zumofen}(1994)}]{klafter1994levy}%
  \BibitemOpen
  \bibfield  {author} {\bibinfo {author} {\bibfnamefont {J.}~\bibnamefont
  {Klafter}}\ and\ \bibinfo {author} {\bibfnamefont {G.}~\bibnamefont
  {Zumofen}},\ }\bibfield  {title} {\bibinfo {title} {L\'evy statistics in a
  {H}amiltonian system},\ }\href {https://doi.org/10.1103/PhysRevE.49.4873}
  {\bibfield  {journal} {\bibinfo  {journal} {Phys. Rev. E}\ }\textbf {\bibinfo
  {volume} {49}},\ \bibinfo {pages} {4873} (\bibinfo {year}
  {1994})}\BibitemShut {NoStop}%
\bibitem [{\citenamefont {Massignan}\ \emph {et~al.}(2014)\citenamefont
  {Massignan}, \citenamefont {Manzo}, \citenamefont {Torre{\~n}o-Pina},
  \citenamefont {Garcia-Parajo}, \citenamefont {Lewenstein},\ and\
  \citenamefont {Lapeyre}}]{massignan2014nonergodic}%
  \BibitemOpen
  \bibfield  {author} {\bibinfo {author} {\bibfnamefont {P.}~\bibnamefont
  {Massignan}}, \bibinfo {author} {\bibfnamefont {C.}~\bibnamefont {Manzo}},
  \bibinfo {author} {\bibfnamefont {J.}~\bibnamefont {Torre{\~n}o-Pina}},
  \bibinfo {author} {\bibfnamefont {M.}~\bibnamefont {Garcia-Parajo}}, \bibinfo
  {author} {\bibfnamefont {M.}~\bibnamefont {Lewenstein}},\ and\ \bibinfo
  {author} {\bibfnamefont {G.}~\bibnamefont {Lapeyre}},\ }\bibfield  {title}
  {\bibinfo {title} {Nonergodic subdiffusion from {B}rownian motion in an
  inhomogeneous medium.},\ }\href@noop {} {\bibfield  {journal} {\bibinfo
  {journal} {Phys. Rev. Lett.}\ }\textbf {\bibinfo {volume} {112}},\ \bibinfo
  {pages} {150603} (\bibinfo {year} {2014})}\BibitemShut {NoStop}%
\bibitem [{\citenamefont {Mandelbrot}\ and\ \citenamefont
  {Van~Ness}(1968)}]{mandelbrot1968fractional}%
  \BibitemOpen
  \bibfield  {author} {\bibinfo {author} {\bibfnamefont {B.}~\bibnamefont
  {Mandelbrot}}\ and\ \bibinfo {author} {\bibfnamefont {J.}~\bibnamefont
  {Van~Ness}},\ }\bibfield  {title} {\bibinfo {title} {Fractional {B}rownian
  motions, fractional noises and applications},\ }\href
  {https://doi.org/10.1137/1010093} {\bibfield  {journal} {\bibinfo  {journal}
  {SIAM Rev.}\ }\textbf {\bibinfo {volume} {10}},\ \bibinfo {pages} {422}
  (\bibinfo {year} {1968})}\BibitemShut {NoStop}%
\bibitem [{\citenamefont {Jeon}\ and\ \citenamefont
  {Metzler}(2010{\natexlab{a}})}]{jeon2010fractional}%
  \BibitemOpen
  \bibfield  {author} {\bibinfo {author} {\bibfnamefont {J.}~\bibnamefont
  {Jeon}}\ and\ \bibinfo {author} {\bibfnamefont {R.}~\bibnamefont {Metzler}},\
  }\bibfield  {title} {\bibinfo {title} {Fractional {B}rownian motion and
  motion governed by the fractional {L}angevin equation in confined
  geometries},\ }\href@noop {} {\bibfield  {journal} {\bibinfo  {journal}
  {Phys. Rev. E}\ }\textbf {\bibinfo {volume} {81}},\ \bibinfo {pages} {021103}
  (\bibinfo {year} {2010}{\natexlab{a}})}\BibitemShut {NoStop}%
\bibitem [{\citenamefont {Lim}\ and\ \citenamefont
  {Muniandy}(2002)}]{lim2002self-similar}%
  \BibitemOpen
  \bibfield  {author} {\bibinfo {author} {\bibfnamefont {S.}~\bibnamefont
  {Lim}}\ and\ \bibinfo {author} {\bibfnamefont {S.}~\bibnamefont {Muniandy}},\
  }\bibfield  {title} {\bibinfo {title} {Self-similar {G}aussian processes for
  modeling anomalous diffusion},\ }\href@noop {} {\bibfield  {journal}
  {\bibinfo  {journal} {Phys. Rev. E}\ }\textbf {\bibinfo {volume} {66}},\
  \bibinfo {pages} {021114} (\bibinfo {year} {2002})}\BibitemShut {NoStop}%
\bibitem [{\citenamefont {Oliveira}\ \emph {et~al.}(2019)\citenamefont
  {Oliveira}, \citenamefont {Ferreira}, \citenamefont {Lapas},\ and\
  \citenamefont {Vainstein}}]{oliveira2019anomalous}%
  \BibitemOpen
  \bibfield  {author} {\bibinfo {author} {\bibfnamefont {F.}~\bibnamefont
  {Oliveira}}, \bibinfo {author} {\bibfnamefont {R.}~\bibnamefont {Ferreira}},
  \bibinfo {author} {\bibfnamefont {L.}~\bibnamefont {Lapas}},\ and\ \bibinfo
  {author} {\bibfnamefont {M.}~\bibnamefont {Vainstein}},\ }\bibfield  {title}
  {\bibinfo {title} {Anomalous diffusion: A basic mechanism for the evolution
  of inhomogeneous systems},\ }\href@noop {} {\bibfield  {journal} {\bibinfo
  {journal} {Front. Phys.}\ }\textbf {\bibinfo {volume} {7}},\ \bibinfo {pages}
  {18} (\bibinfo {year} {2019})}\BibitemShut {NoStop}%
\bibitem [{\citenamefont {Klafter}\ and\ \citenamefont
  {Sokolov}(2005)}]{klafter2005anomalous}%
  \BibitemOpen
  \bibfield  {author} {\bibinfo {author} {\bibfnamefont {J.}~\bibnamefont
  {Klafter}}\ and\ \bibinfo {author} {\bibfnamefont {I.}~\bibnamefont
  {Sokolov}},\ }\bibfield  {title} {\bibinfo {title} {Anomalous diffusion
  spreads its wings},\ }\href@noop {} {\bibfield  {journal} {\bibinfo
  {journal} {Phy. World}\ }\textbf {\bibinfo {volume} {18}},\ \bibinfo {pages}
  {29} (\bibinfo {year} {2005})}\BibitemShut {NoStop}%
\bibitem [{\citenamefont {Sagi}\ \emph {et~al.}(2012)\citenamefont {Sagi},
  \citenamefont {Brook}, \citenamefont {Almog},\ and\ \citenamefont
  {Davidson}}]{sagi2012observation}%
  \BibitemOpen
  \bibfield  {author} {\bibinfo {author} {\bibfnamefont {Y.}~\bibnamefont
  {Sagi}}, \bibinfo {author} {\bibfnamefont {M.}~\bibnamefont {Brook}},
  \bibinfo {author} {\bibfnamefont {I.}~\bibnamefont {Almog}},\ and\ \bibinfo
  {author} {\bibfnamefont {N.}~\bibnamefont {Davidson}},\ }\bibfield  {title}
  {\bibinfo {title} {Observation of anomalous diffusion and fractional
  self-similarity in one dimension},\ }\href
  {https://doi.org/10.1103/PhysRevLett.108.093002} {\bibfield  {journal}
  {\bibinfo  {journal} {Phys. Rev. Lett.}\ }\textbf {\bibinfo {volume} {108}},\
  \bibinfo {pages} {093002} (\bibinfo {year} {2012})}\BibitemShut {NoStop}%
\bibitem [{\citenamefont {Dechant}\ \emph {et~al.}(2019)\citenamefont
  {Dechant}, \citenamefont {Kindermann}, \citenamefont {Widera},\ and\
  \citenamefont {Lutz}}]{dechant2019continous}%
  \BibitemOpen
  \bibfield  {author} {\bibinfo {author} {\bibfnamefont {A.}~\bibnamefont
  {Dechant}}, \bibinfo {author} {\bibfnamefont {F.}~\bibnamefont {Kindermann}},
  \bibinfo {author} {\bibfnamefont {A.}~\bibnamefont {Widera}},\ and\ \bibinfo
  {author} {\bibfnamefont {E.}~\bibnamefont {Lutz}},\ }\bibfield  {title}
  {\bibinfo {title} {Continuous-time random walk for a particle in a periodic
  potential},\ }\bibfield  {journal} {\bibinfo  {journal} {Phys. Rev. Lett.}\
  }\textbf {\bibinfo {volume} {123}},\ \href
  {https://doi.org/10.1103/physrevlett.123.070602}
  {10.1103/physrevlett.123.070602} (\bibinfo {year} {2019})\BibitemShut
  {NoStop}%
\bibitem [{\citenamefont {Kindermann}\ \emph {et~al.}(2017)\citenamefont
  {Kindermann}, \citenamefont {Dechant}, \citenamefont {Dipl}, \citenamefont
  {Lausch}, \citenamefont {Mayer}, \citenamefont {Schmidt}, \citenamefont
  {Lutz},\ and\ \citenamefont {Widera}}]{kindermann2017nonergodic}%
  \BibitemOpen
  \bibfield  {author} {\bibinfo {author} {\bibfnamefont {F.}~\bibnamefont
  {Kindermann}}, \bibinfo {author} {\bibfnamefont {A.}~\bibnamefont {Dechant}},
  \bibinfo {author} {\bibfnamefont {I.}~\bibnamefont {Dipl}, \bibfnamefont
  {M.~Hohmann}}, \bibinfo {author} {\bibfnamefont {T.}~\bibnamefont {Lausch}},
  \bibinfo {author} {\bibfnamefont {D.}~\bibnamefont {Mayer}}, \bibinfo
  {author} {\bibfnamefont {F.}~\bibnamefont {Schmidt}}, \bibinfo {author}
  {\bibfnamefont {E.}~\bibnamefont {Lutz}},\ and\ \bibinfo {author}
  {\bibfnamefont {A.}~\bibnamefont {Widera}},\ }\bibfield  {title} {\bibinfo
  {title} {Nonergodic diffusion of single atoms in a periodic potential},\
  }\href@noop {} {\bibfield  {journal} {\bibinfo  {journal} {Nature Phys.}\
  }\textbf {\bibinfo {volume} {13}},\ \bibinfo {pages} {137} (\bibinfo {year}
  {2017})}\BibitemShut {NoStop}%
\bibitem [{\citenamefont {Krapf}\ \emph {et~al.}(2019)\citenamefont {Krapf},
  \citenamefont {Lukat}, \citenamefont {Marinari}, \citenamefont {Metzler},
  \citenamefont {Oshanin}, \citenamefont {Selhuber-Unkel}, \citenamefont
  {Squarcini}, \citenamefont {Stadler}, \citenamefont {Weiss},\ and\
  \citenamefont {Xu}}]{krapf2019spectral}%
  \BibitemOpen
  \bibfield  {author} {\bibinfo {author} {\bibfnamefont {D.}~\bibnamefont
  {Krapf}}, \bibinfo {author} {\bibfnamefont {N.}~\bibnamefont {Lukat}},
  \bibinfo {author} {\bibfnamefont {E.}~\bibnamefont {Marinari}}, \bibinfo
  {author} {\bibfnamefont {R.}~\bibnamefont {Metzler}}, \bibinfo {author}
  {\bibfnamefont {G.}~\bibnamefont {Oshanin}}, \bibinfo {author} {\bibfnamefont
  {C.}~\bibnamefont {Selhuber-Unkel}}, \bibinfo {author} {\bibfnamefont
  {A.}~\bibnamefont {Squarcini}}, \bibinfo {author} {\bibfnamefont
  {L.}~\bibnamefont {Stadler}}, \bibinfo {author} {\bibfnamefont
  {M.}~\bibnamefont {Weiss}},\ and\ \bibinfo {author} {\bibfnamefont
  {X.}~\bibnamefont {Xu}},\ }\bibfield  {title} {\bibinfo {title} {Spectral
  content of a single non-{B}rownian trajectory},\ }\href
  {https://doi.org/10.1103/PhysRevX.9.011019} {\bibfield  {journal} {\bibinfo
  {journal} {Phys. Rev. X}\ }\textbf {\bibinfo {volume} {9}},\ \bibinfo {pages}
  {011019} (\bibinfo {year} {2019})}\BibitemShut {NoStop}%
\bibitem [{\citenamefont {Bronstein}\ \emph {et~al.}(2009)\citenamefont
  {Bronstein}, \citenamefont {Israel}, \citenamefont {Kepten}, \citenamefont
  {Mai}, \citenamefont {Shav-Tal}, \citenamefont {Barkai},\ and\ \citenamefont
  {Garini}}]{bronstein2009transient}%
  \BibitemOpen
  \bibfield  {author} {\bibinfo {author} {\bibfnamefont {I.}~\bibnamefont
  {Bronstein}}, \bibinfo {author} {\bibfnamefont {Y.}~\bibnamefont {Israel}},
  \bibinfo {author} {\bibfnamefont {E.}~\bibnamefont {Kepten}}, \bibinfo
  {author} {\bibfnamefont {S.}~\bibnamefont {Mai}}, \bibinfo {author}
  {\bibfnamefont {Y.}~\bibnamefont {Shav-Tal}}, \bibinfo {author}
  {\bibfnamefont {E.}~\bibnamefont {Barkai}},\ and\ \bibinfo {author}
  {\bibfnamefont {Y.}~\bibnamefont {Garini}},\ }\bibfield  {title} {\bibinfo
  {title} {Transient anomalous diffusion of telomeres in the nucleus of
  mammalian cells},\ }\href {https://doi.org/10.1103/PhysRevLett.103.018102}
  {\bibfield  {journal} {\bibinfo  {journal} {Phys. Rev. Lett.}\ }\textbf
  {\bibinfo {volume} {103}},\ \bibinfo {pages} {018102} (\bibinfo {year}
  {2009})}\BibitemShut {NoStop}%
\bibitem [{\citenamefont {Stadler}\ and\ \citenamefont
  {Weiss}(2017)}]{stadler2017non-equilibrium}%
  \BibitemOpen
  \bibfield  {author} {\bibinfo {author} {\bibfnamefont {L.}~\bibnamefont
  {Stadler}}\ and\ \bibinfo {author} {\bibfnamefont {M.}~\bibnamefont
  {Weiss}},\ }\bibfield  {title} {\bibinfo {title} {Non-equilibrium forces
  drive the anomalous diffusion of telomeres in the nucleus of mammalian
  cells},\ }\href {https://doi.org/10.1088/1367-2630/aa8fe1} {\bibfield
  {journal} {\bibinfo  {journal} {New J. Phys.}\ }\textbf {\bibinfo {volume}
  {19}},\ \bibinfo {pages} {113048} (\bibinfo {year} {2017})}\BibitemShut
  {NoStop}%
\bibitem [{\citenamefont {Manzo}\ and\ \citenamefont
  {Garcia-Parajo}(2015)}]{manzo2015review}%
  \BibitemOpen
  \bibfield  {author} {\bibinfo {author} {\bibfnamefont {C.}~\bibnamefont
  {Manzo}}\ and\ \bibinfo {author} {\bibfnamefont {M.}~\bibnamefont
  {Garcia-Parajo}},\ }\bibfield  {title} {\bibinfo {title} {A review of
  progress in single particle tracking: from methods to biophysical insights},\
  }\href@noop {} {\bibfield  {journal} {\bibinfo  {journal} {Rep. Prog. Phys.}\
  }\textbf {\bibinfo {volume} {78}},\ \bibinfo {pages} {124601} (\bibinfo
  {year} {2015})}\BibitemShut {NoStop}%
\bibitem [{\citenamefont {Caspi}\ \emph {et~al.}(2000)\citenamefont {Caspi},
  \citenamefont {Rony~Granek},\ and\ \citenamefont
  {Elbaum}}]{caspi2000enhanced}%
  \BibitemOpen
  \bibfield  {author} {\bibinfo {author} {\bibfnamefont {A.}~\bibnamefont
  {Caspi}}, \bibinfo {author} {\bibfnamefont {R.}~\bibnamefont {Rony~Granek}},\
  and\ \bibinfo {author} {\bibfnamefont {M.}~\bibnamefont {Elbaum}},\
  }\bibfield  {title} {\bibinfo {title} {Enhanced diffusion in active
  intracellular transport},\ }\href@noop {} {\bibfield  {journal} {\bibinfo
  {journal} {Phys. Rev. Lett.}\ }\textbf {\bibinfo {volume} {85}},\ \bibinfo
  {pages} {5655} (\bibinfo {year} {2000})}\BibitemShut {NoStop}%
\bibitem [{\citenamefont {Weber}\ \emph {et~al.}(2010)\citenamefont {Weber},
  \citenamefont {Spakowitz},\ and\ \citenamefont
  {Theriot}}]{weber2010bacterial}%
  \BibitemOpen
  \bibfield  {author} {\bibinfo {author} {\bibfnamefont {S.}~\bibnamefont
  {Weber}}, \bibinfo {author} {\bibfnamefont {A.}~\bibnamefont {Spakowitz}},\
  and\ \bibinfo {author} {\bibfnamefont {J.}~\bibnamefont {Theriot}},\
  }\bibfield  {title} {\bibinfo {title} {Bacterial chromosomal loci move
  subdiffusively through a viscoelastic cytoplasm.},\ }\href@noop {} {\bibfield
   {journal} {\bibinfo  {journal} {Phys. Rev. Lett.}\ }\textbf {\bibinfo
  {volume} {104}},\ \bibinfo {pages} {238102} (\bibinfo {year}
  {2010})}\BibitemShut {NoStop}%
\bibitem [{\citenamefont {Regner}\ \emph {et~al.}(2013)\citenamefont {Regner},
  \citenamefont {Vucinic}, \citenamefont {Domnisoru}, \citenamefont {Bartol},
  \citenamefont {Hetzer}, \citenamefont {Tartakovsky},\ and\ \citenamefont
  {Sejnowski}}]{regner2013anomalous}%
  \BibitemOpen
  \bibfield  {author} {\bibinfo {author} {\bibfnamefont {B.}~\bibnamefont
  {Regner}}, \bibinfo {author} {\bibfnamefont {D.}~\bibnamefont {Vucinic}},
  \bibinfo {author} {\bibfnamefont {C.}~\bibnamefont {Domnisoru}}, \bibinfo
  {author} {\bibfnamefont {T.}~\bibnamefont {Bartol}}, \bibinfo {author}
  {\bibfnamefont {M.~W.}\ \bibnamefont {Hetzer}}, \bibinfo {author}
  {\bibfnamefont {D.}~\bibnamefont {Tartakovsky}},\ and\ \bibinfo {author}
  {\bibfnamefont {T.}~\bibnamefont {Sejnowski}},\ }\bibfield  {title} {\bibinfo
  {title} {Anomalous diffusion of single particles in cytoplasm},\ }\href@noop
  {} {\bibfield  {journal} {\bibinfo  {journal} {Biophys. J.}\ }\textbf
  {\bibinfo {volume} {104 8}},\ \bibinfo {pages} {1652} (\bibinfo {year}
  {2013})}\BibitemShut {NoStop}%
\bibitem [{\citenamefont {Weigel}\ \emph {et~al.}(2011)\citenamefont {Weigel},
  \citenamefont {Simon}, \citenamefont {Tamkun},\ and\ \citenamefont
  {Krapf}}]{weigel2011ergodic}%
  \BibitemOpen
  \bibfield  {author} {\bibinfo {author} {\bibfnamefont {A.}~\bibnamefont
  {Weigel}}, \bibinfo {author} {\bibfnamefont {B.}~\bibnamefont {Simon}},
  \bibinfo {author} {\bibfnamefont {M.}~\bibnamefont {Tamkun}},\ and\ \bibinfo
  {author} {\bibfnamefont {D.}~\bibnamefont {Krapf}},\ }\bibfield  {title}
  {\bibinfo {title} {Ergodic and nonergodic processes coexist in the plasma
  membrane as observed by single-molecule tracking},\ }\href@noop {} {\bibfield
   {journal} {\bibinfo  {journal} {Proc. Natl. Acad. Sci.}\ }\textbf {\bibinfo
  {volume} {108}},\ \bibinfo {pages} {6438} (\bibinfo {year}
  {2011})}\BibitemShut {NoStop}%
\bibitem [{\citenamefont {Manzo}\ \emph {et~al.}(2015)\citenamefont {Manzo},
  \citenamefont {Torreno-Pina}, \citenamefont {Massignan}, \citenamefont
  {Lapeyre}, \citenamefont {Lewenstein},\ and\ \citenamefont
  {Garcia~Parajo}}]{manzo2015weak}%
  \BibitemOpen
  \bibfield  {author} {\bibinfo {author} {\bibfnamefont {C.}~\bibnamefont
  {Manzo}}, \bibinfo {author} {\bibfnamefont {J.}~\bibnamefont {Torreno-Pina}},
  \bibinfo {author} {\bibfnamefont {P.}~\bibnamefont {Massignan}}, \bibinfo
  {author} {\bibfnamefont {G.~J.}\ \bibnamefont {Lapeyre}}, \bibinfo {author}
  {\bibfnamefont {M.}~\bibnamefont {Lewenstein}},\ and\ \bibinfo {author}
  {\bibfnamefont {M.}~\bibnamefont {Garcia~Parajo}},\ }\bibfield  {title}
  {\bibinfo {title} {Weak ergodicity breaking of receptor motion in living
  cells stemming from random diffusivity},\ }\href
  {https://doi.org/10.1103/PhysRevX.5.011021} {\bibfield  {journal} {\bibinfo
  {journal} {Phys. Rev. X}\ }\textbf {\bibinfo {volume} {5}},\ \bibinfo {pages}
  {011021} (\bibinfo {year} {2015})}\BibitemShut {NoStop}%
\bibitem [{\citenamefont {Tolic-N{\o}rrelykke}\ \emph
  {et~al.}(2004)\citenamefont {Tolic-N{\o}rrelykke}, \citenamefont {Munteanu},
  \citenamefont {Thon}, \citenamefont {Broeng~Oddershede},\ and\ \citenamefont
  {Berg-S{\o}rensen}}]{tolic-norrelykke2004anomalous}%
  \BibitemOpen
  \bibfield  {author} {\bibinfo {author} {\bibfnamefont {I.}~\bibnamefont
  {Tolic-N{\o}rrelykke}}, \bibinfo {author} {\bibfnamefont {E.}~\bibnamefont
  {Munteanu}}, \bibinfo {author} {\bibfnamefont {G.}~\bibnamefont {Thon}},
  \bibinfo {author} {\bibfnamefont {L.}~\bibnamefont {Broeng~Oddershede}},\
  and\ \bibinfo {author} {\bibfnamefont {K.}~\bibnamefont {Berg-S{\o}rensen}},\
  }\bibfield  {title} {\bibinfo {title} {Anomalous diffusion in living yeast
  cells},\ }\href@noop {} {\bibfield  {journal} {\bibinfo  {journal} {Phys.
  Rev. Lett.}\ }\textbf {\bibinfo {volume} {93 7}},\ \bibinfo {pages} {078102}
  (\bibinfo {year} {2004})}\BibitemShut {NoStop}%
\bibitem [{\citenamefont {J.H.}\ \emph {et~al.}(2013)\citenamefont {J.H.},
  \citenamefont {Leijnse}, \citenamefont {Oddershede},\ and\ \citenamefont
  {Metzler}}]{jeon2013anomalous}%
  \BibitemOpen
  \bibfield  {author} {\bibinfo {author} {\bibfnamefont {J.}~\bibnamefont
  {J.H.}}, \bibinfo {author} {\bibfnamefont {N.}~\bibnamefont {Leijnse}},
  \bibinfo {author} {\bibfnamefont {L.}~\bibnamefont {Oddershede}},\ and\
  \bibinfo {author} {\bibfnamefont {R.}~\bibnamefont {Metzler}},\ }\bibfield
  {title} {\bibinfo {title} {Anomalous diffusion and power-law relaxation of
  the time averaged mean squared displacement in worm-like micellar
  solutions},\ }\href {https://doi.org/10.1088/1367-2630/15/4/045011}
  {\bibfield  {journal} {\bibinfo  {journal} {New J. Phys.}\ }\textbf {\bibinfo
  {volume} {15}},\ \bibinfo {pages} {045011} (\bibinfo {year}
  {2013})}\BibitemShut {NoStop}%
\bibitem [{\citenamefont {Ozarslan}\ \emph {et~al.}(2006)\citenamefont
  {Ozarslan}, \citenamefont {Basser}, \citenamefont {Shepherd}, \citenamefont
  {Thelwall}, \citenamefont {Vemuri},\ and\ \citenamefont
  {Blackband}}]{ozarslan2006anomalous}%
  \BibitemOpen
  \bibfield  {author} {\bibinfo {author} {\bibfnamefont {E.}~\bibnamefont
  {Ozarslan}}, \bibinfo {author} {\bibfnamefont {P.}~\bibnamefont {Basser}},
  \bibinfo {author} {\bibfnamefont {T.}~\bibnamefont {Shepherd}}, \bibinfo
  {author} {\bibfnamefont {P.}~\bibnamefont {Thelwall}}, \bibinfo {author}
  {\bibfnamefont {B.}~\bibnamefont {Vemuri}},\ and\ \bibinfo {author}
  {\bibfnamefont {S.}~\bibnamefont {Blackband}},\ }\bibfield  {title} {\bibinfo
  {title} {Observation of anomalous diffusion in excised tissue by
  characterizing the diffusion-time dependence of the {MR} signal},\ }\href
  {https://doi.org/10.1016/j.jmr.2006.08.009} {\bibfield  {journal} {\bibinfo
  {journal} {J. Magn. Reson.}\ }\textbf {\bibinfo {volume} {183}},\ \bibinfo
  {pages} {315} (\bibinfo {year} {2006})}\BibitemShut {NoStop}%
\bibitem [{\citenamefont {Magin}\ \emph {et~al.}(2013)\citenamefont {Magin},
  \citenamefont {Ingo}, \citenamefont {Colon-Perez}, \citenamefont {Triplett},\
  and\ \citenamefont {Mareci}}]{magin2013anomalous}%
  \BibitemOpen
  \bibfield  {author} {\bibinfo {author} {\bibfnamefont {R.}~\bibnamefont
  {Magin}}, \bibinfo {author} {\bibfnamefont {C.}~\bibnamefont {Ingo}},
  \bibinfo {author} {\bibfnamefont {L.}~\bibnamefont {Colon-Perez}}, \bibinfo
  {author} {\bibfnamefont {W.}~\bibnamefont {Triplett}},\ and\ \bibinfo
  {author} {\bibfnamefont {T.}~\bibnamefont {Mareci}},\ }\bibfield  {title}
  {\bibinfo {title} {Anomalous diffusion in porous biological tissues using
  fractional order derivatives and entropy},\ }\href
  {https://doi.org/doi.org/10.1016/j.micromeso.2013.02.054} {\bibfield
  {journal} {\bibinfo  {journal} {Microporous Mesoporous Mater.}\ }\textbf
  {\bibinfo {volume} {178}},\ \bibinfo {pages} {39–43} (\bibinfo {year}
  {2013})}\BibitemShut {NoStop}%
\bibitem [{\citenamefont {Zhang}\ and\ \citenamefont
  {Angst}(2020)}]{zhang2020dual-permeability}%
  \BibitemOpen
  \bibfield  {author} {\bibinfo {author} {\bibfnamefont {Z.}~\bibnamefont
  {Zhang}}\ and\ \bibinfo {author} {\bibfnamefont {U.}~\bibnamefont {Angst}},\
  }\bibfield  {title} {\bibinfo {title} {A dual-permeability approach to study
  anomalous moisture transport properties of cement-based materials},\ }\href
  {https://doi.org/10.1007/s11242-020-01469-y} {\bibfield  {journal} {\bibinfo
  {journal} {Transp. Porous Media}\ }\textbf {\bibinfo {volume} {135}},\
  \bibinfo {pages} {59} (\bibinfo {year} {2020})}\BibitemShut {NoStop}%
\bibitem [{\citenamefont {Vilk}\ \emph
  {et~al.}(2022{\natexlab{b}})\citenamefont {Vilk}, \citenamefont {Aghion},
  \citenamefont {Avgar}, \citenamefont {Beta}, \citenamefont {Nagel},
  \citenamefont {Sabri}, \citenamefont {Sarfati}, \citenamefont {Schwartz},
  \citenamefont {Weiss}, \citenamefont {Krapf} \emph
  {et~al.}}]{vilk2021unravelling}%
  \BibitemOpen
  \bibfield  {author} {\bibinfo {author} {\bibfnamefont {O.}~\bibnamefont
  {Vilk}}, \bibinfo {author} {\bibfnamefont {E.}~\bibnamefont {Aghion}},
  \bibinfo {author} {\bibfnamefont {T.}~\bibnamefont {Avgar}}, \bibinfo
  {author} {\bibfnamefont {C.}~\bibnamefont {Beta}}, \bibinfo {author}
  {\bibfnamefont {O.}~\bibnamefont {Nagel}}, \bibinfo {author} {\bibfnamefont
  {A.}~\bibnamefont {Sabri}}, \bibinfo {author} {\bibfnamefont
  {R.}~\bibnamefont {Sarfati}}, \bibinfo {author} {\bibfnamefont {D.~K.}\
  \bibnamefont {Schwartz}}, \bibinfo {author} {\bibfnamefont {M.}~\bibnamefont
  {Weiss}}, \bibinfo {author} {\bibfnamefont {D.}~\bibnamefont {Krapf}}, \emph
  {et~al.},\ }\bibfield  {title} {\bibinfo {title} {Unravelling the origins of
  anomalous diffusion: from molecules to migrating storks},\ }\href@noop {}
  {\bibfield  {journal} {\bibinfo  {journal} {Physical Review Research}\
  }\textbf {\bibinfo {volume} {4}},\ \bibinfo {pages} {033055} (\bibinfo {year}
  {2022}{\natexlab{b}})}\BibitemShut {NoStop}%
\bibitem [{\citenamefont {Magdziarz}\ \emph {et~al.}(2009)\citenamefont
  {Magdziarz}, \citenamefont {Weron}, \citenamefont {Burnecki},\ and\
  \citenamefont {Klafter}}]{magdziarz2009fractional}%
  \BibitemOpen
  \bibfield  {author} {\bibinfo {author} {\bibfnamefont {M.}~\bibnamefont
  {Magdziarz}}, \bibinfo {author} {\bibfnamefont {A.}~\bibnamefont {Weron}},
  \bibinfo {author} {\bibfnamefont {K.}~\bibnamefont {Burnecki}},\ and\
  \bibinfo {author} {\bibfnamefont {J.}~\bibnamefont {Klafter}},\ }\bibfield
  {title} {\bibinfo {title} {Fractional {B}rownian motion versus the
  continuous-time random walk: a simple test for subdiffusive dynamics},\
  }\href@noop {} {\bibfield  {journal} {\bibinfo  {journal} {Phys. Rev. Let.}\
  }\textbf {\bibinfo {volume} {103 18}},\ \bibinfo {pages} {180602} (\bibinfo
  {year} {2009})}\BibitemShut {NoStop}%
\bibitem [{\citenamefont {Golding}\ and\ \citenamefont
  {Cox}(2006)}]{golding2006physical}%
  \BibitemOpen
  \bibfield  {author} {\bibinfo {author} {\bibfnamefont {I.}~\bibnamefont
  {Golding}}\ and\ \bibinfo {author} {\bibfnamefont {E.}~\bibnamefont {Cox}},\
  }\bibfield  {title} {\bibinfo {title} {Physical nature of bacterial
  cytoplasm},\ }\href {https://doi.org/10.1103/PhysRevLett.96.098102}
  {\bibfield  {journal} {\bibinfo  {journal} {Phys. Rev. Lett.}\ }\textbf
  {\bibinfo {volume} {96}},\ \bibinfo {pages} {098102} (\bibinfo {year}
  {2006})}\BibitemShut {NoStop}%
\bibitem [{\citenamefont {Magdziarz}\ and\ \citenamefont
  {Weron}(2011)}]{magdziarz2011anomalous}%
  \BibitemOpen
  \bibfield  {author} {\bibinfo {author} {\bibfnamefont {M.}~\bibnamefont
  {Magdziarz}}\ and\ \bibinfo {author} {\bibfnamefont {A.}~\bibnamefont
  {Weron}},\ }\bibfield  {title} {\bibinfo {title} {Anomalous diffusion:
  {T}esting ergodicity breaking in experimental data},\ }\href@noop {}
  {\bibfield  {journal} {\bibinfo  {journal} {Phys. Rev. E}\ }\textbf {\bibinfo
  {volume} {84}},\ \bibinfo {pages} {051138} (\bibinfo {year}
  {2011})}\BibitemShut {NoStop}%
\bibitem [{\citenamefont {He}\ \emph {et~al.}(2008)\citenamefont {He},
  \citenamefont {Burov}, \citenamefont {Metzler},\ and\ \citenamefont
  {Barkai}}]{he2008random}%
  \BibitemOpen
  \bibfield  {author} {\bibinfo {author} {\bibfnamefont {Y.}~\bibnamefont
  {He}}, \bibinfo {author} {\bibfnamefont {S.}~\bibnamefont {Burov}}, \bibinfo
  {author} {\bibfnamefont {R.}~\bibnamefont {Metzler}},\ and\ \bibinfo {author}
  {\bibfnamefont {E.}~\bibnamefont {Barkai}},\ }\bibfield  {title} {\bibinfo
  {title} {Random time-scale invariant diffusion and transport coefficients},\
  }\href@noop {} {\bibfield  {journal} {\bibinfo  {journal} {Phys. Rev. Let.}\
  }\textbf {\bibinfo {volume} {101}},\ \bibinfo {pages} {058101} (\bibinfo
  {year} {2008})}\BibitemShut {NoStop}%
\bibitem [{\citenamefont {Molina-Garc{\'\i}a}\ \emph
  {et~al.}(2016)\citenamefont {Molina-Garc{\'\i}a}, \citenamefont {Pham},
  \citenamefont {Paradisi}, \citenamefont {Manzo},\ and\ \citenamefont
  {Pagnini}}]{molina-garcia2016fractional}%
  \BibitemOpen
  \bibfield  {author} {\bibinfo {author} {\bibfnamefont {D.}~\bibnamefont
  {Molina-Garc{\'\i}a}}, \bibinfo {author} {\bibfnamefont {T.}~\bibnamefont
  {Pham}}, \bibinfo {author} {\bibfnamefont {P.}~\bibnamefont {Paradisi}},
  \bibinfo {author} {\bibfnamefont {C.}~\bibnamefont {Manzo}},\ and\ \bibinfo
  {author} {\bibfnamefont {G.}~\bibnamefont {Pagnini}},\ }\bibfield  {title}
  {\bibinfo {title} {Fractional kinetics emerging from ergodicity breaking in
  random media},\ }\href@noop {} {\bibfield  {journal} {\bibinfo  {journal}
  {Phys. Rev. E}\ }\textbf {\bibinfo {volume} {94}},\ \bibinfo {pages} {052147}
  (\bibinfo {year} {2016})}\BibitemShut {NoStop}%
\bibitem [{\citenamefont {Thiel}\ and\ \citenamefont
  {Sokolov}(2014)}]{thiel2013-weak-ergodicity-breaking}%
  \BibitemOpen
  \bibfield  {author} {\bibinfo {author} {\bibfnamefont {F.}~\bibnamefont
  {Thiel}}\ and\ \bibinfo {author} {\bibfnamefont {I.~M.}\ \bibnamefont
  {Sokolov}},\ }\bibfield  {title} {\bibinfo {title} {Weak ergodicity breaking
  in an anomalous diffusion process of mixed origins},\ }\href
  {https://doi.org/10.1103/PhysRevE.89.012136} {\bibfield  {journal} {\bibinfo
  {journal} {Phys. Rev. E}\ }\textbf {\bibinfo {volume} {89}},\ \bibinfo
  {pages} {012136} (\bibinfo {year} {2014})}\BibitemShut {NoStop}%
\bibitem [{\citenamefont {Godec}\ and\ \citenamefont
  {Metzler}(2013{\natexlab{a}})}]{godec2013finite-time}%
  \BibitemOpen
  \bibfield  {author} {\bibinfo {author} {\bibfnamefont {A.}~\bibnamefont
  {Godec}}\ and\ \bibinfo {author} {\bibfnamefont {R.}~\bibnamefont
  {Metzler}},\ }\bibfield  {title} {\bibinfo {title} {Finite-time effects and
  ultraweak ergodicity breaking in superdiffusive dynamics},\ }\href@noop {}
  {\bibfield  {journal} {\bibinfo  {journal} {Phys. Rev. Lett.}\ }\textbf
  {\bibinfo {volume} {110 2}},\ \bibinfo {pages} {020603} (\bibinfo {year}
  {2013}{\natexlab{a}})}\BibitemShut {NoStop}%
\bibitem [{\citenamefont {Godec}\ and\ \citenamefont
  {Metzler}(2013{\natexlab{b}})}]{godec2013linear}%
  \BibitemOpen
  \bibfield  {author} {\bibinfo {author} {\bibfnamefont {A.}~\bibnamefont
  {Godec}}\ and\ \bibinfo {author} {\bibfnamefont {R.}~\bibnamefont
  {Metzler}},\ }\bibfield  {title} {\bibinfo {title} {Linear response,
  fluctuation-dissipation, and finite-system-size effects in superdiffusion},\
  }\href {https://doi.org/10.1103/PhysRevE.88.012116} {\bibfield  {journal}
  {\bibinfo  {journal} {Phys. Rev. E}\ }\textbf {\bibinfo {volume} {88}},\
  \bibinfo {pages} {012116} (\bibinfo {year} {2013}{\natexlab{b}})}\BibitemShut
  {NoStop}%
\bibitem [{\citenamefont {Deng}\ and\ \citenamefont
  {Barkai}(2009)}]{deng2009ergodic}%
  \BibitemOpen
  \bibfield  {author} {\bibinfo {author} {\bibfnamefont {W.}~\bibnamefont
  {Deng}}\ and\ \bibinfo {author} {\bibfnamefont {E.}~\bibnamefont {Barkai}},\
  }\bibfield  {title} {\bibinfo {title} {Ergodic properties of fractional
  {B}rownian-{L}angevin motion.},\ }\href@noop {} {\bibfield  {journal}
  {\bibinfo  {journal} {Phys. Rev. E}\ }\textbf {\bibinfo {volume} {79}},\
  \bibinfo {pages} {011112} (\bibinfo {year} {2009})}\BibitemShut {NoStop}%
\bibitem [{\citenamefont {Schwarzl}\ \emph {et~al.}(2017)\citenamefont
  {Schwarzl}, \citenamefont {Godec},\ and\ \citenamefont
  {Metzler}}]{schwarzl2017quantifying}%
  \BibitemOpen
  \bibfield  {author} {\bibinfo {author} {\bibfnamefont {M.}~\bibnamefont
  {Schwarzl}}, \bibinfo {author} {\bibfnamefont {A.}~\bibnamefont {Godec}},\
  and\ \bibinfo {author} {\bibfnamefont {R.}~\bibnamefont {Metzler}},\
  }\bibfield  {title} {\bibinfo {title} {Quantifying non-ergodicity of
  anomalous diffusion with higher order moments.},\ }\href
  {https://doi.org/10.1038/s41598-017-03712-x} {\bibfield  {journal} {\bibinfo
  {journal} {Sci. Rep.}\ }\textbf {\bibinfo {volume} {7}},\ \bibinfo {pages}
  {3878} (\bibinfo {year} {2017})}\BibitemShut {NoStop}%
\bibitem [{\citenamefont {Mardoukhi}\ \emph {et~al.}(2020)\citenamefont
  {Mardoukhi}, \citenamefont {Chechkin},\ and\ \citenamefont
  {Metzler}}]{mardoukhi2020spurious}%
  \BibitemOpen
  \bibfield  {author} {\bibinfo {author} {\bibfnamefont {Y.}~\bibnamefont
  {Mardoukhi}}, \bibinfo {author} {\bibfnamefont {A.}~\bibnamefont
  {Chechkin}},\ and\ \bibinfo {author} {\bibfnamefont {M.}~\bibnamefont
  {Metzler}},\ }\bibfield  {title} {\bibinfo {title} {Spurious ergodicity
  breaking in normal and fractional {O}rnstein--{U}hlenbeck process},\
  }\href@noop {} {\bibfield  {journal} {\bibinfo  {journal} {New J. Phys.}\
  }\textbf {\bibinfo {volume} {22}},\ \bibinfo {pages} {073012} (\bibinfo
  {year} {2020})}\BibitemShut {NoStop}%
\bibitem [{\citenamefont {Chenouard}\ \emph {et~al.}(2014)\citenamefont
  {Chenouard}, \citenamefont {Smal}, \citenamefont {de~Chaumont}, \citenamefont
  {Ma\v{s}ka}, \citenamefont {Sbalzarini}, \citenamefont {Gong}, \citenamefont
  {Cardinale}, \citenamefont {Carthel}, \citenamefont {Coraluppi},
  \citenamefont {Winter}, \citenamefont {Cohen},\ and\ \citenamefont
  {et~al}}]{chenouard2014objective}%
  \BibitemOpen
  \bibfield  {author} {\bibinfo {author} {\bibfnamefont {N.}~\bibnamefont
  {Chenouard}}, \bibinfo {author} {\bibfnamefont {I.}~\bibnamefont {Smal}},
  \bibinfo {author} {\bibfnamefont {F.}~\bibnamefont {de~Chaumont}}, \bibinfo
  {author} {\bibfnamefont {M.}~\bibnamefont {Ma\v{s}ka}}, \bibinfo {author}
  {\bibfnamefont {I.}~\bibnamefont {Sbalzarini}}, \bibinfo {author}
  {\bibfnamefont {Y.}~\bibnamefont {Gong}}, \bibinfo {author} {\bibfnamefont
  {J.}~\bibnamefont {Cardinale}}, \bibinfo {author} {\bibfnamefont {C.~A.}\
  \bibnamefont {Carthel}}, \bibinfo {author} {\bibfnamefont {S.~P.}\
  \bibnamefont {Coraluppi}}, \bibinfo {author} {\bibfnamefont {M.~R.}\
  \bibnamefont {Winter}}, \bibinfo {author} {\bibfnamefont {A.~R.}\
  \bibnamefont {Cohen}},\ and\ \bibinfo {author} {\bibnamefont {et~al}},\
  }\bibfield  {title} {\bibinfo {title} {Objective comparison of particle
  tracking methods},\ }\href@noop {} {\bibfield  {journal} {\bibinfo  {journal}
  {Nature Meth.}\ }\textbf {\bibinfo {volume} {11}},\ \bibinfo {pages} {281 }
  (\bibinfo {year} {2014})}\BibitemShut {NoStop}%
\bibitem [{\citenamefont {Jeon}\ \emph {et~al.}(2013)\citenamefont {Jeon},
  \citenamefont {Barkai},\ and\ \citenamefont {Ralf~Metzler}}]{jeon2013noisy}%
  \BibitemOpen
  \bibfield  {author} {\bibinfo {author} {\bibfnamefont {J.}~\bibnamefont
  {Jeon}}, \bibinfo {author} {\bibfnamefont {E.}~\bibnamefont {Barkai}},\ and\
  \bibinfo {author} {\bibfnamefont {R.}~\bibnamefont {Ralf~Metzler}},\
  }\bibfield  {title} {\bibinfo {title} {Noisy continuous time random walks},\
  }\href@noop {} {\bibfield  {journal} {\bibinfo  {journal} {J. Chem. Phys.}\
  }\textbf {\bibinfo {volume} {139}},\ \bibinfo {pages} {121916} (\bibinfo
  {year} {2013})}\BibitemShut {NoStop}%
\bibitem [{\citenamefont {Benmehdi}\ \emph {et~al.}(2011)\citenamefont
  {Benmehdi}, \citenamefont {Makarava}, \citenamefont {Benhamidouche},\ and\
  \citenamefont {Holschneider}}]{benmehdi2011bayesian}%
  \BibitemOpen
  \bibfield  {author} {\bibinfo {author} {\bibfnamefont {S.}~\bibnamefont
  {Benmehdi}}, \bibinfo {author} {\bibfnamefont {N.}~\bibnamefont {Makarava}},
  \bibinfo {author} {\bibfnamefont {N.}~\bibnamefont {Benhamidouche}},\ and\
  \bibinfo {author} {\bibfnamefont {M.}~\bibnamefont {Holschneider}},\
  }\bibfield  {title} {\bibinfo {title} {Bayesian estimation of the
  self-similarity exponent of the {N}ile {R}iver fluctuation},\ }\href
  {https://doi.org/10.5194/npg-18-441-2011} {\bibfield  {journal} {\bibinfo
  {journal} {Nonlinear Process. Geophys.}\ }\textbf {\bibinfo {volume} {18}},\
  \bibinfo {pages} {441} (\bibinfo {year} {2011})}\BibitemShut {NoStop}%
\bibitem [{\citenamefont {Hinsen}\ and\ \citenamefont
  {Kneller}(2016)}]{hinsen2016communication}%
  \BibitemOpen
  \bibfield  {author} {\bibinfo {author} {\bibfnamefont {K.}~\bibnamefont
  {Hinsen}}\ and\ \bibinfo {author} {\bibfnamefont {G.}~\bibnamefont
  {Kneller}},\ }\bibfield  {title} {\bibinfo {title} {Communication: {A}
  multiscale {B}ayesian inference approach to analyzing subdiffusion in
  particle trajectories},\ }\href@noop {} {\bibfield  {journal} {\bibinfo
  {journal} {J. Chem. Phys.}\ }\textbf {\bibinfo {volume} {145}},\ \bibinfo
  {pages} {151101} (\bibinfo {year} {2016})}\BibitemShut {NoStop}%
\bibitem [{\citenamefont {Agliari}\ \emph {et~al.}(2020)\citenamefont
  {Agliari}, \citenamefont {S{\'a}ez}, \citenamefont {Barra}, \citenamefont
  {Piel}, \citenamefont {Vargas},\ and\ \citenamefont
  {Castellana}}]{agliari2020statistical}%
  \BibitemOpen
  \bibfield  {author} {\bibinfo {author} {\bibfnamefont {E.}~\bibnamefont
  {Agliari}}, \bibinfo {author} {\bibfnamefont {P.}~\bibnamefont {S{\'a}ez}},
  \bibinfo {author} {\bibfnamefont {A.}~\bibnamefont {Barra}}, \bibinfo
  {author} {\bibfnamefont {M.}~\bibnamefont {Piel}}, \bibinfo {author}
  {\bibfnamefont {P.}~\bibnamefont {Vargas}},\ and\ \bibinfo {author}
  {\bibfnamefont {M.}~\bibnamefont {Castellana}},\ }\bibfield  {title}
  {\bibinfo {title} {A statistical inference approach to reconstruct
  intercellular interactions in cell migration experiments},\ }\href@noop {}
  {\bibfield  {journal} {\bibinfo  {journal} {Sci. Adv.}\ }\textbf {\bibinfo
  {volume} {6}},\ \bibinfo {pages} {eaay2103} (\bibinfo {year} {2020})},\
  \Eprint
  {https://arxiv.org/abs/https://www.science.org/doi/pdf/10.1126/sciadv.aay2103}
  {https://www.science.org/doi/pdf/10.1126/sciadv.aay2103} \BibitemShut
  {NoStop}%
\bibitem [{\citenamefont {Burnecki}\ \emph {et~al.}(2015)\citenamefont
  {Burnecki}, \citenamefont {Kepten}, \citenamefont {Y.}, \citenamefont
  {Sikora},\ and\ \citenamefont {A.}}]{burnecki2015estimating}%
  \BibitemOpen
  \bibfield  {author} {\bibinfo {author} {\bibfnamefont {K.}~\bibnamefont
  {Burnecki}}, \bibinfo {author} {\bibfnamefont {E.}~\bibnamefont {Kepten}},
  \bibinfo {author} {\bibfnamefont {G.}~\bibnamefont {Y.}}, \bibinfo {author}
  {\bibfnamefont {G.}~\bibnamefont {Sikora}},\ and\ \bibinfo {author}
  {\bibfnamefont {W.}~\bibnamefont {A.}},\ }\bibfield  {title} {\bibinfo
  {title} {Estimating the anomalous diffusion exponent for single particle
  tracking data with measurement errors - {A}n alternative approach},\
  }\href@noop {} {\bibfield  {journal} {\bibinfo  {journal} {Sci. Rep.}\
  }\textbf {\bibinfo {volume} {5}},\ \bibinfo {pages} {11306} (\bibinfo {year}
  {2015})}\BibitemShut {NoStop}%
\bibitem [{\citenamefont {Krapf}\ \emph {et~al.}(2018)\citenamefont {Krapf},
  \citenamefont {Marinari}, \citenamefont {Metzler}, \citenamefont {Oshanin},
  \citenamefont {Xu},\ and\ \citenamefont {Squarcini}}]{krapf2018power}%
  \BibitemOpen
  \bibfield  {author} {\bibinfo {author} {\bibfnamefont {D.}~\bibnamefont
  {Krapf}}, \bibinfo {author} {\bibfnamefont {E.}~\bibnamefont {Marinari}},
  \bibinfo {author} {\bibfnamefont {R.}~\bibnamefont {Metzler}}, \bibinfo
  {author} {\bibfnamefont {G.}~\bibnamefont {Oshanin}}, \bibinfo {author}
  {\bibfnamefont {X.}~\bibnamefont {Xu}},\ and\ \bibinfo {author}
  {\bibfnamefont {A.}~\bibnamefont {Squarcini}},\ }\bibfield  {title} {\bibinfo
  {title} {Power spectral density of a single {B}rownian trajectory: what one
  can and cannot learn from it},\ }\href@noop {} {\bibfield  {journal}
  {\bibinfo  {journal} {New J. Phys.}\ }\textbf {\bibinfo {volume} {20}},\
  \bibinfo {pages} {023029} (\bibinfo {year} {2018})}\BibitemShut {NoStop}%
\bibitem [{\citenamefont {Thapa}\ \emph {et~al.}(2018)\citenamefont {Thapa},
  \citenamefont {Lomholt}, \citenamefont {Krog}, \citenamefont {Cherstvy},\
  and\ \citenamefont {Metzler}}]{thapa2018bayesian}%
  \BibitemOpen
  \bibfield  {author} {\bibinfo {author} {\bibfnamefont {S.}~\bibnamefont
  {Thapa}}, \bibinfo {author} {\bibfnamefont {M.}~\bibnamefont {Lomholt}},
  \bibinfo {author} {\bibfnamefont {J.}~\bibnamefont {Krog}}, \bibinfo {author}
  {\bibfnamefont {A.}~\bibnamefont {Cherstvy}},\ and\ \bibinfo {author}
  {\bibfnamefont {R.}~\bibnamefont {Metzler}},\ }\bibfield  {title} {\bibinfo
  {title} {Bayesian analysis of single-particle tracking data using the
  nested-sampling algorithm: maximum-likelihood model selection applied to
  stochastic-diffusivity data},\ }\href {https://doi.org/10.1039/C8CP04043E}
  {\bibfield  {journal} {\bibinfo  {journal} {Phys. Chem. Chem. Phys.}\
  }\textbf {\bibinfo {volume} {20}},\ \bibinfo {pages} {29018} (\bibinfo {year}
  {2018})}\BibitemShut {NoStop}%
\bibitem [{\citenamefont {Jeon}\ and\ \citenamefont
  {Metzler}(2010{\natexlab{b}})}]{jeon2010analysis}%
  \BibitemOpen
  \bibfield  {author} {\bibinfo {author} {\bibfnamefont {J.}~\bibnamefont
  {Jeon}}\ and\ \bibinfo {author} {\bibfnamefont {R.}~\bibnamefont {Metzler}},\
  }\bibfield  {title} {\bibinfo {title} {Analysis of short subdiffusive time
  series: scatter of the time-averaged mean-squared displacement},\ }\href@noop
  {} {\bibfield  {journal} {\bibinfo  {journal} {J. Phys. A: Math. Theor.}\
  }\textbf {\bibinfo {volume} {43}},\ \bibinfo {pages} {252001} (\bibinfo
  {year} {2010}{\natexlab{b}})}\BibitemShut {NoStop}%
\bibitem [{\citenamefont {Wagner}\ \emph {et~al.}(2016)\citenamefont {Wagner},
  \citenamefont {Kroll}, \citenamefont {Wiemann},\ and\ \citenamefont
  {Lipinski}}]{wagner2016classification}%
  \BibitemOpen
  \bibfield  {author} {\bibinfo {author} {\bibfnamefont {T.}~\bibnamefont
  {Wagner}}, \bibinfo {author} {\bibfnamefont {A.}~\bibnamefont {Kroll}},
  \bibinfo {author} {\bibfnamefont {M.}~\bibnamefont {Wiemann}},\ and\ \bibinfo
  {author} {\bibfnamefont {H.}~\bibnamefont {Lipinski}},\ }\bibfield  {title}
  {\bibinfo {title} {{Classification of nanoparticle diffusion processes in
  vital cells by a multifeature random forests approach: application to
  simulated data, darkfield, and confocal laser scanning microscopy}},\ }in\
  \href {https://doi.org/10.1117/12.2227499} {\emph {\bibinfo {booktitle}
  {Biophotonics: Photonic Solutions for Better Health Care V}}},\ Vol.\
  \bibinfo {volume} {9887},\ \bibinfo {editor} {edited by\ \bibinfo {editor}
  {\bibfnamefont {J.}~\bibnamefont {Popp}}, \bibinfo {editor} {\bibfnamefont
  {V.}~\bibnamefont {Tuchin}}, \bibinfo {editor} {\bibfnamefont
  {D.}~\bibnamefont {Matthews}},\ and\ \bibinfo {editor} {\bibfnamefont
  {F.}~\bibnamefont {Pavone}}},\ \bibinfo {organization} {International Society
  for Optics and Photonics}\ (\bibinfo  {publisher} {SPIE},\ \bibinfo {year}
  {2016})\ p.\ \bibinfo {pages} {988722}\BibitemShut {NoStop}%
\bibitem [{\citenamefont {Mu{\~n}oz-Gil}\ \emph {et~al.}(2020)\citenamefont
  {Mu{\~n}oz-Gil}, \citenamefont {Garcia-March}, \citenamefont {Manzo},
  \citenamefont {Mart{\'\i}n-Guerrero},\ and\ \citenamefont
  {Lewenstein}}]{munoz-gil2020single}%
  \BibitemOpen
  \bibfield  {author} {\bibinfo {author} {\bibfnamefont {G.}~\bibnamefont
  {Mu{\~n}oz-Gil}}, \bibinfo {author} {\bibfnamefont {M.}~\bibnamefont
  {Garcia-March}}, \bibinfo {author} {\bibfnamefont {C.}~\bibnamefont {Manzo}},
  \bibinfo {author} {\bibfnamefont {J.}~\bibnamefont {Mart{\'\i}n-Guerrero}},\
  and\ \bibinfo {author} {\bibfnamefont {M.}~\bibnamefont {Lewenstein}},\
  }\bibfield  {title} {\bibinfo {title} {Single trajectory characterization via
  machine learning},\ }\href@noop {} {\bibfield  {journal} {\bibinfo  {journal}
  {New J. Phys.}\ }\textbf {\bibinfo {volume} {22}},\ \bibinfo {pages} {013010}
  (\bibinfo {year} {2020})}\BibitemShut {NoStop}%
\bibitem [{\citenamefont {Janczura}\ \emph {et~al.}(2020)\citenamefont
  {Janczura}, \citenamefont {Kowalek}, \citenamefont {Loch-Olszewska},
  \citenamefont {Szwabi{\'n}ski},\ and\ \citenamefont
  {Weron}}]{janczura2020classification}%
  \BibitemOpen
  \bibfield  {author} {\bibinfo {author} {\bibfnamefont {J.}~\bibnamefont
  {Janczura}}, \bibinfo {author} {\bibfnamefont {P.}~\bibnamefont {Kowalek}},
  \bibinfo {author} {\bibfnamefont {H.}~\bibnamefont {Loch-Olszewska}},
  \bibinfo {author} {\bibfnamefont {J.}~\bibnamefont {Szwabi{\'n}ski}},\ and\
  \bibinfo {author} {\bibfnamefont {A.}~\bibnamefont {Weron}},\ }\bibfield
  {title} {\bibinfo {title} {Classification of particle trajectories in living
  cells: {M}achine learning versus statistical testing hypothesis for
  fractional anomalous diffusion},\ }\href@noop {} {\bibfield  {journal}
  {\bibinfo  {journal} {Phys. Rev. E}\ }\textbf {\bibinfo {volume} {102}},\
  \bibinfo {pages} {032402} (\bibinfo {year} {2020})}\BibitemShut {NoStop}%
\bibitem [{\citenamefont {Loch-Olszewska}\ and\ \citenamefont
  {Szwabi{\'n}ski}(2020)}]{loch-olszewska2020impact}%
  \BibitemOpen
  \bibfield  {author} {\bibinfo {author} {\bibfnamefont {H.}~\bibnamefont
  {Loch-Olszewska}}\ and\ \bibinfo {author} {\bibfnamefont {J.}~\bibnamefont
  {Szwabi{\'n}ski}},\ }\bibfield  {title} {\bibinfo {title} {Impact of feature
  choice on machine learning classification of fractional anomalous
  diffusion},\ }\bibfield  {journal} {\bibinfo  {journal} {Entropy}\ }\textbf
  {\bibinfo {volume} {22}},\ \href {https://doi.org/10.3390/e22121436}
  {10.3390/e22121436} (\bibinfo {year} {2020})\BibitemShut {NoStop}%
\bibitem [{\citenamefont {Kowalek}\ \emph {et~al.}(2019)\citenamefont
  {Kowalek}, \citenamefont {Loch-Olszewska},\ and\ \citenamefont
  {Szwabi{\'n}ski}}]{kowalek2019classification}%
  \BibitemOpen
  \bibfield  {author} {\bibinfo {author} {\bibfnamefont {P.}~\bibnamefont
  {Kowalek}}, \bibinfo {author} {\bibfnamefont {H.}~\bibnamefont
  {Loch-Olszewska}},\ and\ \bibinfo {author} {\bibfnamefont {J.}~\bibnamefont
  {Szwabi{\'n}ski}},\ }\bibfield  {title} {\bibinfo {title} {Classification of
  diffusion modes in single-particle tracking data: {F}eature-based versus
  deep-learning approach},\ }\href@noop {} {\bibfield  {journal} {\bibinfo
  {journal} {Phys. Rev. E}\ }\textbf {\bibinfo {volume} {100}},\ \bibinfo
  {pages} {032410} (\bibinfo {year} {2019})}\BibitemShut {NoStop}%
\bibitem [{\citenamefont {Gentili}\ and\ \citenamefont
  {Volpe}(2021)}]{gentili2021characterization}%
  \BibitemOpen
  \bibfield  {author} {\bibinfo {author} {\bibfnamefont {A.}~\bibnamefont
  {Gentili}}\ and\ \bibinfo {author} {\bibfnamefont {G.}~\bibnamefont
  {Volpe}},\ }\bibfield  {title} {\bibinfo {title} {Characterization of
  anomalous diffusion classical statistics powered by deep learning
  {(CONDOR)}},\ }\href@noop {} {\bibfield  {journal} {\bibinfo  {journal} {J.
  Phys. A: Math. Theor.}\ }\textbf {\bibinfo {volume} {54}} (\bibinfo {year}
  {2021})}\BibitemShut {NoStop}%
\bibitem [{\citenamefont {Bo}\ \emph {et~al.}(2019)\citenamefont {Bo},
  \citenamefont {Schmidt}, \citenamefont {Eichhorn},\ and\ \citenamefont
  {Volpe}}]{bo2019measurement}%
  \BibitemOpen
  \bibfield  {author} {\bibinfo {author} {\bibfnamefont {S.}~\bibnamefont
  {Bo}}, \bibinfo {author} {\bibfnamefont {F.}~\bibnamefont {Schmidt}},
  \bibinfo {author} {\bibfnamefont {R.}~\bibnamefont {Eichhorn}},\ and\
  \bibinfo {author} {\bibfnamefont {G.}~\bibnamefont {Volpe}},\ }\bibfield
  {title} {\bibinfo {title} {Measurement of anomalous diffusion using recurrent
  neural networks},\ }\href@noop {} {\bibfield  {journal} {\bibinfo  {journal}
  {Phys. Rev. E}\ }\textbf {\bibinfo {volume} {100}},\ \bibinfo {pages}
  {010102} (\bibinfo {year} {2019})}\BibitemShut {NoStop}%
\bibitem [{\citenamefont {Argun}\ \emph {et~al.}(2021)\citenamefont {Argun},
  \citenamefont {Volpe},\ and\ \citenamefont {Bo}}]{argun2021classification}%
  \BibitemOpen
  \bibfield  {author} {\bibinfo {author} {\bibfnamefont {A.}~\bibnamefont
  {Argun}}, \bibinfo {author} {\bibfnamefont {G.}~\bibnamefont {Volpe}},\ and\
  \bibinfo {author} {\bibfnamefont {S.}~\bibnamefont {Bo}},\ }\bibfield
  {title} {\bibinfo {title} {Classification, inference and segmentation of
  anomalous diffusion with recurrent neural networks},\ }\href@noop {}
  {\bibfield  {journal} {\bibinfo  {journal} {J. Phys. A: Math. Theor.}\
  }\textbf {\bibinfo {volume} {54}},\ \bibinfo {pages} {294003} (\bibinfo
  {year} {2021})}\BibitemShut {NoStop}%
\bibitem [{\citenamefont {Firbas}\ \emph {et~al.}(pear)\citenamefont {Firbas},
  \citenamefont {Garibo-i Orts}, \citenamefont {Garcia-March},\ and\
  \citenamefont {Conejero}}]{firbas2022transformers}%
  \BibitemOpen
  \bibfield  {author} {\bibinfo {author} {\bibfnamefont {N.}~\bibnamefont
  {Firbas}}, \bibinfo {author} {\bibfnamefont {{\`O}.}~\bibnamefont {Garibo-i
  Orts}}, \bibinfo {author} {\bibfnamefont {M.}~\bibnamefont {Garcia-March}},\
  and\ \bibinfo {author} {\bibfnamefont {J.}~\bibnamefont {Conejero}},\
  }\bibfield  {title} {\bibinfo {title} {Characterization of anomalous
  diffusion through convolutional transformers},\ }\href@noop {} {\bibfield
  {journal} {\bibinfo  {journal} {J. Phys. A: Math. Theor.}\ } (\bibinfo {year}
  {To appear})}\BibitemShut {NoStop}%
\bibitem [{\citenamefont {Gram}(1883)}]{gram1883ueber}%
  \BibitemOpen
  \bibfield  {author} {\bibinfo {author} {\bibfnamefont {J.~P.}\ \bibnamefont
  {Gram}},\ }\bibfield  {title} {\bibinfo {title} {Ueber die entwickelung
  reeller functionen in reihen mittelst der methode der kleinsten quadrate.},\
  }\href@noop {} {\  (\bibinfo {year} {1883})}\BibitemShut {NoStop}%
\bibitem [{\citenamefont {Wang}\ and\ \citenamefont
  {Oates}(2015)}]{wang2015imaging}%
  \BibitemOpen
  \bibfield  {author} {\bibinfo {author} {\bibfnamefont {Z.}~\bibnamefont
  {Wang}}\ and\ \bibinfo {author} {\bibfnamefont {T.}~\bibnamefont {Oates}},\
  }\bibfield  {title} {\bibinfo {title} {Imaging time-series to improve
  classification and imputation},\ }in\ \href@noop {} {\emph {\bibinfo
  {booktitle} {Twenty-Fourth International Joint Conference on Artificial
  Intelligence}}}\ (\bibinfo {year} {2015})\BibitemShut {NoStop}%
\bibitem [{\citenamefont {Hong}\ \emph {et~al.}(2020)\citenamefont {Hong},
  \citenamefont {Martinez},\ and\ \citenamefont {Fajardo}}]{hong2020day-ahead}%
  \BibitemOpen
  \bibfield  {author} {\bibinfo {author} {\bibfnamefont {Y.}~\bibnamefont
  {Hong}}, \bibinfo {author} {\bibfnamefont {J.}~\bibnamefont {Martinez}},\
  and\ \bibinfo {author} {\bibfnamefont {A.}~\bibnamefont {Fajardo}},\
  }\bibfield  {title} {\bibinfo {title} {Day-ahead solar irradiation
  forecasting utilizing gramian angular field and convolutional long short-term
  memory},\ }\href {https://doi.org/10.1109/ACCESS.2020.2967900} {\bibfield
  {journal} {\bibinfo  {journal} {IEEE Access}\ }\textbf {\bibinfo {volume}
  {8}},\ \bibinfo {pages} {18741} (\bibinfo {year} {2020})}\BibitemShut
  {NoStop}%
\bibitem [{\citenamefont {Zhang}\ \emph {et~al.}(2019)\citenamefont {Zhang},
  \citenamefont {Si}, \citenamefont {Wang}, \citenamefont {Yang},\ and\
  \citenamefont {Sun}}]{Zhang2019Automated}%
  \BibitemOpen
  \bibfield  {author} {\bibinfo {author} {\bibfnamefont {G.}~\bibnamefont
  {Zhang}}, \bibinfo {author} {\bibfnamefont {Y.}~\bibnamefont {Si}}, \bibinfo
  {author} {\bibfnamefont {D.}~\bibnamefont {Wang}}, \bibinfo {author}
  {\bibfnamefont {W.}~\bibnamefont {Yang}},\ and\ \bibinfo {author}
  {\bibfnamefont {Y.}~\bibnamefont {Sun}},\ }\bibfield  {title} {\bibinfo
  {title} {Automated detection of myocardial infarction using a gramian angular
  field and principal component analysis network},\ }\href
  {https://doi.org/10.1109/ACCESS.2019.2955555} {\bibfield  {journal} {\bibinfo
   {journal} {IEEE Access}\ }\textbf {\bibinfo {volume} {7}},\ \bibinfo {pages}
  {171570} (\bibinfo {year} {2019})}\BibitemShut {NoStop}%
\bibitem [{\citenamefont {Thanaraj}\ \emph {et~al.}(2020)\citenamefont
  {Thanaraj}, \citenamefont {Parvathavarthini}, \citenamefont {Tanik},
  \citenamefont {Rajinikanth}, \citenamefont {Kadry},\ and\ \citenamefont
  {Kamalanand}}]{Palani2020Implementation}%
  \BibitemOpen
  \bibfield  {author} {\bibinfo {author} {\bibfnamefont {K.~P.}\ \bibnamefont
  {Thanaraj}}, \bibinfo {author} {\bibfnamefont {B.}~\bibnamefont
  {Parvathavarthini}}, \bibinfo {author} {\bibfnamefont {U.~J.}\ \bibnamefont
  {Tanik}}, \bibinfo {author} {\bibfnamefont {V.}~\bibnamefont {Rajinikanth}},
  \bibinfo {author} {\bibfnamefont {S.}~\bibnamefont {Kadry}},\ and\ \bibinfo
  {author} {\bibfnamefont {K.}~\bibnamefont {Kamalanand}},\ }\href@noop {}
  {\bibinfo {title} {Implementation of deep neural networks to classify eeg
  signals using gramian angular summation field for epilepsy diagnosis}}
  (\bibinfo {year} {2020}),\ \Eprint {https://arxiv.org/abs/2003.04534}
  {arXiv:2003.04534 [cs.CV]} \BibitemShut {NoStop}%
\bibitem [{\citenamefont {Xu}\ \emph {et~al.}(2020)\citenamefont {Xu},
  \citenamefont {Li}, \citenamefont {Yuan}, \citenamefont {Liu}, \citenamefont
  {Fan}, \citenamefont {Li},\ and\ \citenamefont {Sun}}]{Xu2020HumanActivity}%
  \BibitemOpen
  \bibfield  {author} {\bibinfo {author} {\bibfnamefont {H.}~\bibnamefont
  {Xu}}, \bibinfo {author} {\bibfnamefont {J.}~\bibnamefont {Li}}, \bibinfo
  {author} {\bibfnamefont {H.}~\bibnamefont {Yuan}}, \bibinfo {author}
  {\bibfnamefont {Q.}~\bibnamefont {Liu}}, \bibinfo {author} {\bibfnamefont
  {S.}~\bibnamefont {Fan}}, \bibinfo {author} {\bibfnamefont {T.}~\bibnamefont
  {Li}},\ and\ \bibinfo {author} {\bibfnamefont {X.}~\bibnamefont {Sun}},\
  }\bibfield  {title} {\bibinfo {title} {Human activity recognition based on
  gramian angular field and deep convolutional neural network},\ }\href
  {https://doi.org/10.1109/ACCESS.2020.3032699} {\bibfield  {journal} {\bibinfo
   {journal} {IEEE Access}\ }\textbf {\bibinfo {volume} {8}},\ \bibinfo {pages}
  {199393} (\bibinfo {year} {2020})}\BibitemShut {NoStop}%
\bibitem [{\citenamefont {Wickramaratne}\ and\ \citenamefont
  {Mahmud}(2021)}]{Wickramaratne2021ADeep}%
  \BibitemOpen
  \bibfield  {author} {\bibinfo {author} {\bibfnamefont {S.~D.}\ \bibnamefont
  {Wickramaratne}}\ and\ \bibinfo {author} {\bibfnamefont {M.~S.}\ \bibnamefont
  {Mahmud}},\ }\bibfield  {title} {\bibinfo {title} {A deep learning based
  ternary task classification system using gramian angular summation field in
  fnirs neuroimaging data},\ }in\ \href
  {https://doi.org/10.1109/HEALTHCOM49281.2021.9398993} {\emph {\bibinfo
  {booktitle} {2020 IEEE International Conference on E-health Networking,
  Application Services (HEALTHCOM)}}}\ (\bibinfo {year} {2021})\ pp.\ \bibinfo
  {pages} {1--4}\BibitemShut {NoStop}%
\bibitem [{\citenamefont {Liu}\ \emph {et~al.}(2022)\citenamefont {Liu},
  \citenamefont {Wang}, \citenamefont {Hu},\ and\ \citenamefont
  {Bi}}]{LIU2022Determination}%
  \BibitemOpen
  \bibfield  {author} {\bibinfo {author} {\bibfnamefont {S.}~\bibnamefont
  {Liu}}, \bibinfo {author} {\bibfnamefont {S.}~\bibnamefont {Wang}}, \bibinfo
  {author} {\bibfnamefont {C.}~\bibnamefont {Hu}},\ and\ \bibinfo {author}
  {\bibfnamefont {W.}~\bibnamefont {Bi}},\ }\bibfield  {title} {\bibinfo
  {title} {Determination of alcohols-diesel oil by near infrared spectroscop
  based on gramian angular field image coding and deep learning},\ }\href
  {https://doi.org/https://doi.org/10.1016/j.fuel.2021.122121} {\bibfield
  {journal} {\bibinfo  {journal} {Fuel}\ }\textbf {\bibinfo {volume} {309}},\
  \bibinfo {pages} {122121} (\bibinfo {year} {2022})}\BibitemShut {NoStop}%
\bibitem [{\citenamefont {Liu}\ \emph {et~al.}(2020)\citenamefont {Liu},
  \citenamefont {Hu}, \citenamefont {Yuan},\ and\ \citenamefont
  {Yang}}]{Liu2020MotionArtifact}%
  \BibitemOpen
  \bibfield  {author} {\bibinfo {author} {\bibfnamefont {X.}~\bibnamefont
  {Liu}}, \bibinfo {author} {\bibfnamefont {Q.}~\bibnamefont {Hu}}, \bibinfo
  {author} {\bibfnamefont {H.}~\bibnamefont {Yuan}},\ and\ \bibinfo {author}
  {\bibfnamefont {C.}~\bibnamefont {Yang}},\ }\bibfield  {title} {\bibinfo
  {title} {Motion artifact detection in ppg signals based on gramian angular
  field and 2-d-cnn},\ }in\ \href
  {https://doi.org/10.1109/CISP-BMEI51763.2020.9263630} {\emph {\bibinfo
  {booktitle} {2020 13th International Congress on Image and Signal Processing,
  BioMedical Engineering and Informatics (CISP-BMEI)}}}\ (\bibinfo {year}
  {2020})\ pp.\ \bibinfo {pages} {743--747}\BibitemShut {NoStop}%
\bibitem [{\citenamefont {Faouzi}\ and\ \citenamefont
  {Janati}(2020)}]{faouzi2020pyts}%
  \BibitemOpen
  \bibfield  {author} {\bibinfo {author} {\bibfnamefont {J.}~\bibnamefont
  {Faouzi}}\ and\ \bibinfo {author} {\bibfnamefont {H.}~\bibnamefont
  {Janati}},\ }\bibfield  {title} {\bibinfo {title} {pyts: A python package for
  time series classification},\ }\href {http://jmlr.org/papers/v21/19-763.html}
  {\bibfield  {journal} {\bibinfo  {journal} {J. Mach. Learn. Res.}\ }\textbf
  {\bibinfo {volume} {21}},\ \bibinfo {pages} {1} (\bibinfo {year}
  {2020})}\BibitemShut {NoStop}%
\bibitem [{\citenamefont {He}\ \emph {et~al.}(2016)\citenamefont {He},
  \citenamefont {Zhang}, \citenamefont {Ren},\ and\ \citenamefont
  {Sun}}]{he2015resnet}%
  \BibitemOpen
  \bibfield  {author} {\bibinfo {author} {\bibfnamefont {K.}~\bibnamefont
  {He}}, \bibinfo {author} {\bibfnamefont {X.}~\bibnamefont {Zhang}}, \bibinfo
  {author} {\bibfnamefont {S.}~\bibnamefont {Ren}},\ and\ \bibinfo {author}
  {\bibfnamefont {J.}~\bibnamefont {Sun}},\ }\bibfield  {title} {\bibinfo
  {title} {Deep residual learning for image recognition},\ }in\ \href@noop {}
  {\emph {\bibinfo {booktitle} {Proceedings of the IEEE conference on computer
  vision and pattern recognition}}}\ (\bibinfo {year} {2016})\ pp.\ \bibinfo
  {pages} {770--778}\BibitemShut {NoStop}%
\bibitem [{\citenamefont {Howard}\ \emph {et~al.}(2017)\citenamefont {Howard},
  \citenamefont {Zhu}, \citenamefont {Chen}, \citenamefont {Kalenichenko},
  \citenamefont {Wang}, \citenamefont {Weyand}, \citenamefont {Andreetto},\
  and\ \citenamefont {Adam}}]{howard2017mobilenet}%
  \BibitemOpen
  \bibfield  {author} {\bibinfo {author} {\bibfnamefont {A.}~\bibnamefont
  {Howard}}, \bibinfo {author} {\bibfnamefont {M.}~\bibnamefont {Zhu}},
  \bibinfo {author} {\bibfnamefont {B.}~\bibnamefont {Chen}}, \bibinfo {author}
  {\bibfnamefont {D.}~\bibnamefont {Kalenichenko}}, \bibinfo {author}
  {\bibfnamefont {W.}~\bibnamefont {Wang}}, \bibinfo {author} {\bibfnamefont
  {T.}~\bibnamefont {Weyand}}, \bibinfo {author} {\bibfnamefont
  {M.}~\bibnamefont {Andreetto}},\ and\ \bibinfo {author} {\bibfnamefont
  {H.}~\bibnamefont {Adam}},\ }\href@noop {} {\bibinfo {title} {Mobilenets:
  {E}fficient convolutional neural networks for mobile vision applications}}
  (\bibinfo {year} {2017}),\ \Eprint {https://arxiv.org/abs/1704.04861}
  {arXiv:1704.04861 [cs.CV]} \BibitemShut {NoStop}%
\bibitem [{\citenamefont {Kowalek}\ \emph {et~al.}(2022)\citenamefont
  {Kowalek}, \citenamefont {Loch-Olszewska}, \citenamefont {Łukasz Łaszczuk},
  \citenamefont {Opała},\ and\ \citenamefont
  {Szwabiński}}]{kowalek2022boosting_performance}%
  \BibitemOpen
  \bibfield  {author} {\bibinfo {author} {\bibfnamefont {P.}~\bibnamefont
  {Kowalek}}, \bibinfo {author} {\bibfnamefont {H.}~\bibnamefont
  {Loch-Olszewska}}, \bibinfo {author} {\bibnamefont {Łukasz Łaszczuk}},
  \bibinfo {author} {\bibfnamefont {J.}~\bibnamefont {Opała}},\ and\ \bibinfo
  {author} {\bibfnamefont {J.}~\bibnamefont {Szwabiński}},\ }\bibfield
  {title} {\bibinfo {title} {Boosting the performance of anomalous diffusion
  classifiers with the proper choice of features},\ }\href
  {https://doi.org/10.1088/1751-8121/ac6d2a} {\bibfield  {journal} {\bibinfo
  {journal} {Journal of Physics A: Mathematical and Theoretical}\ }\textbf
  {\bibinfo {volume} {55}},\ \bibinfo {pages} {244005} (\bibinfo {year}
  {2022})}\BibitemShut {NoStop}%
\bibitem [{\citenamefont {Gajowczyk}\ and\ \citenamefont
  {Szwabiński}(2021)}]{gajowczyk2021detecting_anomalous}%
  \BibitemOpen
  \bibfield  {author} {\bibinfo {author} {\bibfnamefont {M.}~\bibnamefont
  {Gajowczyk}}\ and\ \bibinfo {author} {\bibfnamefont {J.}~\bibnamefont
  {Szwabiński}},\ }\bibfield  {title} {\bibinfo {title} {Detection of
  anomalous diffusion with deep residual networks},\ }\bibfield  {journal}
  {\bibinfo  {journal} {Entropy}\ }\textbf {\bibinfo {volume} {23}},\ \href
  {https://doi.org/10.3390/e23060649} {10.3390/e23060649} (\bibinfo {year}
  {2021})\BibitemShut {NoStop}%
\bibitem [{\citenamefont {Bai}\ \emph {et~al.}(2018)\citenamefont {Bai},
  \citenamefont {Kolter},\ and\ \citenamefont {Koltun}}]{bai2018empirical}%
  \BibitemOpen
  \bibfield  {author} {\bibinfo {author} {\bibfnamefont {S.}~\bibnamefont
  {Bai}}, \bibinfo {author} {\bibfnamefont {J.}~\bibnamefont {Kolter}},\ and\
  \bibinfo {author} {\bibfnamefont {V.}~\bibnamefont {Koltun}},\ }\bibfield
  {title} {\bibinfo {title} {An empirical evaluation of generic convolutional
  and recurrent networks for sequence modeling},\ }\href@noop {} {\bibfield
  {journal} {\bibinfo  {journal} {arXiv preprint arXiv:1803.01271}\ } (\bibinfo
  {year} {2018})}\BibitemShut {NoStop}%
\bibitem [{\citenamefont {Granik}\ \emph {et~al.}(2019)\citenamefont {Granik},
  \citenamefont {Weiss}, \citenamefont {Nehme}, \citenamefont {Levin},
  \citenamefont {Chein}, \citenamefont {Perlson}, \citenamefont {Roichman},\
  and\ \citenamefont {Shechtman}}]{granik2019single-particle}%
  \BibitemOpen
  \bibfield  {author} {\bibinfo {author} {\bibfnamefont {N.}~\bibnamefont
  {Granik}}, \bibinfo {author} {\bibfnamefont {L.}~\bibnamefont {Weiss}},
  \bibinfo {author} {\bibfnamefont {E.}~\bibnamefont {Nehme}}, \bibinfo
  {author} {\bibfnamefont {M.}~\bibnamefont {Levin}}, \bibinfo {author}
  {\bibfnamefont {M.}~\bibnamefont {Chein}}, \bibinfo {author} {\bibfnamefont
  {E.}~\bibnamefont {Perlson}}, \bibinfo {author} {\bibfnamefont
  {Y.}~\bibnamefont {Roichman}},\ and\ \bibinfo {author} {\bibfnamefont
  {Y.}~\bibnamefont {Shechtman}},\ }\bibfield  {title} {\bibinfo {title}
  {Single-particle diffusion characterization by deep learning},\ }\href
  {https://doi.org/https://doi.org/10.1016/j.bpj.2019.06.015} {\bibfield
  {journal} {\bibinfo  {journal} {Biophys. J.}\ }\textbf {\bibinfo {volume}
  {117}},\ \bibinfo {pages} {185} (\bibinfo {year} {2019})}\BibitemShut
  {NoStop}%
\end{thebibliography}%

\end{document}